\documentclass[10pt]{IEEEtran}
\usepackage[LGR,T1]{fontenc}
\usepackage[latin9]{inputenc}
\usepackage{color}
\usepackage{amsmath}
\usepackage{amsthm}
\usepackage{amssymb}
\usepackage{graphicx}
\usepackage[unicode=true,
 bookmarks=true,bookmarksnumbered=true,bookmarksopen=true,bookmarksopenlevel=3,
 breaklinks=false,pdfborder={0 0 1},backref=false,colorlinks=true]
 {hyperref}
\hypersetup{pdftitle={Renyi},
 pdfauthor={Lei Yu},
 pdfborderstyle=,pdfborderstyle={},pdfborderstyle={},pdfborderstyle={},pdfborderstyle={},pdfborderstyle={},pdfborderstyle={},pdfborderstyle={},pdfborderstyle={},pdfborderstyle={},pdfborderstyle={},pdfborderstyle={},pdfborderstyle={},pdfborderstyle={},pdfborderstyle={},pdfborderstyle={},pdfborderstyle={},pdfborderstyle={},pdfborderstyle={},pdfborderstyle={},pdfpagelayout=OneColumn,pdfnewwindow=true,pdfstartview=XYZ,plainpages=false,linkcolor=blue,urlcolor=blue,citecolor=red,anchorcolor=blue,linkcolor=blue,urlcolor=blue,citecolor=red,anchorcolor=blue}

\makeatletter
\theoremstyle{plain}
\newtheorem{thm}{\protect\theoremname}
\theoremstyle{definition}
\newtheorem{defn}{\protect\definitionname}
\theoremstyle{plain}
\newtheorem{lem}{\protect\lemmaname}
\theoremstyle{remark}
\newtheorem{rem}{\protect\remarkname}
\theoremstyle{plain}
\newtheorem{prop}{\protect\propositionname}
\theoremstyle{definition}
\newtheorem{example}{\protect\examplename}
\theoremstyle{plain}
\newtheorem{cor}{\protect\corollaryname}
\theoremstyle{plain}
\newtheorem{conjecture}{\protect\conjecturename}
\theoremstyle{remark}
\newtheorem{claim}{\protect\claimname}

\usepackage{cite}

\allowdisplaybreaks[1]

\newcommand{\rme}{\mathrm{e}}

\newcommand{\calM}{\mathcal{M}}

\newcommand{\calX}{\mathcal{X}}

\newcommand{\bbN}{\mathbb{N}}

\DeclareMathAlphabet{\mathbsf}{OT1}{cmss}{bx}{n}
\DeclareMathAlphabet{\mathssf}{OT1}{cmss}{m}{sl}

\DeclareSymbolFont{bsfletters}{OT1}{cmss}{bx}{n}
\DeclareSymbolFont{ssfletters}{OT1}{cmss}{m}{n}
\DeclareMathSymbol{\bsfGamma}{0}{bsfletters}{'000}
\DeclareMathSymbol{\ssfGamma}{0}{ssfletters}{'000}
\DeclareMathSymbol{\bsfDelta}{0}{bsfletters}{'001}
\DeclareMathSymbol{\ssfDelta}{0}{ssfletters}{'001}
\DeclareMathSymbol{\bsfTheta}{0}{bsfletters}{'002}
\DeclareMathSymbol{\ssfTheta}{0}{ssfletters}{'002}
\DeclareMathSymbol{\bsfLambda}{0}{bsfletters}{'003}
\DeclareMathSymbol{\ssfLambda}{0}{ssfletters}{'003}
\DeclareMathSymbol{\bsfXi}{0}{bsfletters}{'004}
\DeclareMathSymbol{\ssfXi}{0}{ssfletters}{'004}
\DeclareMathSymbol{\bsfPi}{0}{bsfletters}{'005}
\DeclareMathSymbol{\ssfPi}{0}{ssfletters}{'005}
\DeclareMathSymbol{\bsfSigma}{0}{bsfletters}{'006}
\DeclareMathSymbol{\ssfSigma}{0}{ssfletters}{'006}
\DeclareMathSymbol{\bsfUpsilon}{0}{bsfletters}{'007}
\DeclareMathSymbol{\ssfUpsilon}{0}{ssfletters}{'007}
\DeclareMathSymbol{\bsfPhi}{0}{bsfletters}{'010}
\DeclareMathSymbol{\ssfPhi}{0}{ssfletters}{'010}
\DeclareMathSymbol{\bsfPsi}{0}{bsfletters}{'011}
\DeclareMathSymbol{\ssfPsi}{0}{ssfletters}{'011}
\DeclareMathSymbol{\bsfOmega}{0}{bsfletters}{'012}
\DeclareMathSymbol{\ssfOmega}{0}{ssfletters}{'012}

\DeclareMathOperator*{\argmin}{arg\,min}

\def\dotle{\mathrel{\dot{\le}}}

\def\dotge{\mathrel{\dot{\ge}}}

  \providecommand{\definitionname}{Definition}
  \providecommand{\lemmaname}{Lemma}
  \providecommand{\remarkname}{Remark}
\providecommand{\theoremname}{Theorem}

\allowdisplaybreaks[1]
\providecommand{\conjecturename}{Conjecture}
\providecommand{\corollaryname}{Corollary}

\providecommand{\examplename}{Example}

\providecommand{\propositionname}{Proposition}

\makeatother

\providecommand{\claimname}{Claim}
\providecommand{\conjecturename}{Conjecture}
\providecommand{\corollaryname}{Corollary}
\providecommand{\definitionname}{Definition}
\providecommand{\examplename}{Example}
\providecommand{\lemmaname}{Lemma}
\providecommand{\propositionname}{Proposition}
\providecommand{\remarkname}{Remark}
\providecommand{\theoremname}{Theorem}

\begin{document}
\title{Asymptotic Coupling and Its Applications in Information Theory }

\maketitle
\author{Lei Yu and Vincent Y. F. Tan, \IEEEmembership{Senior Member,~IEEE}
\thanks{Manuscript received December 21, 2017; revised July 15, 2018; accepted
July 15, 2018. This work was supported by a Singapore National Research
Foundation (NRF) National Cybersecurity R\&D Grant (R-263-000-C74-281
and NRF2015NCR-NCR003-006). The first author was also supported by
a National Natural Science Foundation of China (NSFC) under Grant
(61631017). 
This paper was presented in part at the 2018 IEEE International Symposium
on Information Theory (ISIT) \cite{yu2018coupling}.} \thanks{ L.~Yu is with the Department of Electrical and Computer Engineering,
National University of Singapore (NUS), Singapore 117583 (e-mail:
leiyu@nus.edu.sg). V.~Y.~F.~Tan is with the with the Department
of Electrical and Computer Engineering and the Department of Mathematics,
NUS, Singapore 119076 (e-mail: vtan@nus.edu.sg).} \thanks{ Communicated by I. Kontoyiannis, Associate Editor at Large. }
\thanks{Copyright (c) 2017 IEEE. Personal use of this material is permitted.
However, permission to use this material for any other purposes must
be obtained from the IEEE by sending a request to pubs-permissions@ieee.org.}}
\begin{abstract}
A coupling of two distributions $P_{X}$ and $P_{Y}$ is a joint distribution
$P_{XY}$ with marginal distributions equal to $P_{X}$ and $P_{Y}$.
Given marginals $P_{X}$ and $P_{Y}$ and a real-valued function $f$
of the joint distribution $P_{XY}$, what is its minimum over all
couplings $P_{XY}$ of $P_{X}$ and $P_{Y}$? We study the asymptotics
of such coupling problems with different $f$'s and with $X$ and
$Y$ replaced by $X^{n}=(X_{1},\ldots,X_{n})$ and $Y^{n}=(Y_{1},\ldots,Y_{n})$
where $X_{i}$ and $Y_{i}$ are i.i.d.\ copies of random variables
$X$ and $Y$ with distributions $P_{X}$ and $P_{Y}$ respectively.
These include the maximal coupling, minimum distance coupling, maximal
guessing coupling, and minimum entropy coupling problems. We characterize
the limiting values of these coupling problems as $n$ tends to infinity.
We show that they typically converge at least exponentially fast to
their limits. Moreover, for the problems of maximal coupling and minimum
excess-distance probability coupling, we also characterize (or bound)
the optimal convergence rates (exponents). Furthermore, for the maximal
guessing coupling problem we show that it is equivalent to the distribution
approximation problem. Therefore, some existing results for the latter
problem can be used to derive the asymptotics of the maximal guessing
coupling problem. We also study the asymptotics of the maximal guessing
coupling problem for two \emph{general} sources and a generalization
of this problem, named the \emph{maximal guessing coupling through
a channel problem}. We apply the preceding results to several new
information-theoretic problems, including exact intrinsic randomness,
exact resolvability, channel capacity with input distribution constraint,
and perfect stealth and secrecy communication. 
\end{abstract}


\section{\label{sec:Introduction}Introduction}

A coupling of two probability distributions $P_{X}$ and $P_{Y}$
is a joint distribution $P_{XY}$ such that the marginals on ${\cal X}$
and ${\cal Y}$ are $P_{X}$ and $P_{Y}$ respectively. Given two
marginal distributions $P_{X}$ and $P_{Y}$ and a function $f$ of
the joint distribution $P_{XY}$, what is the minimum of $f(P_{XY})$
over all couplings $P_{XY}$ of $P_{X}$ and $P_{Y}$? This problem
has been studied for different functions $f$ \cite{thorisson2000coupling,villani2008optimal,kovavcevic2015entropy,steinberg1996simulation}.
In this work, we investigate asymptotics of several coupling problems
for product marginal distributions $P_{X^{n}}=P_{X}^{n}$ and $P_{Y^{n}}=P_{Y}^{n}$,
when the dimension of the distributions $n$ tends to infinity. These
problems include the maximal coupling problem, the minimum distance
coupling problem, the maximal guessing coupling problem, and the minimum
entropy coupling problem (or the maximum mutual information coupling
problem). Our results have several applications in information theory,
including the following: 

{[}leftmargin={*}{]} 
\begin{enumerate}
\item Exact intrinsic randomness: The \emph{intrinsic randomness} is the
problem of determining the amount of randomness contained in a source
\cite{vembu1995generating}. Given an arbitrary general source $\boldsymbol{X}=\left\{ X^{n}\right\} _{n=1}^{\infty}$
(usually called the coin source), we try to approximate, by using
$\boldsymbol{X}=\left\{ X^{n}\right\} _{n=1}^{\infty}$, a uniform
random number with as large rates as possible. Vembu and Verdú \cite{vembu1995generating}
and Han \cite{Han03} determined the supremum of achievable uniform
random number generation rates, by invoking the information spectrum
method. In this paper, we consider a new variation of this problem,
named the \emph{exact intrinsic randomness}. We require the output
to be \emph{exactly} a uniform random number. Since in general there
is no function satisfying such a requirement, we relax the mapping
to be an \emph{asymptotic} function (i.e., the mapping asymptotically
almost surely approaches some target function as the blocklength tends
to infinity; see Definition \ref{def:asymptoticfunction}), instead
of a function. 
\item Exact resolvability: The {\em channel resolvability problem} is
the problem of determining how much information is needed to simulate
a random process through a given channel so that it approximates a
target output distribution. This problem was first studied by Han
and Verdú \cite{Han}. In \cite{Han}, the total variation (TV) distance
and the normalized relative entropy (Kullback-Leibler divergence)
were used to measure the level of approximation. The resolvability
problem with the \emph{unnormalized} relative entropy was studied
by Hayashi \cite{Hayashi06,Hayashi11}. Recently, Liu, Cuff, and Verdú
\cite{Liu} and Yu and Tan \cite{Yu} extended the theory of resolvability
by respectively using the so-called $E_{\gamma}$ metric with $\gamma\geq1$
and various Rényi divergences to measure the level of approximation.
In this paper, we define a new variation of the channel resolvability
problem, named\emph{ exact channel resolvability}. We now require
the output to \emph{exactly} match the target distribution. Again
since in general there is no function satisfying such requirement,
we relax the mapping to be an asymptotic function. A related problem
named \emph{exact common information} was studied by Kumar, Li, and
Gamal \cite{kumar2014exact}, where differently from our definition,
they required the mapping to be a function and variable-length codes
were allowed. For their problem, to obtain the exact output distribution,
the input, in general, does not follow the uniform distribution. Hence
Kumar, Li, and Gamal's definition is input-distribution sensitive,
in contrast to our definition here. 
\item Perfect stealth and secrecy communication: In \cite{hou2014effective},
Hou and Kramer defined a new security measure---\emph{effective secrecy}---for
wiretap channels that incorporates into its framework not only reliability
and secrecy but also {\em stealth}. The signal overheard by the
eavesdropper from her channel is forced to be close to a target distribution
(i.e., the output distribution of the channel when there is no useful
information transmitted). Hou and Kramer used ideas from channel resolvability
to study the \emph{effective secrecy capacity}{} (the maximum rate
which can be transmitted in a stealthy, secret, and reliable way)
of wiretap channels, where they used the relative entropy to measure
the level of secrecy and stealth. Furthermore, if we set the target
distribution as the channel output distribution induced by some fixed
channel input $x_{0}$ (the channel input symbol when the channel
is idle), then the communication problem with stealth reduces to the
so-called {\em covert communication problem}. In the covert communication
problem, a sender Alice wishes to reliably transmit a message to a
receiver Bob over a wiretap channel, while simultaneously ensuring
that her transmission cannot be detected by an eavesdropper Eve, who
observes the transmitted signal through the wiretap channel. Most
researchers focused on the regime that Eve is\emph{ asymptotically}{}
unable to detect the transmission, i.e., the probability of detection
vanishes as the blocklength tends to infinity. For such a scenario,
Bash \emph{et al.}{} \cite{bash2012limits,bash2015quantum}, Wang
\emph{et al.}{} \cite{wang2016fundamental}, and Bloch \cite{bloch2016covert}
showed that for Gaussian or discrete memoryless wiretap channels the
number of bits that can be reliably and covertly transmitted over
$n$ channel uses scales as $\Theta(\sqrt{n})$, as long as the no-input
symbol is not redundant, i.e., the output distribution at the eavesdropper
induced by the no-input symbol is not a mixture of the output distributions
induced by other input symbols. This is colloquially known as the
``square root law''. On the other hand, if the no-input symbol is
redundant, and the secret key length shared by Alice and Bob is sufficiently
long, then the number of bits that can be reliably and covertly transmitted
over $n$ channel uses linearly increases as $n$ goes to infinity
\cite{wang2016fundamental,bloch2016covert}. In contrast to Hou and
Kramer's work \cite{hou2014effective}, we generalize the effective
secrecy problem by forcing the channel output to \emph{exactly}{}
match the target distribution rather than approximately. Hence, the
problem studied here can be termed as a \emph{perfectly}{} stealthy
and secret communication problem. Furthermore, if we set the target
distribution to be the channel output distribution induced by a channel
input fixed to be $x_{0}$, then our problem reduces to the \emph{perfectly}{}
covert and secret communication problem. 
\end{enumerate}
Furthermore, maximal couplings have been widely studied in probability
theory and information theory; see, e.g., \cite{strassen1965existence,zhang2007estimating,marton1966simple,sason2013entropy,prelov2015coupling}
and references therein. The main difference between our work and these
works is that we consider the asymptotic scenario when $X$ and $Y$
are replaced by $X^{n}=(X_{1},\ldots,X_{n})$ and $Y^{n}=(Y_{1},\ldots,Y_{n})$
where $X_{i}$ and $Y_{i}$ are i.i.d.\ copies of random variables
$X$ and $Y$ with distributions $P_{X}$ and $P_{Y}$ respectively
and $n$ tends to infinity. In all these papers, the authors consider
the finite length (typically one-shot) case. 
Furthermore, most of these works are only concerned with {\em maximal
couplings}, i.e., couplings that maximize $\mathbb{P}\left\{ X=Y\right\} $
whereas we are interested in several more general functionals of $P_{XY}$.
Besides these works, \cite{steinberg1996simulation} used several
distance measures between distributions to study the source resolvability
problem (and also the source coding problem), where the definitions
of those measures involve optimization over couplings. In the source
resolvability problem, the target distribution is fixed but the generated
(code-induced) distribution is not. Hence one of the marginal distributions
of couplings in the optimization problems involved in \cite{steinberg1996simulation}
is fixed, but the other marginal distribution is not fixed. However,
in this paper, {\em both} of the marginal distributions are fixed.

\subsection{Main Contributions}

Our main contributions are as follows: 

{[}leftmargin={*}{]} 
\begin{enumerate}
\item We study the asymptotics of several coupling problems, including the
problems of maximal coupling, minimum distance coupling, maximal guessing
coupling, and minimum entropy coupling (or maximum mutual information
coupling). We characterize the limiting values of these coupling problems
as the dimension goes to infinity. We show that they typically converge
at least exponentially fast to their limits. Moreover, for the maximal
coupling and minimum excess-distance probability coupling problems,
we also characterize the optimal convergence rates of these two coupling
problems. Interestingly, product couplings achieve the optimal limiting
values of these coupling problems, but they cannot achieve the optimal
convergence rates. Hence, for these two problems, non-product couplings
strictly outperform product couplings in the exponent sense. Furthermore,
we show that the maximal guessing coupling problem is equivalent to
the traditional distribution approximation problem \cite[Sec. 2.1]{Han03}.
Therefore, some existing results on the latter problem can be used
to derive asymptotic results on the former problem. 
\item We also consider the asymptotics of the maximal guessing coupling
problem for two \emph{general} sources and a generalization of this
problem, named as the \emph{maximal guessing coupling through a channel
problem}. We derive upper and lower bounds on the fundamental limits
of these two problems. As a by-product, these upper bounds and lower
bounds are also bounds on the fundamental limits of the general source-channel
resolvability problem, in which the source and channel are general
and the source is a part of the channel input. 
\item We apply the preceding results to several novel information-theoretic
problems, including the exact intrinsic randomness, exact resolvability,
channel capacity with input distribution constraint, and perfect stealth
and secrecy communication problems. For the exact intrinsic randomness
and exact source resolvability problems, we show that they are respectively
equivalent to the traditional (approximate) intrinsic randomness and
source resolvability problems. For the exact resolvability problem,
we completely characterize the optimal rate for full-rank channels.
For the problem of channel capacity with an input distribution constraint,
we show that the channel capacity under condition that the input distribution
is constrained to be some product distribution is the Gács-Körner
common information between the channel input and the channel output.
For perfect stealth and secrecy communication, we show that 1) the
perfect stealth-secrecy capacity is positive if and only if the wiretap
channel is a $P_{Z}$-redundant channel; 2) for full-rank wiretap
channels, the perfect stealth-secrecy capacity is zero, and the perfect
stealth/covertness capacity (the maximum rate can be transmitted in
the perfectly stealthy or covert way) is the Gács-Körner common information
$C_{\mathsf{GK}}(X;Y)$, where $P_{X}$ is the unique distribution
that induces $P_{Z}$ through $P_{Z|X}$. 
\end{enumerate}
Our initial motivation of studying these coupling problems stems from
the fact that perfect stealth and secrecy communication problems are
of great practical significance. We show that the maximal guessing
coupling problem is of crucial importance to solving these problems
communication problems. Furthermore, as by-products of applying our
results on coupling problems to the perfect stealth and secrecy communication
problem, we also obtain some intermediate and interesting results,
e.g., the channel capacity with input distribution constraint problem,
the exact intrinsic randomness problem, and the exact resolvability
problem.

\subsection{Notation }

We use $P_{X}(x)$ to denote the probability distribution of a random
variable $X$, which is also shortly denoted as $P(x)$ (when the
random variable $X$ is clear from the context). We also use $P_{X}$,
$\widetilde{P}_{X}$, and $Q_{X}$ to denote various probability distributions
with alphabet $\mathcal{X}$. The set of probability distributions
on $\mathcal{X}$ is denoted as $\mathcal{P}\left(\mathcal{X}\right)$,
and the set of conditional probability distributions on $\mathcal{Y}$
given a variable in $\mathcal{X}$ is denoted as $\mathcal{P}\left(\mathcal{Y}|\mathcal{X}\right):=\left\{ P_{Y|X}:P_{Y|X}\left(\cdot|x\right)\in\mathcal{P}\left(\mathcal{Y}\right),x\in\mathcal{X}\right\} $.
Given $P_{X}$ and $P_{Y|X}$, we write $[P_{Y|X}\circ P_{X}](y):=\sum_{x}P_{Y|X}(y|x)P_{X}(x)$.
For simplicity, all the alphabets involved in this paper are assumed
to be finite, unless stated explicitly.

We use $T_{x^{n}}\left(x\right):=\frac{1}{n}\sum_{i=1}^{n}1\left\{ x_{i}=x\right\} $
to denote the type (empirical distribution) of a sequence $x^{n}$,
and $T_{X}$ to denote a type of sequences in $\mathcal{X}^{n}$,
where the indicator function $1\{A\}$ equals $1$ if the clause $A$
is true and $0$ otherwise. For a type $T_{X}$, the type class (set
of sequences having the same type $T_{X}$) is denoted by $\mathcal{T}(T_{X})$.
The set of types of sequences in $\mathcal{X}^{n}$ is denoted as
$\mathcal{P}_{n}\left(\mathcal{X}\right):=\left\{ T_{x^{n}}:x^{n}\in\mathcal{X}^{n}\right\} $.
The $\epsilon$-typical set relative to $Q_{X}$ is denoted as $\mathcal{T}_{\epsilon}^{n}\left(Q_{X}\right):=\left\{ x^{n}\in\mathcal{X}^{n}:\left|T_{x^{n}}\left(x\right)-Q_{X}\left(x\right)\right|\leq\epsilon Q_{X}\left(x\right),\forall x\in\mathcal{X}\right\} $.
For brevity, we sometimes write $\mathcal{T}_{\epsilon}^{n}\left(Q_{X}\right)$
as $\mathcal{T}_{\epsilon}^{n}$. Other notation generally follow
the book by Csiszár and Körner~\cite{Csi97}.

The total variation distance between two probability mass functions
$P$ and $Q$ with a common alphabet $\calX$ is defined by 
\begin{equation}
|P-Q|:=\frac{1}{2}\sum_{x\in\calX}|P(x)-Q(x)|.
\end{equation}
By the definition of $\epsilon$-typical set, we have that for any
$x^{n}\in\mathcal{T}_{\epsilon}^{n}\left(Q_{X}\right)$, $\left|T_{x^{n}}-Q_{X}\right|\leq\frac{\epsilon}{2}$.

We use $\boldsymbol{P}_{X}$ or $\boldsymbol{P}_{Y|X}$ to denote
the vector or matrix form of $P_{X}$ or $P_{Y|X}$. We use $\boldsymbol{P}^{\otimes n}$
to denote $n$-fold Kronecker product of a vector or matrix $\boldsymbol{P}$.

We use $\boldsymbol{Z}=\{Z^{n}\}_{n=1}^{\infty}$ to denote a general
source, and $P_{\boldsymbol{Y}|\boldsymbol{X}}=\{P_{Y^{n}|X^{n}}\}_{n=1}^{\infty}$
to denote a general channel \cite{Han03}. For any given sequence
of random variables $\{Z_{n}\}_{n=1}^{\infty}$, we introduce quantities
which play an important role in \emph{information spectrum analysis}
\cite{Han03}. For $\delta\in[0,1]$, the $\delta$-limit superior
in probability is defined as 
\begin{align}
\delta\mbox{-}\mathrm{p}\mbox{-}\limsup_{n\rightarrow\infty}Z_{n} & \!:=\!\inf\left\{ \alpha:\limsup_{n\rightarrow\infty}\mathbb{P}\{Z_{n}\!>\!\alpha\}\!\le\!\delta\right\} .\label{eq:quantity1}
\end{align}
For $\delta=0$, 
\begin{equation}
\mathrm{p}\mbox{-}\limsup_{n\rightarrow\infty}Z_{n}:=0\mbox{-}\mathrm{p}\mbox{-}\limsup_{n\rightarrow\infty}Z_{n}
\end{equation}
and 
\begin{equation}
\mathrm{p}\mbox{-}\liminf_{n\rightarrow\infty}Z_{n}:=-\mathrm{p}\mbox{-}\limsup_{n\rightarrow\infty}(-Z_{n}).
\end{equation}
Furthermore, $\imath_{X;Y}(x;y):=\log\frac{P_{Y|X}(y|x)}{P_{Y}(y)}$
is the information density\footnote{Unless explicitly stated, the logarithm base can be chosen arbitrarily.
But regardless of the base, $\exp(x)$ or $e^{x}$ always denotes
the inverse of $\log(x)$. }, and $\imath_{X}(x):=\imath_{X;X}(x;x)=\log\frac{1}{P_{X}(x)}$ is
the entropy density. We define the sup- and inf-entropy rates respectively
as 
\begin{align}
\overline{H}(\boldsymbol{Z}) & :=\mathrm{p}\mbox{-}\limsup_{n\rightarrow\infty}\frac{1}{n}\imath_{Z^{n}}(Z^{n})\quad\mbox{and}\\
\underline{H}(\boldsymbol{Z}) & :=\mathrm{p}\mbox{-}\liminf_{n\rightarrow\infty}\frac{1}{n}\imath_{Z^{n}}(Z^{n}).
\end{align}
Finally, we write $f(n)\dotle g(n)$ if $\limsup_{n\to\infty}\frac{1}{n}\log\frac{f(n)}{g(n)}\le0$.
In addition, $f(n)\doteq g(n)$ if and only if $f(n)\dotle g(n)$
and $g(n)\dotle f(n)$.

\subsection{Preliminaries}
\begin{defn}
The \emph{set of couplings} of $P_{X}\in\mathcal{P}\left(\mathcal{X}\right)$
and $P_{Y}\in\mathcal{P}\left(\mathcal{Y}\right)$ is defined as 
\begin{equation}
C(P_{X},P_{Y}):=\left\{ Q_{XY}\in\mathcal{P}\left(\mathcal{X}\times\mathcal{Y}\right):Q_{X}=P_{X},Q_{Y}=P_{Y}\right\} 
\end{equation}
Any $Q_{XY}\in C(P_{X},P_{Y})$ is called a \emph{coupling} of $P_{X},P_{Y}$. 
\end{defn}
\begin{defn}
The \emph{maximal equality-probability }over couplings of two distributions
$P_{X},P_{Y}\in\mathcal{P}\left(\mathcal{X}\right)$ is defined as
\begin{equation}
\mathcal{M}(P_{X},P_{Y}):=\max_{P_{XY}\in C(P_{X},P_{Y})}\mathbb{P}\left\{ Y=X\right\} .
\end{equation}
Any $Q_{XY}\in C(P_{X},P_{Y})$ achieving $\mathcal{M}(P_{X},P_{Y})$
is called a \emph{maximal coupling} of $P_{X},P_{Y}$.

The maximal coupling problem has the following property. 
\end{defn}
\begin{lem}[Maximal Coupling Equality]
\cite{thorisson2000coupling}\label{lem:MaxCoupling} Given two distributions
$P_{X}$ and $P_{Y}$, we have 
\begin{equation}
\mathcal{M}(P_{X},P_{Y})=1-|P_{X}-P_{Y}|.
\end{equation}
\end{lem}
Assume $P_{X},Q_{X}$ are two distributions defined on a set $\mathcal{X}$.
If $P_{X}=Q_{X}$, then obviously, $|P_{X}^{n}-Q_{X}^{n}|=0$ for
all $n\in\bbN$. If $P_{X}\neq Q_{X}$, the following lemma holds. 
\begin{lem}[Asymptotics of Total Variation]
\cite[Theorem 11.9.1]{Cover}\label{lem:AsymptoticsTV} Assume $P_{X},Q_{X}$
are two distinct distributions defined on a set $\mathcal{X}$. Then
$|P_{X}^{n}-Q_{X}^{n}|\to1$ exponentially fast as $n\to\infty$.
More explicitly, the exponent is 
\begin{align}
 & \lim_{n\to\infty}-\frac{1}{n}\log\left(1-|P_{X}^{n}-Q_{X}^{n}|\right)\nonumber \\
 & =\min_{R_{X}\in\mathcal{P}\left(\mathcal{X}\right)}\max\left\{ D(R_{X}\|P_{X}),D(R_{X}\|Q_{X})\right\} \\
 & =B(P_{X},Q_{X}),\label{eq:-44}
\end{align}
where 
\begin{equation}
B(P_{X},Q_{X}):=\max_{0\le\lambda\le1}\left\{ -\log\left(\sum_{x}P_{X}(x)^{\lambda}Q_{X}(x)^{1-\lambda}\right)\right\} \label{eq:-16}
\end{equation}
denotes the Chernoff information between $P_{X}$ and $Q_{X}$. 
\end{lem}
\begin{rem}
Equality \eqref{eq:-44} is justified by the fact that on the one
hand, $1-|P_{X}^{n}-Q_{X}^{n}|$ is the smallest sum of type-I and
type-II error probabilities for a binary hypothesis test between $P_{X}^{n}$
and $Q_{X}^{n}$ (see, for example, \cite[Theorem 13.1.1]{lehmann2006testing});
on the other hand, $B(P_{X},Q_{X})$ is the exponent of this sum of
two error probabilities \cite[Theorem 11.9.1]{Cover}. 
\end{rem}

\section{Maximal Coupling and Minimum Distance Coupling }

In this section, we focus on asymptotic behaviors of two basic coupling
problems: the maximal coupling problem and the minimum distance coupling
problem.

\subsection{Maximal Coupling }

We first consider the asymptotic behavior of maximal equality-probability
$\mathcal{M}(P_{X}^{n},P_{Y}^{n})$. First, it is obvious that if
$P_{X}=P_{Y}$, then $\mathcal{M}(P_{X}^{n},P_{Y}^{n})=1$ for all
$n\in\bbN$. Furthermore, the optimal coupling for this case is $P_{XY}(x,y)=P_{X}(x)1\{y=x\}$.
On the other hand, if $P_{X}\neq P_{Y}$, we have the following theorem. 
\begin{prop}[Maximal Coupling]
\label{prop:MaxCoupling} Assume $P_{X},P_{Y}$ are two distinct
distributions defined on a set $\mathcal{X}$. Then given product
marginal distributions $P_{X}^{n}$ and $P_{Y}^{n}$, we have $\mathcal{M}(P_{X}^{n},P_{Y}^{n})\to0$
exponentially fast as $n\to\infty$. More explicitly, the exponent
is 
\begin{align}
 & \lim_{n\to\infty}-\frac{1}{n}\log\mathcal{M}(P_{X}^{n},P_{Y}^{n})\nonumber \\
 & =\min_{Q}\max\left\{ D(Q\|P_{X}),D(Q\|P_{Y})\right\} \\
 & =B(P_{X},P_{Y}),
\end{align}
where $B(P_{X},P_{Y})$ is defined in \eqref{eq:-16}. 
\end{prop}
\begin{IEEEproof}
We prove this lemma by using a property of the TV distance. According
to the maximal coupling equality (Lemma~\ref{lem:MaxCoupling}) and
Lemma~\ref{lem:AsymptoticsTV}, we have 
\begin{align}
\mathcal{M}(P_{X}^{n},P_{Y}^{n}) & =1-|P_{X}^{n}-P_{Y}^{n}|\\
 & \doteq e^{-n\min_{Q\in\mathcal{P}\left(\mathcal{X}\right)}\max\left\{ D(Q\|P_{X}),D(Q\|P_{Y})\right\} }.
\end{align}
Hence, the optimal exponent is given by $\min_{Q\in\mathcal{P}\left(\mathcal{X}\right)}\max\left\{ D(Q\|P_{X}),D(Q\|P_{Y})\right\} =B(P_{X},P_{Y})$.

For a product coupling $P_{X^{n}Y^{n}}=P_{XY}^{n}$ with $P_{XY}$
achieving $\mathcal{M}(P_{X},P_{Y})$, we have 
\begin{align}
\mathbb{P}\left\{ Y^{n}=X^{n}\right\}  & =\mathbb{P}\left\{ Y=X\right\} ^{n}.
\end{align}
Hence the best exponent for product couplings is $-\log\mathcal{M}(P_{X},P_{Y})=-\log\left(1-|P_{X}-P_{Y}|\right).$ 
\end{IEEEproof}
Note that a product coupling $P_{X^{n}Y^{n}}=P_{XY}^{n}$ with $P_{XY}$
achieving $\mathcal{M}(P_{X},P_{Y})$ only achieves the exponent $-\log\mathcal{M}(P_{X},P_{Y})=-\log\left(1-|P_{X}-P_{Y}|\right)$,
which is suboptimal in general, i.e., 
\begin{equation}
B(P_{X},P_{Y})\leq-\log\left(1-|P_{X}-P_{Y}|\right).\label{eq:}
\end{equation}
The following example shows the inequality in \eqref{eq:} can be
strict. 
\begin{example}
$P_{X}=\{\frac{1}{2},\frac{1}{2}\},P_{Y}=\{\frac{1}{4},\frac{3}{4}\}$
then 
\begin{align}
 & \min_{Q}\max\left\{ D(Q\|P_{X}),D(Q\|P_{Y})\right\} \nonumber \\
 & \leq D(P_{X}\|P_{Y})=\frac{1}{2}\log_{2}\frac{4}{3}\\
 & <-\log\left(1-|P_{X}-P_{Y}|\right)=\log_{2}\frac{4}{3}.
\end{align}
\end{example}

\subsection{Minimum Distance Coupling -- Transportation Theory }

Next we consider the minimum (expected) distance coupling problem,
which is the main problem studied in transportation theory. The Wasserstein
metric is a special case of this coupling problem by specializing
the distance measure to be the quadratic distortion measure.

Define an additive function (general distance or distortion) 
\begin{equation}
d(x^{n},y^{n}):=\frac{1}{n}\sum_{i=1}^{n}d(x_{i},y_{i})
\end{equation}
where $d(x,y)$ is some arbitrary function (distance) of $x,y$. 
\begin{defn}
The \emph{minimum (expected) distance }over couplings of two distributions
$P_{X},P_{Y}$ is defined as 
\begin{equation}
\mathcal{D}(P_{X},P_{Y}):=\min_{P_{XY}\in C(P_{X},P_{Y})}\mathbb{E}d(X,Y).
\end{equation}
Any $Q_{XY}\in C(P_{X},P_{Y})$ achieving $\mathcal{D}(P_{X},P_{Y})$
is called a \emph{minimum (expected) distance coupling} of $P_{X},P_{Y}$. 
\end{defn}
Then given two marginal product distributions $P_{X}^{n}$ and $P_{Y}^{n}$,
the minimum expected distance over couplings of $P_{X},P_{Y}$ is
clearly 
\begin{equation}
\mathcal{D}(P_{X}^{n},P_{Y}^{n})=\mathcal{D}(P_{X},P_{Y}).
\end{equation}
Next we consider another important coupling problem. 
\begin{defn}
The \emph{minimum excess-distance probability }over couplings of two
distributions $P_{X},P_{Y}$ is defined as 
\begin{equation}
\widetilde{\mathcal{D}}_{d}(P_{X},P_{Y}):=\min_{P_{XY}\in C(P_{X},P_{Y})}\mathbb{P}\left\{ d(X,Y)>d\right\} .
\end{equation}
Any $Q_{XY}\in C(P_{X},P_{Y})$ achieving $\widetilde{\mathcal{D}}_{d}(P_{X},P_{Y})$
is called a \emph{minimum excess-distance probability coupling} of
$P_{X},P_{Y}$. 
\end{defn}
The excess-distance probability (or excess-distortion probability)
is an important distortion measure in information theory \cite{steinberg1996simulation,Han03}.
Define the exponents as 
\begin{equation}
\overline{\mathsf{E}}(d):=\liminf_{n\to\infty}-\frac{1}{n}\log\left(1-\widetilde{\mathcal{D}}_{d}(P_{X}^{n},P_{Y}^{n})\right)
\end{equation}
and 
\begin{equation}
\mathsf{E}(d):=\liminf_{n\to\infty}-\frac{1}{n}\log\widetilde{\mathcal{D}}_{d}(P_{X}^{n},P_{Y}^{n}).
\end{equation}
An asymptotic result for the problem of minimum excess-distance probability
coupling is stated in the following theorem. The proof is provided
in Appendix \ref{sec:MinExcessDistProbCoupling}. 
\begin{prop}[Minimum Excess-Distance Probability Coupling]
\label{prop:MinExcessDistProbCoupling} Given two distributions $P_{X}$
and $P_{Y}$, we have: 

{[}leftmargin={*}{]} 
\begin{enumerate}
\item If $\mathcal{D}(P_{X},P_{Y})>d$, then $\widetilde{\mathcal{D}}_{d}(P_{X}^{n},P_{Y}^{n})\to1$
exponentially fast as $n\to\infty$. Moreover, we have 
\begin{equation}
\overline{\mathsf{E}}(d)=\min_{Q_{XY}:\mathbb{E}_{Q}d(X,Y)\leq d}\max\left\{ D(Q_{X}\|P_{X}),D(Q_{Y}\|P_{Y})\right\} .
\end{equation}
\item If $\mathcal{D}(P_{X},P_{Y})<d$, then $\widetilde{\mathcal{D}}_{d}(P_{X}^{n},P_{Y}^{n})\to0$
at least exponentially fast as $n\to\infty$. Moreover, we have 
\begin{equation}
\mathsf{E}(d)\geq\max_{t\geq0}\left(td-\log\mathbb{E}e^{td(X,Y)}\right).
\end{equation}
\item If $\mathcal{D}(P_{X},P_{Y})=d$, then $\frac{1}{2}+O\big(\frac{1}{\sqrt{n}}\big)\leq\widetilde{\mathcal{D}}_{d}(P_{X}^{n},P_{Y}^{n})\leq1$. 
\end{enumerate}
\end{prop}
\begin{rem}
In Statement 1) of Proposition \ref{prop:MinExcessDistProbCoupling},
the exponent is infinity if $\inf\big\{ d:\widetilde{\mathcal{D}}_{d}(P_{X},P_{Y})=1\big\}\le d$. 
\end{rem}
\begin{rem}
If $\mathcal{D}(P_{X},P_{Y})>d$, then an optimal product coupling
$P_{X^{n}Y^{n}}=P_{XY}^{n}$ with $P_{XY}$ achieving $\widetilde{\mathcal{D}}_{d}(P_{X},P_{Y})$
only achieves the exponent 
\begin{equation}
\max_{t\geq0}\left(-td-\log\mathbb{E}e^{-td(X,Y)}\right)\leq\overline{\mathsf{E}}(d).
\end{equation}
If $\mathcal{D}(P_{X},P_{Y})=d$, then such an optimal product coupling
achieves the lower bound $\frac{1}{2}+O\big(\frac{1}{\sqrt{n}}\big)$. 
\end{rem}

\section{Maximal Guessing Coupling }

For the maximal coupling and minimum distance coupling problems, we
showed that product couplings suffice to achieve the optimal limiting
values of maximal equality-probability and minimum excess-distance
probability (although they cannot achieve the optimal exponents).
In the following, we consider several coupling problems for which
product couplings are not optimal in achieving the optimal limiting
values.

\subsection{\label{subsec:Maximal-Guessing-Coupling}Maximal Guessing Coupling:
Memoryless Sources}

Next we define a new coupling problem, named the maximal guessing
coupling problem. 
\begin{defn}
\label{def:maxguessing} The \emph{maximal guessing probability }over
couplings of $P_{X},P_{Y}$ is defined as 
\begin{equation}
\mathcal{G}(P_{X},P_{Y}):=\max_{P_{XY}\in C(P_{X},P_{Y})}\max_{f:{\cal X}\to{\cal Y}}\mathbb{P}\left\{ Y=f(X)\right\} .\label{eq:-45}
\end{equation}
Any $Q_{XY}\in C(P_{X},P_{Y})$ achieving $\mathcal{G}(P_{X},P_{Y})$
is called a \emph{maximal guessing coupling} of $P_{X},P_{Y}$. Moreover,
if a maximal guessing coupling satisfies $\mathcal{G}(P_{X},P_{Y})=1$,
then we call it \emph{deterministic coupling}. Given a sequence of
distribution pairs $(P_{X^{n}},P_{Y^{n}})$, if a sequence of maximal
guessing couplings $\{Q_{X^{n},Y^{n}}\}_{n\in\bbN}$ satisfies $\mathcal{G}(P_{X^{n}},P_{Y^{n}})\rightarrow1$
as $n\to\infty$, then $\{Q_{X^{n},Y^{n}}\}_{n\in\bbN}$ is called
an \emph{asymptotically deterministic coupling.}

Besides, we introduce a new concept, named the \emph{asymptotic function}. 
\end{defn}

\begin{defn}
\label{def:asymptoticfunction} We say $Y^{n}$ is an\emph{ asymptotic
function }of $X^{n}$ if $\lim_{n\to\infty}\mathbb{P}\left\{ Y^{n}=f_{n}(X^{n})\right\} =1$
for some sequence of functions $\{f_{n}\}_{n=1}^{\infty}$. 
\end{defn}
Hence under the asymptotically deterministic coupling $\{Q_{X^{n},Y^{n}}\}_{n\in\bbN}$,
$Y^{n}$ is an asymptotic function of $X^{n}$. Furthermore, the quantity
$\max_{f}\mathbb{P}\left\{ Y=f(X)\right\} $ is called the \emph{guessing
probability}; see \cite{ben1978renyi,berens2013conditional,yu2017shannon,issa2017measuring}.
Note that here and also in these papers, the guessing terminal is
only allowed to guess once; however, in \cite{merhav1999shannon,arikan1998guessing,Schieler,yu2017source,sason2018improved}
it is allowed to guess multiple times.

The deterministic coupling and asymptotically deterministic coupling
are closely related to the distribution matching problem \cite{schulte2016constant,schulte2017divergence},
which is the following. Given a sequence of distribution pairs $(P_{X^{n}},P_{Y^{n}})$,
find a sequence of distributions $P_{W^{n}}$ and a sequence of deterministic
couplings of $(P_{W^{n}},P_{Y^{n}})$ such that $P_{W^{n}}$ and $P_{X^{n}}$
are asymptotically equal under a normalized or unnormalized divergence
measure. If we loosen the requirement to finding a sequence of \emph{asymptotically}
deterministic couplings, and strengthen the constraint on the closeness
of $P_{W^{n}}$ and $P_{X^{n}}$ to be the equality $P_{W^{n}}=P_{X^{n}}$,
then the distribution matching problem becomes the asymptotically
deterministic coupling problem. That is, given a sequence of distribution
pairs $(P_{X^{n}},P_{Y^{n}})$, we would like to find a sequence of
couplings of $(P_{X^{n}},P_{Y^{n}})$ such that $\mathcal{G}(P_{X^{n}},P_{Y^{n}})\rightarrow1$
as $n\to\infty$. Furthermore, our results concerning maximal guessing
couplings or asymptotically deterministic couplings will be applied
to information-theoretic problems in Sections \ref{sec:Exact-Intrinsic-Randomness}--\ref{sec:Perfect-Stealth-}.

By the maximal coupling equality (Lemma \ref{lem:MaxCoupling}), we
can prove the following property of maximal guessing coupling, which
shows the equivalence between the maximal guessing coupling problem
and distribution approximation problem~\cite{Han03}. 
\begin{defn}
\label{def:DefineRenyiCoupling}\cite{berens2013conditional} Define
the \emph{minimum $\alpha$-Rényi conditional entropy }over couplings
of two distributions $P_{X},P_{Y}$ as 
\begin{equation}
\mathcal{H}_{\alpha}^{(c)}(P_{X},P_{Y}):=\min_{P_{XY}\in C(P_{X},P_{Y})}H_{\alpha}(Y|X),
\end{equation}
with the Arimoto-Rényi conditional entropy of order $\alpha\in[0,\infty]$
given by \cite{arimoto1977information,sason2018improved} 
\begin{align*}
 & H_{\alpha}(Y|X):=\\
 & \begin{cases}
\frac{\alpha}{1-\alpha}\log\mathbb{E}\left[\left(\sum_{y}P_{Y|X}^{\alpha}(y|X)\right)^{\frac{1}{\alpha}}\right], & \alpha\in(0,1)\\
 & \qquad\cup(1,\infty)\\
\max_{x\in\mathcal{X}}\log\left|\left\{ y\in\mathcal{Y}:P_{Y|X}(y|x)>0\right\} \right|, & \alpha=0\\
-\mathbb{E}\log P_{Y|X}(Y|X), & \alpha=1\\
-\log\mathbb{E}\left[\max_{y\in\mathcal{Y}}P_{Y|X}(y|X)\right], & \alpha=\infty
\end{cases}.
\end{align*}
We also call the minimum $\infty$-Rényi conditional entropy $\mathcal{H}_{\infty}^{(c)}(P_{X},P_{Y})$
over couplings of $P_{X},P_{Y}$ as \emph{minimum conditional min-entropy},
and the minimum $1$-Rényi conditional entropy $\mathcal{H}_{1}^{(c)}(P_{X},P_{Y})$
over couplings of $P_{X},P_{Y}$ (shortly denoted as $\mathcal{H}^{(c)}(P_{X},P_{Y})$)
as \emph{minimum (Shannon) conditional entropy}. 
\end{defn}
Note that $\mathcal{H}_{\alpha}^{(c)}(P_{X},P_{Y})$ is monotonically
decreasing in $\alpha$ since $H_{\alpha}(Y|X)$ has this monotonicity
property (the latter property was proved in \cite[Proposition~4.6]{fehr2014conditional}
and \cite[Proposition~1]{sason2018arimoto}). 
\begin{thm}[Maximal Guessing Coupling Equality]
\label{thm:equivalence} The maximal guessing coupling problem is
equivalent to the distribution approximation problem. That is, 
\begin{equation}
e^{-\mathcal{H}_{\infty}^{(c)}(P_{X},P_{Y})}=\mathcal{G}(P_{X},P_{Y})=1-\min_{f}|P_{Y}-P_{f(X)}|.\label{eq:-8}
\end{equation}
Moreover, assume that $f$ is an optimal function for the distribution
approximation problem, and $P_{f(X),Y}$ is a maximal coupling of
$P_{f(X)},P_{Y}$, i.e., $f$ is a minimizer of $\min_{f}|P_{Y}-P_{f(X)}|$
and $P_{f(X),Y}$ is a maximizer of the problem 
\begin{equation}
\max_{P_{f(X),Y}\in C(P_{f(X)},P_{Y})}\mathbb{P}\left\{ Y=f(X)\right\} .
\end{equation}
Then $P_{X}P_{Y|f(X)}$ is a maximal guessing coupling of $P_{X},P_{Y}$. 
\end{thm}
\begin{rem}
\eqref{eq:-8} (with $\min$ and $\max$ respectively replaced by
$\inf$ and $\sup$) also holds for general distributions $P_{X},P_{Y}$,
e.g., continuous distributions. 
\end{rem}
\begin{rem}
Since $\mathcal{H}_{\infty}^{(c)}(P_{X},P_{Y})\le H_{0}(P_{Y})=\log|\mathcal{Y}|$,
we have $\mathcal{G}(P_{X},P_{Y})=1-\min_{f}|P_{Y}-P_{f(X)}|\ge\frac{1}{|\mathcal{Y}|}$. 
\end{rem}
\begin{IEEEproof}
Exchanging minimization operations, we have 
\begin{align}
\mathcal{G}(P_{X},P_{Y}) & =\min_{f}\min_{P_{XY}\in C(P_{X},P_{Y})}\mathbb{P}\left\{ Y\neq f(X)\right\} .\label{eq:-7}
\end{align}
Now we prove that given a function $f$, 
\begin{align}
 & \min_{P_{XY}\in C(P_{X},P_{Y})}\mathbb{P}\left\{ Y\neq f(X)\right\} \nonumber \\
 & =\min_{P_{f(X),Y}\in C(P_{f(X)},P_{Y})}\mathbb{P}\left\{ Y\neq f(X)\right\} .\label{eq:-3}
\end{align}

Define 
\begin{align}
 & Q_{X,Y}:=\arg\min_{P_{XY}\in C(P_{X},P_{Y})}\mathbb{P}\left\{ Y\neq f(X)\right\} ,\;\mbox{and}\\
 & Q_{f(X),X,Y}(v,x,y):=Q_{X,Y}(x,y)1\{v=f(x)\}.
\end{align}
Then we have 
\begin{align}
 & \min_{P_{XY}\in C(P_{X},P_{Y})}\mathbb{P}\left\{ Y\neq f(X)\right\} \nonumber \\
 & =\mathbb{P}_{Q_{X,Y}}\left\{ Y\neq f(X)\right\} \\
 & =\mathbb{P}_{Q_{f(X),Y}}\left\{ Y\neq f(X)\right\} \\
 & \geq\min_{P_{f(X),Y}\in C(P_{f(X)},P_{Y})}\mathbb{P}\left\{ Y\neq f(X)\right\} .\label{eq:-4}
\end{align}
On the other hand, denote 
\begin{equation}
Q_{f(X),Y}:=\argmin_{P_{f(X),Y}\in C(P_{f(X)},P_{Y})}\mathbb{P}\left\{ Y\neq f(X)\right\} 
\end{equation}
and 
\begin{align}
 & Q_{f(X),X,Y}(v,x,y)\nonumber \\
 & :=P_{X}(x)1\{v=f(x)\}Q_{Y|f(X)}(y|v)\\
 & =P_{X}(x)Q_{Y|f(X)}(y|f(x))1\{v=f(x)\}.
\end{align}
Then we also have 
\begin{align}
 & \min_{P_{f(X),Y}\in C(P_{f(X)},P_{Y})}\mathbb{P}\left\{ Y\neq f(X)\right\} \nonumber \\
 & =\mathbb{P}_{Q_{f(X),Y}}\left\{ Y\neq f(X)\right\} \\
 & =\mathbb{P}_{Q_{X,Y}}\left\{ Y\neq f(X)\right\} \\
 & \geq\min_{P_{XY}\in C(P_{X},P_{Y})}\mathbb{P}\left\{ Y\neq f(X)\right\} .\label{eq:-6}
\end{align}
Combining \eqref{eq:-4} and \eqref{eq:-6} we have the desired equality
\eqref{eq:-3}.

Substituting \eqref{eq:-3} into \eqref{eq:-7}, we have 
\begin{align}
 & \min_{P_{XY}\in C(P_{X},P_{Y})}\min_{f}\mathbb{P}\left\{ Y\neq f(X)\right\} \nonumber \\
 & =\min_{f}\min_{P_{f(X),Y}\in C(P_{f(X)},P_{Y})}\mathbb{P}\left\{ Y\neq f(X)\right\} \\
 & =\min_{f}|P_{Y}-P_{f(X)}|,\label{eq:-6-1}
\end{align}
where \eqref{eq:-6-1} follows from Lemma \ref{lem:MaxCoupling}.

Furthermore, the first equality of \eqref{eq:-8} follows from the
fact that $\max_{f}\mathbb{P}\left\{ Y=f(X)\right\} =\mathbb{E}\max_{y}P(y|X)=e^{-H_{\infty}(Y|X)}$
\cite[Proposition~4.2]{berens2013conditional}. 
\end{IEEEproof}
By Theorem \ref{thm:equivalence}, to solve the maximal guessing coupling
problem, we only need to compute 
\begin{equation}
\min_{f(x)}\left|P_{Y}-P_{f(X)}\right|.\label{eq:-1}
\end{equation}
Define $\mathcal{A}(y):=\left\{ x:f(x)=y\right\} $. Then \eqref{eq:-1}
is equivalent to 
\begin{equation}
\min_{\left\{ \mathcal{A}(y):y\in\mathcal{Y}\right\} }\sum_{y}\left|P_{Y}(y)-P_{X}(\mathcal{A}(y))\right|,\label{eq:-2}
\end{equation}
where $\left\{ \mathcal{A}(y):y\in\mathcal{Y}\right\} $ is a partition
of $\mathcal{X}$, i.e., $\bigcup_{y\in\mathcal{Y}}\mathcal{A}(y)=\mathcal{X}$
and $\mathcal{A}(y_{1})\cap\mathcal{A}(y_{2})=\emptyset$ for any
$y_{1},y_{2}\in\mathcal{Y}$ and $y_{1}\neq y_{2}$. For any distribution
pair $P_{X},P_{Y}$, is \eqref{eq:-2} equal to zero? This question
is equivalent to the following: Does there exist a partition $\left\{ \mathcal{A}(y):y\in\mathcal{Y}\right\} $
such that $P_{Y}(y)=P_{X}(\mathcal{A}(y))$ for all $y\in\mathcal{Y}$?
This problem involving the search for an optimal partition has been
shown to be NP-hard \cite{garey2002computers}. This implies that
the optimization problem \eqref{eq:-2} is also NP-hard, since in
general, solving the optimization problem \eqref{eq:-2} is strictly
harder than only determining whether \eqref{eq:-2} equals zero. 
\begin{prop}
\cite[p. 223]{garey2002computers} The problem in \eqref{eq:-2} is
NP-hard (more specifically, NP-complete). 
\end{prop}
However, when we consider the asymptotic scenario, the optimal limiting
value of this coupling problem can be easily determined. Furthermore,
we also provide bounds on the rates of convergence of the coupling
problems to their limiting values. Define the optimal exponents as
\begin{equation}
\overline{\mathsf{E}}\left(P_{X},P_{Y}\right):=\liminf_{n\to\infty}-\frac{1}{n}\log\left(1-\mathcal{G}(P_{X}^{n},P_{Y}^{n})\right),\label{eq:optexpforMGC}
\end{equation}
and 
\begin{equation}
\mathsf{E}\left(P_{X},P_{Y}\right):=\liminf_{n\to\infty}-\frac{1}{n}\log\mathcal{G}(P_{X}^{n},P_{Y}^{n}).
\end{equation}
Then we have the following main result. The proof is provided in Appendix
\ref{sec:maxguesscoup}. 
\begin{thm}[Maximal Guessing Coupling]
\label{thm:MaxGuessCoup} Given two product marginal distributions
$P_{X}^{n}$ and $P_{Y}^{n}$, we have: 

{[}leftmargin={*}{]} 
\begin{enumerate}
\item If $H(X)>H(Y)$, then $\mathcal{G}(P_{X}^{n},P_{Y}^{n})\to1$ at least
exponentially fast as $n\to\infty$. Moreover, we have 
\begin{equation}
\overline{\mathsf{E}}\left(P_{X},P_{Y}\right)\geq\frac{1}{2}\max_{t\in[0,1]}t\left(H_{1+t}(X)-H_{1-t}\left(Y\right)\right).\label{eqn:Eiid-2}
\end{equation}
\item If $H(X)<H(Y)$, then $\mathcal{G}(P_{X}^{n},P_{Y}^{n})\to0$ exponentially
fast as $n\to\infty$. Moreover, we have 
\begin{align}
\log|\mathcal{Y}|\ge\mathsf{E}\left(P_{X},P_{Y}\right) & \geq\sup_{\epsilon\in(0,1)}\min\Bigl\{\delta_{\epsilon}(P_{X}),\,\delta_{\epsilon}(P_{Y}),\,\nonumber \\
 & \qquad(1-\epsilon)H(Y)-(1+\epsilon)H(X)\Bigr\},
\end{align}
with $\delta_{\epsilon}(P_{X}):=\frac{1}{3}\epsilon^{2}\min_{x:P_{X}(x)>0}P_{X}(x)$.
\item If $H(X)=H(Y)$, then $\mathcal{G}(P_{X}^{n},P_{Y}^{n})\geq\mathcal{G}(P_{X},P_{Y})^{n}$
for all $n$. 
\end{enumerate}
\end{thm}
\begin{rem}
The exponent whenever $H(X)\geq H(Y)$ is infinity if there exists
a coupling $P_{XY}$ such that $Y$ is expressed as a deterministic
function of $X$. 
\end{rem}
\begin{rem}
By the equivalence between the maximal guessing coupling problem and
the distribution approximation problem (Theorem \ref{thm:equivalence}),
the exponential bounds given in Theorem \ref{thm:MaxGuessCoup} are
also bounds for the distribution approximation problem $\min_{f}|P_{Y}-P_{f(X)}|$. 
\end{rem}
\begin{rem}
Theorem \ref{thm:MaxGuessCoup} implies that given two product distributions
$P_{X}^{n}$ and $P_{Y}^{n}$ with $H(X)>H(Y)$, there exists a joint
distribution $P_{X^{n}Y^{n}}$ satisfying 
\begin{equation}
\lim_{n\to\infty}\min_{\left\{ \mathcal{A}(y^{n}):y^{n}\in\mathcal{Y}^{n}\right\} }\sum_{y^{n}}\left|P_{Y^{n}}(y^{n})-P_{X^{n}}(\mathcal{A}(y^{n}))\right|=0,\label{eq:-22}
\end{equation}
where $\left\{ \mathcal{A}(y^{n}):y^{n}\in\mathcal{Y}^{n}\right\} $
is a partition of $\mathcal{X}^{n}$. Hence the probability values
of $P_{X^{n}}$ asymptotically forms a refinement of the probability
values of $P_{Y^{n}}$ in the sense of \eqref{eq:-22}. This is just
a restatement of the soft-covering lemma \cite{Cuff}. 
\end{rem}
Since we get $e^{-\mathcal{H}_{\infty}^{(c)}(P_{X}^{n},P_{Y}^{n})}=\mathcal{G}(P_{X}^{n},P_{Y}^{n})$
from \eqref{eq:-8}, the following result follows from Theorem \ref{thm:MaxGuessCoup}: 
\begin{cor}
\label{cor:Hinfty}Given two product marginal distributions $P_{X}^{n}$
and $P_{Y}^{n}$, we have: 

{[}leftmargin={*}{]} 
\begin{enumerate}
\item If $H(X)>H(Y)$, then $\mathcal{H}_{\infty}^{(c)}(P_{X}^{n},P_{Y}^{n})\to0$
at least exponentially fast as $n\to\infty$ with exponent $\overline{\mathsf{E}}\left(P_{X},P_{Y}\right)$. 
\item If $H(X)<H(Y)$, then $\mathcal{H}_{\infty}^{(c)}(P_{X}^{n},P_{Y}^{n})\to\infty$
linearly as $n\to\infty$ with scaling factor $\mathsf{E}\left(P_{X},P_{Y}\right)$. 
\item If $H(X)=H(Y)$, then $\mathcal{H}_{\infty}^{(c)}(P_{X}^{n},P_{Y}^{n})\geq n\mathcal{H}_{\infty}^{(c)}(P_{X},P_{Y})$
for all $n$. 
\end{enumerate}
\end{cor}
Theorem \ref{thm:MaxGuessCoup} does not give an asymptotically tight
expression if $H(X)=H(Y)$. However, we conjecture the following: 
\begin{conjecture}[Asymptotically Deterministic Coupling]
\label{conj:AsymptoticallyDeterministicCoupling} Assume $H(X)=H(Y)$.
Then $\mathcal{G}(P_{X}^{n},P_{Y}^{n})\rightarrow1$ if and only if
$\mathcal{G}(P_{X},P_{Y})=1$ (this is also equivalent to the fact
that $P_{X}$ and $P_{Y}$ have the same probability values). 
\end{conjecture}
This conjecture implies when $H(X)=H(Y)$, $\mathcal{G}(P_{X}^{n},P_{Y}^{n})\rightarrow1$
requires some ``matched'' condition on the distributions. In Appendix
\ref{subsec:AsympDetCoupling}, we prove that Conjecture \ref{conj:AsymptoticallyDeterministicCoupling}
is true if $P_{X}$ or $P_{Y}$ is a uniform distribution. 

Similar to the conjecture concerning asymptotically deterministic
couplings, we also have the following conjecture concerning the deterministic
couplings. 
\begin{conjecture}[Deterministic Coupling]
\label{conj:DeterministicCoupling} $\mathcal{G}(P_{X}^{n},P_{Y}^{n})=1$
if and only if $\mathcal{G}(P_{X},P_{Y})=1.$ That is, there exists
a deterministic coupling $P_{X^{n}Y^{n}}\in C(P_{X}^{n},P_{Y}^{n})$
for which $Y^{n}$ is a function of $X^{n}$, if and only if there
exists a deterministic coupling $P_{XY}\in C(P_{X},P_{Y})$ for which
$Y$ is a function of $X$. 
\end{conjecture}
In Appendix \ref{subsec:DetCoupling}, we prove that Conjecture \ref{conj:DeterministicCoupling}
is true for two special cases.

\subsection{\label{sec:Maximal-Guessing-Coupling}Maximal Guessing Coupling:
General Sources and Coupling Through a Channel }

\begin{figure}
\centering\includegraphics[width=0.3\textwidth]{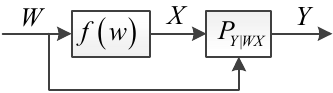}\caption{\label{fig:General-source-channel-resolvabi}General source-channel
resolvability.}
\end{figure}

In the previous subsection, we showed that the maximal guessing coupling
problem is equivalent to the distribution approximation problem. Hence,
to obtain the maximal guessing coupling of a pair of sources, we only
need to solve the problem of probability distribution approximation
for these sources. Here, instead, we consider a more general variation
of distribution approximation problem, called the \emph{general source-channel
resolvability problem}. This is illustrated in Fig.~\ref{fig:General-source-channel-resolvabi},
and will be proven to be equivalent to a maximal guessing coupling
through a channel problem.

Consider a pair of distributions $(P_{W},P_{Z})$ and a channel $P_{Y|WX}$
(this is a source-dependent channel which reduces to a source-independent
channel if we set $P_{Y|WX}=P_{Y|X}$). Denote the output of the channel
$P_{Y|WX}$ with input $X=f(W)$ as $Y_{f}$. Obviously, the distribution
of $Y_{f}$ is 
\begin{equation}
P_{Y_{f}}(y)=\sum_{w}P_{W}(w)P_{Y|WX}(y|w,f(w)).
\end{equation}
If we consider $f(W)$ as a guessing function and $Y_{f}$ as the
final estimate variable of the target variable $Z$, then the optimization
problem $\min_{P_{WZ}\in C(P_{W},P_{Z})}\min_{f}\mathbb{P}\left\{ Z\neq Y_{f}\right\} $
can be seen as the problem of \emph{maximal guessing coupling through
a channel}. It is a generalization of the maximal guessing coupling
problem, since it reduces to the maximal guessing coupling problem
if the channel is set to be the identity channel, i.e., $P_{Y|WX}(y|w,x)=1\{y=x\}$
for all $(w,x,y)$. 
\begin{defn}
Define the \emph{maximal guessing probability through a channel }$P_{Y|X}$
over couplings of $(P_{W},P_{Z})$ as 
\begin{align}
 & \mathcal{G}(P_{W},P_{Z}|P_{Y|X})\nonumber \\
 & :=\max_{P_{WZ}\in C(P_{W},P_{Z})}\max_{f:{\cal W}\to{\cal X}}\nonumber \\
 & \qquad\max_{P_{Y|WZ}:P_{Y|W}(y|w)=P_{Y|WX}(y|w,f(w))}\mathbb{P}\left\{ Z=Y\right\} .
\end{align}
Any $Q_{XY}\in C(P_{X},P_{Y})$ achieving $\mathcal{G}(P_{W},P_{Z}|P_{Y|X})$
is called a \emph{maximal guessing coupling} of $P_{X},P_{Y}$ \emph{through
the channel }$P_{Y|X}$. 
\end{defn}
On the other hand, the \emph{source-channel resolvability} problem
is $\min_{f}|P_{Z}-P_{Y_{f}}|$. Similar to Theorem \ref{thm:equivalence},
the following theorem states the equivalence between the problem of
maximal guessing coupling through a channel and the source-channel
resolvability problem. 
\begin{thm}[Maximal Guessing Coupling Through a Channel]
\label{thm:equivalence-1} The problem of maximal guessing coupling
through a channel is equivalent to the source-channel resolvability
problem. That is, 
\begin{equation}
\mathcal{G}(P_{W},P_{Z}|P_{Y|X})=1-\min_{f}|P_{Z}-P_{Y_{f}}|.\label{eq:-8-1}
\end{equation}
\end{thm}
\begin{IEEEproof}
Exchanging minimization operations, we have 
\begin{align}
 & 1-\mathcal{G}(P_{W},P_{Z}|P_{Y|X})\nonumber \\
 & =\min_{f}\min_{P_{WZ}\in C(P_{W},P_{Z})}\nonumber \\
 & \qquad\min_{P_{Y|WZ}:P_{Y|W}(y|w)=P_{Y|WX}(y|w,f(w))}\mathbb{P}\left\{ Z\neq Y\right\} \\
 & =\min_{f}\min_{\substack{P_{WZY}:P_{WZ}\in C(P_{W},P_{Z}),\\
P_{Y|W}(y|w)=P_{Y|WX}(y|w,f(w))
}
}\mathbb{P}\left\{ Z\neq Y\right\} \\
 & =\min_{f}\min_{\substack{P_{Z|WY}:\sum_{w,y}P_{W}(w)P_{Y|WX}(y|w,f(w))\\
\times P_{Z|WY}(z|w,y)=P_{Z}(z)
}
}\mathbb{P}\left\{ Z\neq Y\right\} \\
 & =\min_{f}\min_{P_{Z|Y}:\sum_{y}P_{Y}(y)P_{Z|Y}(z|y)=P_{Z}(z)}\mathbb{P}\left\{ Z\neq Y\right\} \label{eq:-25}\\
 & =\min_{f}\min_{P_{Y_{f}Z}\in C(P_{Y_{f}},P_{Z})}\mathbb{P}\left\{ Z\neq Y_{f}\right\} \\
 & =\min_{f}|P_{Z}-P_{Y_{f}}|,
\end{align}
where \eqref{eq:-25} follows since the optimized objective $\mathbb{P}\left\{ Z\neq Y\right\} $
depends only on the joint distribution of $Y,Z$. 
\end{IEEEproof}
Note that the coupling $P_{WZ}$ and the channel $P_{Y|X}$ are not
independent, i.e., the channel $P_{Y|X}$ is allowed to be embedded
into the optimal coupling $P_{WZ}$. If such embedding is not allowed,
then the problem reduces to 
\begin{equation}
\widetilde{\mathcal{G}}(P_{W},P_{Z}|P_{Y|X}):=\max_{P_{WZ}\in C(P_{W},P_{Z})}\max_{f}\mathbb{P}\left\{ Z=Y\right\} ,
\end{equation}
where the probability is taken under the distribution $P_{WZ}(w,z)P_{Y|WX}(y|w,f(w))$.
However, for this problem, the equivalence above no longer holds.

\subsubsection{One-shot Bounds}

Next we derive following bounds for the source-channel resolvability
problem. The proof of Theorem \ref{thm:General Source-Channel Resolvability-1}
is provided in Appendix \ref{sec:GeneralSCResolvability}. 
\begin{thm}[General Source-Channel Resolvability]
{} \label{thm:General Source-Channel Resolvability-1} For any distributions
$P_{W}$ and $P_{Z}$, channel $P_{Y|WX}$, and $\tau>0$, we have
\begin{align}
 & \min_{P_{X|W}:P_{WXY}({\cal A}_{0})=0}\left|P_{Y}-P_{Z}\right|\nonumber \\
 & \leq\min_{f:\mathcal{W}\to\mathcal{X}}|P_{Z}-P_{Y_{f}}|\\
 & \leq\min_{P_{X|W}}\left\{ \left|P_{Y}-P_{Z}\right|+P_{WXY}({\cal A}_{\tau})\right\} +\frac{1}{2}e^{\tau/2},\label{eq:-47}
\end{align}
where $Y_{f}$ is the output of the channel $P_{Y|WX}$ with input
$X=f(W)$, and 
\begin{equation}
{\cal A}_{\tau}:=\left\{ (w,x,y)\;:\;\log\frac{P_{W}(w)P_{Y|WX}(y|w,x)}{P_{Y}(y)}>\tau\right\} .
\end{equation}
Furthermore, we have another lower bound 
\begin{align}
 & \min_{f:\mathcal{W}\to\mathcal{X}}|P_{Z}-P_{Y_{f}}|\nonumber \\
 & \ge\min_{P_{WZ}\in C(P_{W},P_{Z})}\min_{P_{X|W}}P_{WXZ}({\cal B}_{\tau})-e^{-\tau},\label{eq:-48}
\end{align}
where 
\begin{equation}
{\cal B}_{\tau}:=\left\{ (w,x,z)\;:\;\log\frac{P_{W}(w)P_{Y|WX}(z|w,x)}{P_{Z}(z)}>\tau\right\} .\label{eq:-10}
\end{equation}
\end{thm}
If an identity channel $P_{Y|WX}(y|w,x)=1\{y=x\}$ is considered,
the source-channel resolvability problem degenerates into the source-source
resolvability problem (using a general source to generate another
general source) or equivalently, the distribution approximation problem.
That is, $\min_{f}|P_{Z}-P_{Y_{f}}|=\min_{f}|P_{Z}-P_{f(W)}|$ where
$P_{W}$ is a source distribution and $P_{Z}$ is a target distribution.
Theorem \ref{thm:General Source-Channel Resolvability-1} results
in the following corollary. 
\begin{cor}[General Source-Source Resolvability: Probability Distribution Approximation]
\label{cor:General Source-Source Resolvability} For any source distribution
$P_{W}$ and target distribution $P_{Z}$, we have 
\begin{align}
 & \min_{P_{WZ}\in C(P_{W},P_{Z})}P({\cal A}_{\tau}')-e^{-\tau}\nonumber \\
 & \leq\min_{f}|P_{Z}-P_{f(W)}|\\
 & \leq\min_{P_{WZ}\in C(P_{W},P_{Z})}P({\cal A}_{\tau}')+\frac{1}{2}e^{\tau/2},
\end{align}
where 
\begin{equation}
{\cal A}_{\tau}':=\left\{ (w,z)\;:\;\log\frac{P_{W}(w)}{P_{Z}(z)}>\tau\right\} .
\end{equation}
\end{cor}

\subsubsection{Asymptotics}

When the asymptotic behavior is considered, Theorem \ref{thm:General Source-Channel Resolvability-1}
results in the following corollary. 
\begin{cor}[General Source-Channel Resolvability]
\label{cor:General Source-Source Resolvability-1} For any source
distribution $P_{\boldsymbol{W}}$, channel $P_{\boldsymbol{Y}|\boldsymbol{WX}}$,
and target distribution $P_{\boldsymbol{Z}}$, we have 
\begin{align}
 & \inf_{\substack{P_{\boldsymbol{X}|\boldsymbol{W}}:\\
\mathrm{p}\mbox{-}\limsup_{n\rightarrow\infty}\left\{ \frac{1}{n}\imath(W^{n}X^{n};Y^{n})-\frac{1}{n}\imath(W^{n})\right\} \leq0
}
}\hspace{-0.3in}\limsup_{n\to\infty}\left|P_{Y^{n}}-P_{Z^{n}}\right|\nonumber \\
 & \quad\leq\limsup_{n\to\infty}\min_{f_{n}}|P_{Z^{n}}-P_{Y_{f_{n}}^{n}}|\label{eq:-26}\\
 & \qquad\leq\hspace{-0.2in}\inf_{\substack{P_{\boldsymbol{X}|\boldsymbol{W}}:\\
\mathrm{p}\mbox{-}\limsup_{n\rightarrow\infty}\left\{ \frac{1}{n}\imath(W^{n}X^{n};Y^{n})-\frac{1}{n}\imath(W^{n})\right\} <0
}
}\hspace{-0.5in}\limsup_{n\to\infty}\left|P_{Y^{n}}-P_{Z^{n}}\right|.\label{eq:-27}
\end{align}
\end{cor}
Moreover, if an identity channel $P_{Y|WX}(y|w,x)=1\{y=x\}$ is considered,
Corollary \ref{cor:General Source-Source Resolvability-1} results
in the following corollary. 
\begin{cor}[General Source-Source Resolvability: Probability Distribution Approximation]
\label{cor:General Source-Source Resolvability-2} For any source
distribution $P_{\boldsymbol{W}}$ and target distribution $P_{\boldsymbol{Z}}$,
we have \eqref{eq:-26}-\eqref{eq:-27} with $\imath(W^{n}X^{n};Y^{n})$
replaced by $\imath(X^{n})$ and $Y^{n}$ replaced by $X^{n}$. Equivalently,
\begin{align}
 & \inf_{\substack{\delta>0,\,P_{\boldsymbol{WZ}}\in C(P_{\boldsymbol{W}},P_{\boldsymbol{Z}}):\\
\delta\mbox{-}\mathrm{p}\mbox{-}\limsup_{n\rightarrow\infty}\left\{ \frac{1}{n}\imath(Z^{n})-\frac{1}{n}\imath(W^{n})\right\} \leq0
}
}\delta\nonumber \\
 & \qquad\leq\limsup_{n\to\infty}\min_{f_{n}}|P_{Z^{n}}-P_{f_{n}(W^{n})}|\\
 & \qquad\qquad\leq\inf_{\substack{\delta>0,\,P_{\boldsymbol{WZ}}\in C(P_{\boldsymbol{W}},P_{\boldsymbol{Z}}):\\
\delta\mbox{-}\mathrm{p}\mbox{-}\limsup_{n\rightarrow\infty}\left\{ \frac{1}{n}\imath(Z^{n})-\frac{1}{n}\imath(W^{n})\right\} <0
}
}\delta,
\end{align}
where $C(P_{\boldsymbol{W}},P_{\boldsymbol{Z}}):=\left\{ P_{\boldsymbol{WZ}}:P_{W^{n}Z^{n}}\in C(P_{W^{n}},P_{Z^{n}}),\forall n\right\} $. 
\end{cor}

\subsubsection{Maximal Guessing Coupling for General Sources and Channels}

According to the equivalence between the maximal guessing coupling
problem and distribution approximation problem (Theorem \ref{thm:equivalence})
and the equivalence between the problem of maximal guessing coupling
through a channel and the problem of source-channel resolvability
(Theorem \ref{thm:equivalence-1}), we have the following conclusions.
The bounds given in Theorem \ref{lem:General Source-Channel Resolvability}
and Corollary \ref{cor:General Source-Source Resolvability-1} are
also bounds for the maximal guessing coupling problem through a channel.
The bounds given in Corollaries \ref{cor:General Source-Source Resolvability}
and \ref{cor:General Source-Source Resolvability-2} are also bounds
for the maximal guessing coupling problem.

\subsection{Application of Maximal Guessing Coupling to Minimum Entropy Coupling}

The problems of minimum entropy coupling and maximum mutual information
coupling were first studied in \cite{kovavcevic2015entropy}. In this
subsection, we study the asymptotics of these coupling problems. In
\cite{kovavcevic2015entropy}, the authors showed that solving the
minimum entropy coupling problem or maximum mutual information coupling
problem is NP-hard. However, in this section, we show that is not
the case for the asymptotic regime. Recall from Definition \ref{def:DefineRenyiCoupling}
the \emph{minimum conditional entropy} $\mathcal{H}^{(c)}(P_{X},P_{Y}):=\min_{P_{XY}\in C(P_{X},P_{Y})}H(Y|X)$
over couplings of $P_{X},P_{Y}$. Then for such a coupling problem,
we have the following result. 
\begin{cor}[Minimum Conditional Entropy Coupling]
\label{cor:mincondentropy} Given two product marginal distributions
$P_{X}^{n}$ and $P_{Y}^{n}$, 
we have 

{[}leftmargin={*}{]} 
\begin{enumerate}
\item $\mathcal{H}^{(c)}(P_{X}^{n},P_{Y}^{n})-n\max\{0,H(Y)-H(X)\}\to0$
at least exponentially fast as $n\to\infty$ if $H(X)\neq H(Y)$; 
\item $\mathcal{H}^{(c)}(P_{X}^{n},P_{Y}^{n})\leq n\mathcal{H}^{(c)}(P_{X},P_{Y})$
for all $n$ if $H(X)=H(Y)$. 
\end{enumerate}
\end{cor}
\begin{IEEEproof}
We only prove Statement 1). Statement 2) is obvious. One simply employs
a product coupling to prove the upper bound.

Denote $p_{\rme}^{(n)}:=\min_{f}\mathbb{P}\left\{ Y^{n}\neq f(X^{n})\right\} $.
Then Fano's inequality \cite[Theorem~2.10.1]{Cover} implies 
\begin{equation}
0\leq H(Y^{n}|X^{n})\leq H(p_{\rme}^{(n)})+np_{\rme}^{(n)}\log|\mathcal{Y}|,\label{eq:-5}
\end{equation}
where $H(p):=-p\log p-(1-p)\log(1-p)$.

From Theorem \ref{thm:MaxGuessCoup}, we know that if $H(X)>H(Y)$,
then there exists a coupling $P_{X^{n}Y^{n}}\in C(P_{X}^{n},P_{Y}^{n})$
such that $p_{\rme}^{(n)}\to0$ at least exponentially fast as $n\to\infty$.
This implies the upper bound also converges to zero at least exponentially
fast. Hence if $H(X)>H(Y)$, then $\mathcal{H}^{(c)}(P_{X}^{n},P_{Y}^{n})\to0$
at least exponentially fast as $n\to\infty$.

On the other hand, we can write $H(Y^{n}|X^{n})=H(X^{n}|Y^{n})+H(Y^{n})-H(X^{n})$.
Since for a coupling $P_{X^{n}Y^{n}}\in C(P_{X}^{n},P_{Y}^{n})$,
$H(Y^{n})-H(X^{n})=n\left(H(Y)-H(X)\right)$, we have $\mathcal{H}^{(c)}(P_{X}^{n},P_{Y}^{n})=\mathcal{H}^{(c)}(P_{Y}^{n},P_{X}^{n})+n\left(H(Y)-H(X)\right)$.
By the argument above, if $H(X)<H(Y)$, then $\mathcal{H}^{(c)}(P_{Y}^{n},P_{X}^{n})\to0$
at least exponentially fast as $n\to\infty$. Hence if $H(X)<H(Y)$,
$\mathcal{H}^{(c)}(P_{X}^{n},P_{Y}^{n})-n\left(H(Y)-H(X)\right)\to0$
at least exponentially fast as $n\to\infty$. 
\end{IEEEproof}
Define the \emph{minimum joint entropy}{} and the \emph{maximum mutual
information}{} over couplings of two distributions $P_{X},P_{Y}$
as 
\begin{align}
\mathcal{H}(P_{X},P_{Y}) & :=\min_{P_{XY}\in C(P_{X},P_{Y})}H(XY),\quad\mbox{and}\\
\mathcal{I}(P_{X},P_{Y}) & :=\max_{P_{XY}\in C(P_{X},P_{Y})}I(X;Y)
\end{align}
respectively. Observe that $H(XY)=H(X)+H(Y|X)$ and $I(X;Y)=H(Y)-H(Y|X)$.
Hence $\mathcal{H}(P_{X},P_{Y})=H(X)+\mathcal{H}^{(c)}(P_{X},P_{Y})$
and $\mathcal{I}(P_{X},P_{Y})=H(Y)-\mathcal{H}^{(c)}(P_{X},P_{Y})$.
Combining these with Corollary \ref{cor:mincondentropy}, we obtain
the following two corollaries. 
\begin{cor}[Minimum Joint Entropy Coupling]
\label{thm:Normalized-1-1-1-1-1} Given two product marginal distributions
$P_{X}^{n}$ and $P_{Y}^{n}$, we have 

{[}leftmargin={*}{]} 
\begin{enumerate}
\item $\mathcal{H}(P_{X}^{n},P_{Y}^{n})-n\max\{H(X),H(Y)\}\to0$ at least
exponentially fast as $n\to\infty$ if $H(X)\neq H(Y)$; 
\item $\mathcal{H}(P_{X}^{n},P_{Y}^{n})\leq n\mathcal{H}(P_{X},P_{Y})$
for all $n$ if $H(X)=H(Y)$. 
\end{enumerate}
\end{cor}
\begin{cor}[Maximum Mutual Information Coupling]
\label{thm:Normalized-1-1-1-1-1-1} Given two product marginal distributions
$P_{X}^{n}$ and $P_{Y}^{n}$, we have 

{[}leftmargin={*}{]} 
\begin{enumerate}
\item $\mathcal{I}(P_{X}^{n},P_{Y}^{n})-n\min\{H(X),H(Y)\}\to0$ at least
exponentially fast as $n\to\infty$ if $H(X)\neq H(Y)$; 
\item $\mathcal{I}(P_{X}^{n},P_{Y}^{n})\geq n\mathcal{I}(P_{X},P_{Y})$
for all $n$ if $H(X)=H(Y)$. 
\end{enumerate}
\end{cor}
Define the \emph{maximum conditional mutual information} over couplings
of two distributions $P_{X},P_{YZ}$ as $\mathcal{I}^{(c)}(P_{X},P_{YZ}):=\max_{P_{XYZ}\in C(P_{X},P_{YZ})}I(X;Y|Z)$. 
\begin{cor}[Maximum Conditional Mutual Information Coupling]
\label{cor:maxcondmic} Given two product marginal distributions
$P_{X}^{n}$ and $P_{YZ}^{n}$, we have 

{[}leftmargin={*}{]} 
\begin{enumerate}
\item $\mathcal{I}^{(c)}(P_{X}^{n},P_{YZ}^{n})-n\min\{H(X),H(Y|Z)\}\to0$
at least exponentially fast as $n\to\infty$ if $H(X)\neq H(Y|Z)$; 
\item $\mathcal{I}^{(c)}(P_{X}^{n},P_{YZ}^{n})\geq n\mathcal{I}^{(c)}(P_{X},P_{YZ})$
for all $n$ if $H(X)=H(Y|Z)$. 
\end{enumerate}
\end{cor}
For Corollary \ref{cor:maxcondmic}, we use $({X}^{n},{Z}^{n})$ with
joint distribution $P_{X}^{n}P_{Z}^{n}$ to guess $({Y}^{n},{Z}^{n})$
with joint distribution $P_{YZ}^{n}$ if $H(X)>H(Y|Z)$, or reversely,
use $({Y}^{n},{Z}^{n})$ to guess $({X}^{n},{Z}^{n})$ if $H(X)<H(Y|Z)$.
The proof is along exactly the same lines as that of Corollary \ref{cor:mincondentropy},
and hence omitted here.

Recall from Definition \ref{def:DefineRenyiCoupling} the \emph{minimum
$\alpha$-Rényi conditional entropy} 
\begin{equation}
\mathcal{H}_{\alpha}^{(c)}(P_{X},P_{Y}):=\min_{P_{XY}\in C(P_{X},P_{Y})}H_{\alpha}(Y|X)
\end{equation}
over couplings of $P_{X},P_{Y}$. We next generalize our result to
the minimum Rényi entropy, and get the following corollary. Statement
2) of Corollary \ref{cor:MinRenyiEntropyCoupling} follows by combining
Corollary \ref{cor:Hinfty} and the fact that $\mathcal{H}_{\alpha}^{(c)}(P_{X}^{n},P_{Y}^{n})$
is non-increasing in $\alpha$. Statement 3) is proven by using product
couplings. The proof of Statement 1) is provided in Appendix \ref{sec:MinRenyiEntropyCoupling}. 
\begin{cor}[Minimum Rényi Conditional Entropy Coupling]
\label{cor:MinRenyiEntropyCoupling} Given two product marginal distributions
$P_{X}^{n}$ and $P_{Y}^{n}$, we have: 

{[}leftmargin={*}{]} 
\begin{enumerate}
\item If $H(X)>H(Y)$, then $\mathcal{H}_{\alpha}^{(c)}(P_{X}^{n},P_{Y}^{n})\to0$
at least exponentially fast as $n\to\infty$ for 
\begin{equation}
\alpha\in\left(\frac{\log|\mathcal{Y}|}{\overline{\mathsf{E}}\left(P_{X},P_{Y}\right)+\log|\mathcal{Y}|},\infty\right],
\end{equation}
where $\overline{\mathsf{E}}\left(P_{X},P_{Y}\right)$ defined in
\eqref{eq:optexpforMGC} denotes the optimal exponent for the maximal
guessing coupling problem; 
\item If $H(X)<H(Y)$, then $\mathcal{H}_{\alpha}^{(c)}(P_{X}^{n},P_{Y}^{n})\to\infty$
linearly fast as $n\to\infty$ for all $\alpha\in[0,\infty]$; 
\item If $H(X)=H(Y)$, then $\mathcal{H}_{\alpha}^{(c)}(P_{X}^{n},P_{Y}^{n})\geq n\mathcal{H}_{\alpha}^{(c)}(P_{X},P_{Y})$
for all $n$ and for all $\alpha\in[0,\infty]$. 
\end{enumerate}
\end{cor}
\begin{defn}
\cite{gacs1973common} The \emph{Gács-Körner (GK) common information}
between two general correlated sources $(\boldsymbol{X},\boldsymbol{Y})$
is defined as 
\begin{align}
 & \overline{C}_{\mathsf{GK}}(\boldsymbol{X};\boldsymbol{Y})\nonumber \\
 & :={\displaystyle \sup_{\{(f_{n},g_{n})\}:\mathbb{P}\left\{ f_{n}\left(X^{n}\right)\neq g_{n}\left(Y^{n}\right)\right\} \to0}\liminf_{n\to\infty}\frac{1}{n}H(f_{n}\left(X^{n}\right))}.\label{eq:-46}
\end{align}
In particular, for two memoryless correlated sources $(X,Y)$, Gács-Körner
showed the GK common information is equal to 
\begin{equation}
C_{\mathsf{GK}}(X;Y):={\displaystyle \sup_{f,g:f\left(X\right)=g\left(Y\right)}H(f\left(X\right))}.\label{eq:-77-2}
\end{equation}
\end{defn}
Define the \emph{maximum GK common information} over couplings of
product distributions of $P_{X},P_{Y}$ as $\mathcal{C}_{\mathsf{GK}}(P_{X},P_{Y}):=\sup_{P_{\boldsymbol{X}\boldsymbol{Y}}:P_{X^{n}Y^{n}}\in C(P_{X}^{n},P_{Y}^{n}),\forall n}\overline{C}_{\mathsf{GK}}(\boldsymbol{X};\boldsymbol{Y})$.
As a consequence of Corollary \ref{cor:mincondentropy}, we have the
following result. 
\begin{cor}[Maximum GK Common Information Coupling]
\label{thm:Normalized} Given two distributions $P_{X}$ and $P_{Y}$,
we have 

{[}leftmargin={*}{]} 
\begin{enumerate}
\item $\mathcal{C}_{\mathsf{GK}}(P_{X},P_{Y})=\min\{H(X),H(Y)\}$ if $H(X)\neq H(Y)$; 
\item $\mathcal{C}_{\mathsf{GK}}(P_{X},P_{Y})\geq\max_{P_{XY}\in C(P_{X},P_{Y})}{C}_{\mathsf{GK}}(P_{X},P_{Y})$
if $H(X)=H(Y)$. 
\end{enumerate}
\end{cor}

\section{\label{sec:Exact-Intrinsic-Randomness}Exact Intrinsic Randomness }

In the next four sections, we apply the results above on the maximal
guessing coupling problem to several information-theoretic problems.
First, we consider a new version of intrinsic randomness problem,
named \emph{exact intrinsic randomness}, and apply our results on
maximal guessing coupling to this problem.

The lossless source coding problem, intrinsic randomness problem,
and source resolvability problem consist of three ingredients: 

{[}leftmargin={*}{]} 
\begin{enumerate}
\item a source distribution $P_{X^{n}}$, 
\item a random variable $M_{n}\in[1:e^{nR}]$, 
\item and a mapping between them $P_{X^{n}|M_{n}}$ or $P_{M_{n}|X^{n}}$. 
\end{enumerate}
Define the uniform distribution as $P_{M_{n}}^{\mathrm{U}}:=\mathsf{Unif}[1:e^{nR}]$.
In the lossless source coding problem, the source distribution $P_{X^{n}}=P_{X}^{n}$
and $X^{n}$ is an asymptotic function of $M_{n}$ under the reconstruction
mapping $P_{X^{n}|M_{n}}$; in the intrinsic randomness problem, the
source distribution $P_{X^{n}}=P_{X}^{n}$, $M_{n}$ is a function
of $X^{n}$ under the randomness extractor $P_{M_{n}|X^{n}}$, and
$P_{M_{n}},P_{M_{n}}^{\mathrm{U}}$ are asymptotically equal under
some distance measure; and in the source resolvability problem, $P_{M_{n}}=P_{M_{n}}^{\mathrm{U}}$,
$X^{n}$ is a deterministic function of $M_{n}$ under the resolvability
code $P_{X^{n}|M_{n}}$, and $P_{X^{n}},P_{X}^{n}$ are asymptotically
equal under some distance measure. However, we usually cannot find
a joint distribution $P_{M_{n}X^{n}}$ such that $P_{X^{n}}=P_{X}^{n}$,
$P_{M_{n}}=P_{M_{n}}^{\mathrm{U}}$, and $X^{n}$ is a function of
$M_{n}$ or $M_{n}$ is a function of $X^{n}$ under $P_{M_{n}X^{n}}$;
see Proposition \ref{prop:DetCoupling}. Therefore, in the traditional
intrinsic randomness problem and source resolvability problem, we
relax the constraint on marginal distributions, i.e., we do not constrain
that $P_{X^{n}}=P_{X}^{n}$ and $P_{M_{n}}=P_{M_{n}}^{\mathrm{U}}$,
but require that $P_{M_{n}},P_{M_{n}}^{\mathrm{U}}$ or $P_{X^{n}},P_{X}^{n}$
are asymptotically equal under some distance measure. In this paper
we define exact intrinsic randomness by relaxing the constraint on
the mapping. Specifically, we require that $P_{X^{n}}=P_{X}^{n}$,
$P_{M_{n}}=P_{M_{n}}^{\mathrm{U}}$, and $M_{n}$ is an asymptotic
function of $X^{n}$. 
\begin{defn}
\label{def:Given-a-memoryless}Given a memoryless source $P_{X}$
and a uniform random variable $M_{n}$ with distribution $P_{M_{n}}^{\mathrm{U}}=\mathsf{Unif}[1:e^{nR}]$,
define the\emph{ exact intrinsic randomness rate} $\overline{S}_{\mathsf{E}}(P_{X})$
as the minimum rate needed to ensure there exists a code $P_{M_{n}|X^{n}}$
such that $P_{M_{n}}=P_{M_{n}}^{\mathrm{U}}$, and $M_{n}$ is an
asymptotic function of $X^{n}$ ($\lim_{n\to\infty}\max_{f_{n}}\mathbb{P}\left\{ M_{n}=f_{n}(X^{n})\right\} =1$).
That is, 
\begin{align}
\overline{S}_{\mathsf{E}}(P_{X}): & =\sup\Bigl\{ R:\exists P_{M_{n}|X^{n}}:P_{M_{n}}=P_{M_{n}}^{\mathrm{U}},\nonumber \\
 & \qquad\lim_{n\to\infty}\max_{f_{n}}\mathbb{P}\left\{ M_{n}=f_{n}(X^{n})\right\} =1\Bigr\},
\end{align}
or equivalently, 
\begin{align}
\overline{S}_{\mathsf{E}}(P_{X}): & =\sup\Bigl\{ R:\exists P_{X^{n}M_{n}}\in C(P_{X}^{n},P_{M_{n}}^{\mathrm{U}}):\nonumber \\
 & \qquad\lim_{n\to\infty}\max_{f_{n}}\mathbb{P}\left\{ M_{n}=f_{n}(X^{n})\right\} =1\Bigr\}.
\end{align}
\end{defn}
From Theorem \ref{thm:equivalence}, we know that the problems of
exact and approximate intrinsic randomness are equivalent. 
\begin{cor}[Equivalence Between Exact and Approximate Intrinsic Randomness]
\label{cor:ExactIR} Given a memoryless source $P_{X}^{n}$ and a
uniform distribution $P_{M_{n}}^{\mathrm{U}}$, 
\begin{align}
 & \min_{P_{X^{n}M_{n}}\in C(P_{X}^{n},P_{M_{n}}^{\mathrm{U}})}\min_{f_{n}}\mathbb{P}\left\{ M_{n}\neq f_{n}(X^{n})\right\} \nonumber \\
 & =\min_{f_{n}}|P_{M_{n}}^{\mathrm{U}}-P_{f_{n}(X^{n})}|.
\end{align}
\end{cor}
Combining Corollary \ref{cor:ExactIR} and existing results on approximate
intrinsic randomness, we completely characterize the exact intrinsic
randomness rate. 
\begin{thm}[Exact Intrinsic Randomness]
\label{thm:ExactIR} 
\begin{equation}
\overline{S}_{\mathsf{E}}(P_{X})=H(X).
\end{equation}
\end{thm}
\begin{rem}
It is easy to verify that Corollary \ref{cor:ExactIR} also holds
for general sources. On the other hand, Vembu and Verdú \cite{vembu1995generating}
showed for a general source $\boldsymbol{X}$, the\emph{ intrinsic
randomness rate }for the approximate intrinsic randomness problem
is $\underline{H}(\boldsymbol{X})$. Hence for a general source $\boldsymbol{X}$,
the\emph{ intrinsic randomness rate }$\overline{S}_{\mathsf{E}}(P_{\boldsymbol{X}})$
for the exact intrinsic randomness problem (defined similarly to the
memoryless case) is $\overline{S}_{\mathsf{E}}(P_{\boldsymbol{X}})=\underline{H}(\boldsymbol{X})$. 
\end{rem}
\begin{IEEEproof}
For the approximate intrinsic randomness problem, Han \cite[Theorem 1.6.1]{Han03}
showed there exists a code for the approximate intrinsic randomness
problem if $R<H(X)$ and only if $R\leq H(X)$. Invoking Corollary
\ref{cor:ExactIR} completes the proof of Theorem \ref{thm:ExactIR}. 
\end{IEEEproof}
\begin{thm}[Second Order Rate]
\label{thm:secondorder} Given a memoryless source $P_{X}^{n}$,
the optimal (maximum) code rate $R_{n}^{*}$ generated under the condition
that the output forms a uniform random variable, i.e., $M_{n}\sim\mathsf{Unif}[1:e^{nR_{n}}]$
and $M_{n}$ is an $\varepsilon$-asymptotic function of the output
$X^{n}$, i.e., $\limsup_{n\to\infty}\min_{f_{n}}\mathbb{P}\left\{ M_{n}\neq f_{n}(X^{n})\right\} \leq\varepsilon$,
satisfies 
\begin{equation}
R_{n}^{*}=H(X)-\sqrt{\frac{V(X)}{n}}{\cal Q}^{-1}(\varepsilon)+o\left(\frac{1}{\sqrt{n}}\right),
\end{equation}
where ${\cal Q}$ is the complementary cumulative distribution function
of a standard Gaussian and $V(X)$ is the variance of $\imath_{X}(X)$. 
\end{thm}
\begin{IEEEproof}
Similarly to the proof of Theorem \ref{thm:ExactIR}, we can prove
Theorem \ref{thm:secondorder} by the equivalence between maximal
guessing coupling problem and source resolvability problem (which
is also approximate intrinsic randomness for this case) (Theorem~\ref{thm:equivalence}),
and the second order rate results for the approximate intrinsic randomness
given by Hayashi \cite{hayashi2008second}. 
\end{IEEEproof}

\section{Exact Resolvability}

The maximal guessing coupling problem through a channel defined in
Section \ref{sec:Maximal-Guessing-Coupling} is the minimization of
the error probability of the channel output $Y^{n}$ and the target
variable $Z^{n}$. Theorem \ref{thm:equivalence-1} shows this problem
is equivalent to the traditional channel resolvability problem (with
the TV distance measure).

In this section, we consider a new channel (or source) resolvability
problem, named \emph{exact channel (or source) resolvability problem}.
In this problem, we require that $P_{Y^{n}}=P_{Y}^{n}$, $P_{M_{n}}=P_{M_{n}}^{\mathrm{U}}$,
and the channel input $X^{n}$ is an asymptotic function of $M_{n}$
($\lim_{n\to\infty}\max_{f_{n}}\mathbb{P}\left\{ X^{n}=f_{n}(M_{n})\right\} =1$). 
\begin{defn}
Given a uniform random variable $M_{n}$ with distribution $P_{M_{n}}^{\mathrm{U}}=\mathsf{Unif}[1:e^{nR}]$
a memoryless channel $P_{Y|X}$, and a target distribution $P_{Y}$,
define the\emph{ exact channel resolvability rate} $\overline{G}_{\mathsf{E}}(P_{Y|X},P_{Y})$
as the minimum rate needed to ensure there exists a code $P_{X^{n}|M_{n}}$
such that $P_{Y^{n}}=P_{Y}^{n}$, and the channel input $X^{n}$ is
an asymptotic function of $M_{n}$ ($\lim_{n\to\infty}\max_{f_{n}}\mathbb{P}\left\{ X^{n}=f_{n}(M_{n})\right\} =1$).
That is,

\begin{align}
\overline{G}_{\mathsf{E}}(P_{Y|X},P_{Y}): & =\inf\Bigl\{ R:\exists P_{X^{n}|M_{n}}:P_{Y^{n}}=P_{Y}^{n},\nonumber \\
 & \qquad\lim_{n\to\infty}\max_{f_{n}}\mathbb{P}\left\{ X^{n}=f_{n}(M_{n})\right\} =1\Bigr\}.
\end{align}
If the channel $P_{Y|X}$ is an identity channel, we define\emph{
exact source resolvability rate} 
\begin{align}
\overline{G}_{\mathsf{E}}(P_{X}): & =\inf\Bigl\{ R:\exists P_{X^{n}|M_{n}}:P_{X^{n}}=P_{X}^{n},\nonumber \\
 & \qquad\lim_{n\to\infty}\max_{f_{n}}\mathbb{P}\left\{ X^{n}=f_{n}(M_{n})\right\} =1\Bigr\},
\end{align}
or equivalently, 
\begin{align}
\overline{G}_{\mathsf{E}}(P_{X}): & =\inf\Bigl\{ R:\exists P_{M_{n}X^{n}}\in C(P_{M_{n}}^{\mathrm{U}},P_{X}^{n}):\nonumber \\
 & \qquad\lim_{n\to\infty}\max_{f_{n}}\mathbb{P}\left\{ X^{n}=f_{n}(M_{n})\right\} =1\Bigr\}.
\end{align}
\end{defn}
\begin{cor}[Source Resolvability]
\label{cor:ExactIR-1-1} Given a memoryless source $P_{X}^{n}$ and
a uniform distribution $P_{M_{n}}^{\mathrm{U}}$, 
\begin{align}
 & \min_{P_{M_{n}X^{n}}\in C(P_{M_{n}}^{\mathrm{U}},P_{X}^{n})}\min_{f_{n}}\mathbb{P}\left\{ X^{n}\neq f_{n}(M_{n})\right\} \nonumber \\
 & \qquad=\min_{f_{n}}|P_{X}^{n}-P_{f_{n}(M_{n})}|.\label{eq:-11}
\end{align}
Furthermore, 
\begin{equation}
\overline{G}_{\mathsf{E}}(P_{X})=H(X).\label{eq:-12}
\end{equation}
\end{cor}
\begin{rem}
\label{rem:It-is-easy}It is easy to verify that the equivalence \eqref{eq:-11}
also holds for general sources. On the other hand, Han and Verdú \cite{Han}
showed for a general source $\boldsymbol{X}$, the resolvability rate
for the approximate source resolvability problem is $\overline{H}(\boldsymbol{X})$.
Hence for a general source $\boldsymbol{X}$, the resolvability rate
$\overline{G}_{\mathsf{E}}(P_{\boldsymbol{X}})$ for the exact source
resolvability problem (defined similarly to the memoryless case) is
$\overline{G}_{\mathsf{E}}(P_{\boldsymbol{X}})=\overline{H}(\boldsymbol{X})$. 
\end{rem}
\begin{IEEEproof}
The equivalence \eqref{eq:-11} follows from Theorem \ref{thm:equivalence}.
Furthermore, Han and Verdú \cite{Han} showed there exists a code
for the approximate source resolvability problem if $R<H(X)$ and
only if $R\leq H(X)$. Combining these two observations yields~\eqref{eq:-12}. 
\end{IEEEproof}
Denote 
\begin{equation}
\mathcal{P}(P_{Y|X},Q_{Y}):=\left\{ P_{X}:P_{Y|X}\circ P_{X}=Q_{Y}\right\} ,\label{eq:-35}
\end{equation}
and assume $\mathcal{P}(P_{Y|X},Q_{Y})\ne\emptyset$. We are now are
ready to establish the following multiletter characterization for
the exact channel resolvability rate. The proof of Proposition \ref{prop:Resolvability}
is given in Appendix \ref{sec:Resolvability}. 
\begin{prop}[Multiletter Characterization of $\overline{G}_{\mathsf{E}}(P_{Y|X},P_{Y})$]
\label{prop:Resolvability} 
\begin{align}
\overline{G}_{\mathsf{E}}(P_{Y|X},P_{Y}) & =\inf_{P_{\boldsymbol{X}}\in\mathcal{P}(P_{\boldsymbol{Y}|\boldsymbol{X}},P_{\boldsymbol{Y}})}\overline{H}(\boldsymbol{X})\\
 & =\lim_{n\to\infty}\min_{P_{X^{n}}\in\mathcal{P}(P_{Y|X}^{n},P_{Y}^{n})}\frac{1}{n}H(X^{n}),\label{eq:-43}
\end{align}
where $\mathcal{P}(P_{\boldsymbol{Y}|\boldsymbol{X}},P_{\boldsymbol{Y}}):=\big\{ P_{\boldsymbol{X}}:P_{X^{n}}\in\mathcal{P}(P_{Y|X}^{n},P_{Y}^{n}),\forall n\big\}$. 
\end{prop}
\begin{rem}
Unlike our definition, Li and El Gamal \cite{li2017distributed} defined
an exact common information rate by considering variable-length coding.
However, their exact common information rate has a similar characterization
as \eqref{eq:-43}, i.e., the exact common information rate 
\begin{equation}
\overline{C}_{\mathsf{E}}(X;Y)=\lim_{n\to\infty}\min_{P_{W_{n}|X^{n}Y^{n}}:X^{n}\to W_{n}\to Y^{n}}\frac{1}{n}H(W_{n}).\label{eq:-40}
\end{equation}
\end{rem}
Furthermore, we can bound $\overline{G}_{\mathsf{E}}(P_{Y|X},P_{Y})$
as follows. 
\begin{prop}
\begin{equation}
G_{\mathsf{TV}}(P_{Y|X},P_{Y})\leq\overline{G}_{\mathsf{E}}(P_{Y|X},P_{Y})\le G_{\mathsf{E}}(P_{Y|X},P_{Y}),
\end{equation}
where $G_{\mathsf{TV}}(P_{Y|X},P_{Y}):=\min_{P_{X}\in\mathcal{P}(P_{Y|X},P_{Y})}I(X;Y)$
denotes the channel resolvability rate under the TV distance measure,
and $G_{\mathsf{E}}(P_{Y|X},P_{Y}):=\min_{P_{X}\in\mathcal{P}(P_{Y|X},P_{Y})}H(P_{X})$. 
\end{prop}
\begin{IEEEproof}
The upper bound is obtained by choosing $X^{n}$ in \eqref{eq:-43}
such that $P_{X^{n}}=P_{X}^{n}$ with $P_{X}\in\mathcal{P}(P_{Y|X},P_{Y})$.
The lower bound is obtained by the following chain of inequalities:
\begin{align}
 & \overline{G}_{\mathsf{E}}(P_{Y|X},P_{Y})\nonumber \\
 & =\lim_{n\to\infty}\min_{P_{X^{n}}\in\mathcal{P}(P_{Y|X}^{n},P_{Y}^{n})}\frac{1}{n}H(X^{n})\\
 & \geq\liminf_{n\to\infty}\min_{P_{X^{n}}\in\mathcal{P}(P_{Y|X}^{n},P_{Y}^{n})}\frac{1}{n}I(X^{n};Y^{n})\\
 & =\liminf_{n\to\infty}\min_{P_{X^{n}}\in\mathcal{P}(P_{Y|X}^{n},P_{Y}^{n})}\frac{1}{n}\left(H(Y^{n})-H(Y^{n}|X^{n})\right)\\
 & =\liminf_{n\to\infty}\min_{P_{X^{n}}\in\mathcal{P}(P_{Y|X}^{n},P_{Y}^{n})}\frac{1}{n}\sum_{i=1}^{n}\left(H(Y_{i})-H(Y_{i}|X_{i})\right)\\
 & =\liminf_{n\to\infty}\min_{P_{X^{n}}\in\mathcal{P}(P_{Y|X}^{n},P_{Y}^{n})}\frac{1}{n}\sum_{i=1}^{n}I(X_{i};Y_{i})\\
 & =\liminf_{n\to\infty}\min_{P_{X^{n}}\in\mathcal{P}(P_{Y|X}^{n},P_{Y}^{n})}I(X_{J};Y_{J}|J)\label{eq:-41}\\
 & \geq\liminf_{n\to\infty}\min_{P_{X^{n}}\in\mathcal{P}(P_{Y|X}^{n},P_{Y}^{n})}I(X_{J};Y_{J})\label{eq:-420}\\
 & =\min_{P_{X}\in\mathcal{P}(P_{Y|X},P_{Y})}I(X;Y),\label{eq:-42}
\end{align}
where in \eqref{eq:-41} $J\sim\mathsf{Unif}[1:n]$ denotes a time-sharing
random variable, \eqref{eq:-41} follows from that $Y_{J}$ is independent
of $J$ since $Y^{n}$ are i.i.d. under $P_{Y}^{n}$, and in \eqref{eq:-42}
$X:=X_{J}$ and $Y:=Y_{J}$. 
\end{IEEEproof}
\begin{prop}
\label{prop:Nether-the-upper}Neither the upper bound $G_{\mathsf{E}}(P_{Y|X},P_{Y})$
nor the lower bound $G_{\mathsf{TV}}(P_{Y|X},P_{Y})$ is tight in
general, i.e., there exists $P_{Y|X},P_{Y}$ such that 
\begin{equation}
\overline{G}_{\mathsf{E}}(P_{Y|X},P_{Y})<G_{\mathsf{E}}(P_{Y|X},P_{Y})
\end{equation}
and also there exists $P_{Y|X},P_{Y}$ such that 
\begin{equation}
G_{\mathsf{TV}}(P_{Y|X},P_{Y})<\overline{G}_{\mathsf{E}}(P_{Y|X},P_{Y}).
\end{equation}
\end{prop}
This proposition implies the exact and approximate channel resolvability
are not equivalent. In general, the exact channel resolvability requires
a larger rate.

\subsection{$P_{Y}$-non-redundant Channel}

Although the upper bound $G_{\mathsf{E}}(P_{Y|X},P_{Y})$ is not tight
in general, we will show it is tight for some special cases, e.g.,
full-rank channels and additive channels. Hence next, we focus on
full-rank channels and additive channels, and prove $\overline{G}_{\mathsf{E}}(P_{Y|X},P_{Y})=G_{\mathsf{E}}(P_{Y|X},P_{Y})$
for these two classes of channels. 
\begin{defn}
\label{def:We-say-}We say $P_{Y|X}$ is a \emph{$P_{Y}$-non-redundant
channel} if given $P_{Y|X}$ and $P_{Y}$, the equation $\boldsymbol{P}_{Y|X}\boldsymbol{P}_{X}=\boldsymbol{P}_{Y}$
has a unique solution $P_{X}$. That is, there exists a unique distribution
$P_{X}$ that induces $P_{Y}$ through $P_{Y|X}$. 
\end{defn}
\begin{defn}
We say $P_{Y|X}$ is a \emph{full-rank channel} if $\textrm{rank}(\boldsymbol{P}_{Y|X})=|\mathcal{X}|$. 
\end{defn}
\begin{defn}
We say $P_{X}$ is a \emph{degenerate distribution} if $P_{X}(x_{0})=1$
for some $x_{0}$ and $P_{X}(x)=0$ for $x\neq x_{0}$. 
\end{defn}
\begin{lem}
\label{lem:Properties:}The following properties hold. 

{[}leftmargin={*}{]} 
\begin{enumerate}
\item If $P_{Y|X}$ is a $P_{Y}$-non-redundant channel, then either $P_{Y|X}$
is a full-rank channel or $P_{X}$ is a degenerate distribution. 
\item For any $n\in\mathbb{N}$, $P_{Y|X}^{n}$ is a $P_{Y}^{n}$-non-redundant
channel, if and only if $P_{Y|X}$ is a $P_{Y}$-non-redundant channel. 
\item Any additive channel $Y=X+Z$ with $Z$ independent of $X$, is a
full-rank channel. 
\item If $P_{X^{n}}\in\mathcal{P}(P_{Y|X}^{n},P_{Y}^{n})$ for some $n\in\mathbb{N}$,
then $\prod_{i=1}^{n}P_{X_{i}}\in\mathcal{P}(P_{Y|X}^{n},P_{Y}^{n})$. 
\end{enumerate}
\end{lem}
\begin{rem}
In general, $P_{X^{n}}\in\mathcal{P}(P_{Y|X}^{n},P_{Y}^{n})$ does
not imply $P_{X^{n}}$ must be a product distribution or that is uniquely
defined. However if $P_{Y|X}$ is a $P_{Y}$-non-redundant channel,
it does imply that $P_{X^{n}}$ must be a product distribution and
that it is unique. 
\end{rem}
\begin{IEEEproof}
Proof of Property 1): Consider the linear equation $\boldsymbol{P}_{Y|X}\boldsymbol{Q}=\boldsymbol{P}_{Y}$
where we do not constrain $\boldsymbol{Q}$ to a probability distribution,
i.e., some components can be negative. We know that it must have no
solution, a unique solution, or infinitely many solutions.

If $\boldsymbol{P}_{X}$ is a probability distribution and the linear
equation $\boldsymbol{P}_{Y|X}\boldsymbol{P}_{X}=\boldsymbol{P}_{Y}$
has a unique solution, then it means that the set of solutions of
$\boldsymbol{P}_{Y|X}\boldsymbol{Q}=\boldsymbol{P}_{Y}$ and the probability
simplex $\left\{ \boldsymbol{P}_{X}:\sum_{x}P_{X}(x)=1,P_{X}(x)\ge0\right\} $
intersect at a single point. Hence either $\boldsymbol{P}_{Y|X}\boldsymbol{Q}=\boldsymbol{P}_{Y}$
has a single unique solution, or it has infinitely many solutions
but they intersect with the probability simplex at the vertices points
of the probability simplex. These two cases respectively correspond
to the case $\textrm{rank}(\boldsymbol{P}_{Y|X})=|\mathcal{X}|$ and
the case where the solution is $P_{X}(x_{0})=1$ for some $x_{0}$
and $P_{X}(x)=0$ for $x\neq x_{0}$.

Property 2) follows from Property 1).

Proof of Property 3): $\boldsymbol{P}_{Y|X}=\boldsymbol{I}_{|\mathcal{X}|}\otimes\boldsymbol{P}_{Z}$,
where $\boldsymbol{I}_{|\mathcal{X}|}$ denotes the identity matrix
with size $|\mathcal{X}|$. Hence $\textrm{rank}(\boldsymbol{P}_{Y|X})=\textrm{rank}(\boldsymbol{I}_{|\mathcal{X}|})\textrm{rank}(\boldsymbol{P}_{Z})=|\mathcal{X}|$.

Property 4) is obvious. 
\end{IEEEproof}
\begin{thm}
If the channel $P_{Y|X}$ is a $P_{Y}$-non-redundant channel, then
\begin{equation}
\overline{G}_{\mathsf{E}}(P_{Y|X},P_{Y})=G_{\mathsf{E}}(P_{Y|X},P_{Y})=H(P_{X}),\label{eq:-1-1}
\end{equation}
where $P_{X}$ is the unique distribution that induces $P_{Y}$ through
$P_{Y|X}$. 
\end{thm}
For an AWGN (additive white Gaussian noise) channel $P_{Y|X}$ and
a Gaussian distribution $P_{Y}$, we have that $P_{X}\in\mathcal{P}\left(P_{Y|X},P_{Y}\right)$
is also Gaussian and unique. So for this case, we get the following
result. 
\begin{prop}
For an AWGN channel $P_{Y|X}$ and a Gaussian distribution $P_{Y}$,
we have 
\begin{equation}
\overline{G}_{\mathsf{E}}(P_{Y|X},P_{Y})=G_{\mathsf{E}}(P_{Y|X},P_{Y})=H(P_{X})=\infty.
\end{equation}
\end{prop}
\begin{rem}
The exact channel resolvability rate is infinite, although the approximate
channel resolvability rate $G_{\mathsf{TV}}(P_{Y|X},P_{Y})=\frac{1}{2}\log\frac{N_{Y}}{N_{Z}}$
is finite. This point is different from the exact common information.
Li and El Gamal \cite{li2017distributed} showed the exact common
information $C_{\mathsf{E}}(X;Y)$ satisfies 
\[
I(X;Y)\le C_{\mathsf{Wyner}}(X;Y)\le\overline{C}_{\mathsf{E}}(X;Y)\le I(X;Y)+24\log2.
\]
where $C_{\mathsf{Wyner}}(X;Y)$ is Wyner's common information, and
$\overline{C}_{\mathsf{E}}(X;Y)$ is the exact common information.
Applying this result to two jointly Gaussian random variables shows
that only a finite amount of common randomness is needed for simulating
them in a distributed manner.

Next we consider the second-order rate for the exact channel resolvability
problem. Given a memoryless channel $P_{Y|X}$, define $R_{n}^{*}$
as the optimal (minimum) code rate needed to ensure the channel output
follows distribution $P_{Y}^{n}$ and $X^{n}$ is an $\varepsilon$-asymptotic
function of the output $M_{n}$, i.e., $\limsup_{n\to\infty}\min_{f_{n}}\mathbb{P}\left\{ X^{n}\neq f_{n}(M_{n})\right\} \leq\varepsilon$. 
\end{rem}
\begin{thm}[Second Order Rate for $P_{Y}$-non-redundant Channels]
\label{theorem second order-1-3} Given a memoryless $P_{Y}$-non-redundant
channel, we have 
\begin{equation}
R_{n}^{*}=H(X)+\sqrt{\frac{V(X)}{n}}{\cal Q}^{-1}(\varepsilon)+o\left(\frac{1}{\sqrt{n}}\right),\label{eqn:rn_star}
\end{equation}
where $P_{X}$ is the unique distribution that induces $P_{Y}$ through
$P_{Y|X}$. 
\end{thm}
\begin{IEEEproof}
For $P_{Y}$-non-redundant channels, the channel input distribution
is unique and equal to $P_{X}$. Hence for this case, the exact channel
resolvability problem is equivalent to the exact source resolvability
problem. On the other hand, by Corollary \ref{cor:ExactIR-1-1} we
know that the exact source resolvability problem is also equivalent
to the approximate source resolvability problem. Hence the exact channel
resolvability problem is equivalent to the approximate source resolvability
problem. Furthermore, for the latter problem, Nomura and Han \cite[Theorem 1.6.1]{nomura2013second}
showed that the optimal rate is as in \eqref{eqn:rn_star}. 
\end{IEEEproof}

\section{Channel Capacity With Input Distribution Constraint }
\begin{defn}
Given a distribution $P_{X}$, the channel capacity with input distribution
constraint $P_{X}$ is defined as the maximum rate $R$ such that
there exists a sequence of codes $(P_{X^{n}|M_{n}},P_{\widehat{M}_{n}|Y^{n}})_{n=1}^{\infty}$
satisfying $P_{X^{n}}=P_{X}^{n}$ and $\lim_{n\to\infty}\mathbb{P}\{M_{n}=\widehat{M}_{n}\}=1$
with $M_{n}\sim\mathsf{Unif}[1:e^{nR}]$. That is, 
\begin{align}
C\left(P_{X}\right): & =\sup\Bigl\{ R:\exists(P_{X^{n}|M_{n}},P_{\widehat{M}_{n}|Y^{n}})_{n=1}^{\infty}:\nonumber \\
 & \qquad P_{X^{n}}=P_{X}^{n},\lim_{n\to\infty}\mathbb{P}\left\{ M_{n}=\widehat{M}_{n}\right\} =1\Bigr\}.
\end{align}
\end{defn}
\begin{thm}
\label{thm:capacity}$C\left(P_{X}\right)=C_{\mathsf{GK}}(X;Y)$,
where $C_{\mathsf{GK}}(X;Y)$ denotes the GK common information between
$X$ and $Y$ (under the distribution $P_{X}P_{Y|X}$). 
\end{thm}
\begin{rem}
$C\left(P_{X}\right)\leq I(X;Y)\leq C$, where $C$ denotes the traditional
Shannon capacity (i.e., the channel capacity without the input distribution
constraint). 
\end{rem}
\begin{IEEEproof}
Assume $W$ is a common part of $X$ and $Y$ (under distribution
$P_{X}P_{Y|X}$) (i.e., $W=g(X)=h(Y)$ a.s. for some functions $g$
and $h$). If $R<H(W)$, then according to Theorem \ref{thm:MaxGuessCoup}
there exists a maximal guessing coupling $P_{M_{n}W^{n}}$ such that
$P_{M_{n}}=\mathsf{Unif}[1:e^{nR}]$ and $\max_{f_{n}}\mathbb{P}\left\{ M_{n}=f_{n}(W^{n})\right\} \to1$.
Assume $f_{n}$ is a maximizing function of $\max_{f_{n}}\mathbb{P}\left\{ M_{n}=f_{n}(W^{n})\right\} $.
Apply $P_{W^{n}|M_{n}}(w^{n}|m)P_{X|W}^{n}(x^{n}|w^{n})$ as the encoder,
and $P_{W|Y}^{n}(w^{n}|y^{n})\cdot1\left\{ \widehat{m}=f_{n}\left(w^{n}\right)\right\} $
as the decoder. Then $P_{X^{n}}=P_{X}^{n}$ and $\lim_{n\to\infty}\mathbb{P}\{M_{n}=\widehat{M}_{n}\}=1$.
Hence $C\left(P_{X}\right)\geq C_{\mathsf{GK}}(X;Y)$.

On the other hand, we can convert a code for the problem of channel
capacity with input distribution constraint $P_{X}$ into a code for
the GK common information problem. For any code $(P_{X^{n}|M_{n}},P_{\widehat{M}_{n}|Y^{n}})$
satisfying $P_{X^{n}}=P_{X}^{n}$ and $\lim_{n\to\infty}\mathbb{P}\{M_{n}=\widehat{M}_{n}\}=1$,
the induced joint distribution of $X^{n}$ and $Y^{n}$ is the product
distribution $P_{XY}^{n}$. Hence $(P_{M_{n}|X^{n}},P_{\widehat{M}_{n}|Y^{n}})$
forms a code for the GK common information problem \cite{gacs1973common}.
According to the converse for GK common information problem, we conclude
that the code rate is not larger than $C_{\mathsf{GK}}(X;Y)$. 
\end{IEEEproof}
Next we consider the second-order rate. Given a distribution $P_{X}$,
define $R_{n}^{*}$ as the optimal (maximum) code rate needed to ensure
that there exists a sequence of codes satisfying $P_{X^{n}}=P_{X}^{n}$
and $\limsup_{n\to\infty}\mathbb{P}\{M_{n}\neq\widehat{M}_{n}\}\leq\varepsilon$.
In order to present the bounds on the second-order rate on $R_{n}^{*}$,
we need define some quantities. Given a distribution $P_{X}$, let
$W$ be a common random variable of $X$ and $Y$ (under the distribution
$P_{X}P_{Y|X}$), i.e., $W=f(X)=g(Y)$ for some functions $f$ and
$g$ achieving $C_{\mathsf{GK}}(X;Y)$ in \eqref{eq:-77-2} (where
the $\sup$ is a $\max$ for finite-valued $X$ and $Y$). Let $\rho_{\mathrm{m}}(X;Y|W)$
denote the conditional maximal correlation \cite{csiszar2000common,yu2016generalized}
between $X$ and $Y$ given the common random variable $W$ defined
as 
\begin{align}
 & \rho_{\mathrm{m}}(X;Y|W)\nonumber \\
 & :=\sup\left\{ \frac{\mathbb{E}[\mathrm{cov}(g(X,W),h(Y,W)|W)]}{\sqrt{\mathbb{E}[\mathrm{var}(g(X,W)|W)]}\sqrt{\mathbb{E}[\mathrm{var}(h(Y,W)|W)]}}\right\} 
\end{align}
where the supremum extends over all functions $g:\mathcal{X}\times\mathcal{W}\to\mathbb{R}$
and $h:\mathcal{Y}\times\mathcal{W}\to\mathbb{R}$ satisfying 
\begin{align}
\mathbb{E}[\mathrm{var}(g(X,W)|W)]>0,\quad\mbox{and}\quad\mathbb{E}[\mathrm{var}(h(Y,W)|W)]>0.
\end{align}
Denote 
\begin{equation}
\varepsilon^{*}:=\frac{-\frac{1}{3}+\sqrt{1+(1-\rho_{\mathrm{m}}(X;Y|W))^{-2}}}{2\Bigl(1+\frac{9}{8}(1-\rho_{\mathrm{m}}(X;Y|W))^{-2}\Bigr)}.\label{eq:epsilonstar}
\end{equation}
For $\theta\in(0,1)$ such that $\frac{\varepsilon}{\theta}<\min\{\varepsilon^{*},1/3\}$,
let $\xi^{*}(\frac{\varepsilon}{\theta})$ is the unique solution
on $(\frac{\varepsilon}{\theta},1/3)$ to the equation 
\begin{equation}
2(1-\rho_{\mathrm{m}}(X;Y|W))\sqrt{\xi(1-\xi)(\xi-\frac{\varepsilon}{\theta})(1-\xi+\frac{\varepsilon}{\theta})}=\frac{\varepsilon}{\theta}\label{eq:equation}
\end{equation}
with $\xi$ unknown.  For $0<\varepsilon<\min\{\varepsilon^{*},1/3\}$,
denote 
\begin{align}
\mu(\varepsilon):= & \inf_{\max\{3\varepsilon,\frac{\varepsilon}{\varepsilon^{*}}\}\le\theta<1}\left\{ 1-\left(1-\theta\right)\left(1-\xi^{*}(\frac{\varepsilon}{\theta})\right)\right\} .\label{eq:theta}
\end{align}

\begin{thm}[Second Order Rate]
\label{thm:secondorder-1} Given a distribution $P_{X}$, we have
for $0<\varepsilon<1$, 
\begin{align}
 & R_{n}^{*}\ge C_{\mathsf{GK}}(X;Y)-\sqrt{\frac{V(W)}{n}}{\cal Q}^{-1}(\varepsilon)+o\left(\frac{1}{\sqrt{n}}\right),\label{eq:-58}
\end{align}
and for $0<\varepsilon<\min\{\varepsilon^{*},1/3\}$, 
\begin{align}
 & R_{n}^{*}\leq C_{\mathsf{GK}}(X;Y)-\sqrt{\frac{V(W)}{n}}{\cal Q}^{-1}(\mu(\varepsilon))+o\left(\frac{1}{\sqrt{n}}\right).\label{eq:gk}
\end{align}
where $W$ is a common random variable of $X$ and $Y$, $\varepsilon^{*}$
is defined in \eqref{eq:epsilonstar}, and $\mu(\varepsilon)$ is
defined in \eqref{eq:theta}. 
\end{thm}
\begin{IEEEproof}
Achievability (Lower Bound): Consider the coding scheme used in the
proof of Theorem \ref{thm:capacity}. By the achievability part of
Theorem \ref{thm:secondorder}, we have that if 
\begin{equation}
\limsup_{n\to\infty}\sqrt{n}(R_{n}-H(W))<-\sqrt{V(W)}{\cal Q}^{-1}(\varepsilon),
\end{equation}
then there exists a maximal guessing coupling $P_{M_{n}W^{n}}$ such
that $P_{M_{n}}=\mathsf{Unif}[1:e^{nR_{n}}]$ and 
\begin{equation}
\limsup_{n\to\infty}\min_{f_{n}}\mathbb{P}\left\{ M_{n}\neq f_{n}(W^{n})\right\} \leq\varepsilon.\label{eqn:limsup_min}
\end{equation}
On the other hand, the legitimate user first recovers $W^{n}$ losslessly
and then reconstructs $M_{n}$ as $\widehat{M}_{n}=f_{n}\left(W^{n}\right)$.
Hence \eqref{eqn:limsup_min} implies that $\limsup_{n\to\infty}\mathbb{P}\{M_{n}\neq\widehat{M}_{n}\}\leq\varepsilon$.
That is, \eqref{eq:-58} holds. 

Converse (Upper Bound): To show converse, we need the following lemma,
which is a quantitative version of \cite[Lemma~1.1]{csiszar2000common}. 
\begin{lem}
\cite[Lemma 1.1]{csiszar2000common} \label{lem:csiszar} Given a
pair of random variables $(X,Y)$, let $W$ be a common random variable
of $X$ and $Y$. Let $U,V$ be two random variables such that $U\to X\to Y\to V$
and 
\begin{equation}
\mathbb{P}\left\{ U\neq V\right\} \leq\varepsilon\label{eq:UV}
\end{equation}
for some $0<\varepsilon<\min\{\varepsilon^{*},1/3\}$, where $\varepsilon^{*}$
is defined in \eqref{eq:epsilonstar}. Then 
\begin{equation}
\inf_{h:\mathcal{W}\to\mathcal{U}}\mathbb{P}\left\{ U\neq h(W)\right\} \leq\mu(\varepsilon),
\end{equation}
where $\mu(\varepsilon)$ is defined in \eqref{eq:theta}. 
\end{lem}
For completeness, we provide the proof of Lemma \ref{lem:csiszar}
at the end of this proof. Applying this lemma to our setting by the
identification $(X^{n},Y^{n},M_{n},\widehat{M}_{n})$ as $(X,Y,U,V)$,
we have 
\begin{equation}
\inf_{f}\mathbb{P}\left\{ M_{n}\neq f(W^{n})\right\} \leq\mu\left(\mathbb{P}\left\{ M_{n}\neq\widehat{M}_{n}\right\} \right).
\end{equation}
Taking limsup's, we have 
\begin{align}
 & \limsup_{n\to\infty}\inf_{f}\mathbb{P}\left\{ M_{n}\neq f(W^{n})\right\} \nonumber \\
 & \leq\limsup_{n\to\infty}\mu\left(\mathbb{P}\left\{ M_{n}\neq\widehat{M}_{n}\right\} \right)\\
 & \leq\mu(\varepsilon),\label{eq:-33}
\end{align}
where \eqref{eq:-33} follows since $\mu(\varepsilon)$ is continuous
and non-decreasing in $\varepsilon$. Then by the converse part of
Theorem \ref{thm:secondorder}, we have that any achievable $\{R_{n}\}_{n=1}^{\infty}$
must satisfy 
\begin{equation}
\limsup_{n\to\infty}\sqrt{n}(R_{n}-H(W))<-\sqrt{{V(W)}}{\cal Q}^{-1}(\mu(\varepsilon)),
\end{equation}
which completes the proof of the upper bound in~\eqref{eq:gk}. 
\end{IEEEproof}
\begin{IEEEproof}[Proof of Lemma \ref{lem:csiszar}]
We first make the following claim. 
\begin{claim}
\label{claim:It-holds-that} If additionally, $C_{\mathsf{GK}}(X;Y)=0$,
then $p_{\max}:=\max_{u}P_{U}(u)\ge1-\xi^{*}(\varepsilon)$ where
$\xi^{*}(\varepsilon)$ is the unique solution on $(\varepsilon,1/3)$
to the equation \eqref{eq:equation} with $\theta=1$. 
\end{claim}
We now prove this claim. By assumption, $C_{\mathsf{GK}}(X;Y)=0$,
i.e., any common random variables $W$ are constant. Denote 
\begin{equation}
\varphi(\xi):=2(1-\rho_{\mathrm{m}}(X;Y|W))\sqrt{\xi(1-\xi)(\xi-\varepsilon)(1-\xi+\varepsilon)}.
\end{equation}
Obviously, $\varphi(\xi)$ is continuous and increasing in $\xi\in(\varepsilon,1/3)$.
Moreover, $\varphi(\varepsilon)=0<\varepsilon$ and $\varphi(1/3)>\varepsilon$.
The latter inequality follows by the assumption $0<\varepsilon<\min\{\varepsilon^{*},1/3\}$.
Hence, there is a unique solution $\xi^{*}(\varepsilon)\in(\varepsilon,1/3)$
to the equation $\varphi(\xi)=\varepsilon$. For brevity, we denote
$\xi^{*}:=\xi^{*}(\varepsilon)$. Suppose instead that $p_{\max}<1-\xi^{*}$.
Then, there exists a set $\mathcal{A}\subseteq\mathcal{U}$ such that
\begin{equation}
\xi^{*}<P_{U}(\mathcal{A})<1-\xi^{*}.\label{eq:A}
\end{equation}
(Sort elements in $\mathcal{U}$ as $u_{1},u_{2},...,u_{m}$ such
that $P_{U}(u_{1})\ge P_{U}(u_{2})\ge...\ge P_{U}(u_{m})$. If $p_{\max}=P_{U}(u_{1})>\xi^{*}$,
then $\mathcal{A}$ can be chosen as $\{u_{1}\}$. If $p_{\max}<\xi^{*}$,
then $\mathcal{A}$ can be chosen as $\{u_{1},u_{2},...,u_{k}\}$
for some $1\le k\le m$ such that \eqref{eq:A} holds. The existence
of such $k$ follows since $\xi^{*}<1/3$.) 

Observe that \eqref{eq:UV} implies that 
\begin{equation}
\mathbb{P}\left\{ 1_{\mathcal{A}}(U)\neq1_{\mathcal{A}}(V)\right\} \leq\varepsilon.\label{eq:UV-1}
\end{equation}
By \eqref{eq:A} and \eqref{eq:UV-1}, we get
\begin{equation}
\xi^{*}<\mathbb{P}\left\{ 1_{\mathcal{A}}(U)=1\right\} <1-\xi^{*}\label{eq:UV-1-1}
\end{equation}
and 
\begin{equation}
\xi^{*}-\varepsilon<\mathbb{P}\left\{ 1_{\mathcal{A}}(V)=1\right\} <1-\xi^{*}+\varepsilon.\label{eq:UV-1-1-1}
\end{equation}

By \cite[Theorem 2]{witsenhausen1975sequences}, 
\begin{align}
 & 2(1-\rho_{\mathrm{m}}(X;Y))\sqrt{\mathbb{P}\left\{ 1_{\mathcal{A}}(U)=1\right\} \mathbb{P}\left\{ 1_{\mathcal{A}}(U)=0\right\} }\nonumber \\
 & \qquad\times\sqrt{\mathbb{P}\left\{ 1_{\mathcal{A}}(V)=1\right\} \mathbb{P}\left\{ 1_{\mathcal{A}}(V)=0\right\} }\nonumber \\
 & \leq\mathbb{P}\left\{ 1_{\mathcal{A}}(U)\neq1_{\mathcal{A}}(V)\right\} ,
\end{align}
which implies that 
\begin{equation}
\varphi(\xi^{*})<\varepsilon.\label{eq:eps}
\end{equation}
This contradicts with the assumption $\varphi(\xi^{*})=\varepsilon$.
Hence, $p_{\max}\ge1-\xi^{*}(\varepsilon)$, i.e., Claim \ref{claim:It-holds-that}
holds. 

We now turn back to prove Lemma \ref{lem:csiszar}. Note that as assumed,
$C_{\mathsf{GK}}(X;Y)>0$. For each $w$, denote $\rho_{\mathrm{m}}(X;Y|W=w)=\rho_{\mathrm{m}}(X';Y')$
where $(X',Y')\sim P_{XY|W=w}$. Then, $\rho_{\mathrm{m}}(X;Y|W)=\max_{w}\rho_{\mathrm{m}}(X;Y|W=w)$.
 Let $\theta$ be such that $\frac{\varepsilon}{\theta}<\min\{\varepsilon^{*},1/3\}$,
which implies that $\frac{\varepsilon}{\theta}<\varepsilon_{w}^{*}:=\frac{-\frac{1}{3}+\sqrt{1+(1-\rho_{\mathrm{m}}(X;Y|W=w))^{-2}}}{2\Bigl(1+\frac{9}{8}(1-\rho_{\mathrm{m}}(X;Y|W=w))^{-2}\Bigr)}$
for all $w$ since $\varepsilon_{w}^{*}\le\varepsilon^{*}$. Denote
$\mathcal{B}$ as the set of $w$ such that 
\begin{equation}
\mathbb{P}\left\{ U\neq V|W=w\right\} \leq\frac{\varepsilon}{\theta}.\label{eq:UV-2}
\end{equation}
By definition, given $W=w$, $C_{\mathsf{GK}}(X';Y')=0$ for $(X',Y')\sim P_{XY|W=w}$.
Then, applying Claim \ref{claim:It-holds-that} to $(X',Y')\sim P_{XY|W=w}$,
we have 
\begin{equation}
p_{\max}^{(w)}:=\max_{u}P_{U|W}(u|w)\ge1-\xi_{w}^{*}(\frac{\varepsilon}{\theta}),\label{eq:pmax_w}
\end{equation}
where $\xi_{w}^{*}(\frac{\varepsilon}{\theta})$ is the unique solution
on $(\frac{\varepsilon}{\theta},1/3)$ to the equation $\varphi_{w}(\rho_{\mathrm{m}}(X;Y|W=w),\xi)=\frac{\varepsilon}{\theta}$
with $\xi$ unknown and with 
\begin{equation}
\varphi_{w}(s,\xi):=2(1-s)\sqrt{\xi(1-\xi)(\xi-\frac{\varepsilon}{\theta})(1-\xi+\frac{\varepsilon}{\theta})}.
\end{equation}
Since $\varphi_{w}(\rho_{\mathrm{m}}(X;Y|W=w),\xi)\ge\varphi_{w}(\rho_{\mathrm{m}}(X;Y|W),\xi)$,
we have $\xi_{w}^{*}(\frac{\varepsilon}{\theta})\le\xi^{*}(\frac{\varepsilon}{\theta})$.
Therefore, 
\begin{equation}
p_{\max}^{(w)}\ge1-\xi^{*}(\frac{\varepsilon}{\theta}).\label{eq:pmax_w-1}
\end{equation}

 On the other hand, observe that 
\begin{align}
\varepsilon\ge\mathbb{P}\left\{ U\neq V\right\} & \ge\mathbb{P}\left\{ W\in\mathcal{B}^{c}\right\} \mathbb{P}\left\{ U\neq V|W\in\mathcal{B}^{c}\right\} \\
& \ge P_{W}(\mathcal{B}^{c})\frac{\varepsilon}{\theta}.\label{eq:UV-2-1-2}
\end{align}
Hence, 
\begin{equation}
P_{W}(\mathcal{B}^{c})\le\theta,
\end{equation}
i.e., 
\begin{equation}
P_{W}(\mathcal{B})\ge1-\theta.
\end{equation}
Therefore, 
\begin{align}
 & \sup_{h:\mathcal{W}\to\mathcal{U}}\mathbb{P}\left\{ U=h(W)\right\} \nonumber \\
 & =\sum_{w}P_{W}(w)\max_{u}P_{U|W}(u|w)\\
 & \ge(1-\theta)(1-\xi^{*}(\frac{\varepsilon}{\theta})),
\end{align}
where the last line follows from \eqref{eq:pmax_w-1}. 
\end{IEEEproof}

\section{\label{sec:Perfect-Stealth-}Perfect Stealth and Secrecy Communication}

In this section, we apply the preceding results on exact resolvability
to the perfectly stealthy (or covert) and secret communication over
the discrete memoryless wiretap channel \cite{Wyner75,Csiszar78}.
Stealth or covert communication was studied by Hou and Kramer \cite{hou2014effective},
Yu and Tan \cite{Yu}, Bash \emph{et al.} \cite{bash2012limits,bash2015quantum},
Wang \emph{et al.} \cite{wang2016fundamental}, and Bloch \cite{bloch2016covert},
where the relative entropy and the Rényi divergence were used to measure
the level of stealth (or covertness) of communication. In this paper,
we consider a \emph{perfectly} stealthy (or covert) and secret communication
system, where the eavesdropper is forced to observe a channel output
exactly, rather than approximately, following a target distribution
and, at the same time, the secret part of transmitted messages is
independent of the eavesdropper's observation. For this new problem,
we aim at characterizing the rate region of secret and non-secret
parts of the transmitted messages.

Consider a discrete memoryless wiretap channel $P_{YZ|X}$, and two
messages $\left(M_{0},M_{1}\right)$ that are uniformly distributed
over $\calM_{0}:=[1:e^{nR_{0}}]$ and $\calM_{1}:=[1:e^{nR_{1}}]$
respectively. A sender wants to transmit the pair $\left(M_{0},M_{1}\right)$
to a legitimate user reliably, and, at the same time, ensure that
$M_{1}$ is independent of the eavesdropper's observation $Z^{n}$. 
\begin{defn}
An $\left(n,R_{0},R_{1}\right)$ secrecy code is defined by two stochastic
mappings $P_{X^{n}|M_{0}M_{1}}:\mathcal{M}_{0}\times\mathcal{M}_{1}\to\mathcal{X}^{n}$
and $P_{\widehat{M}_{0}\widehat{M}_{1}|Y^{n}}:\mathcal{Y}^{n}\to\mathcal{M}_{0}\times\mathcal{M}_{1}$. 
\end{defn}
Given a target distribution $P_{Z}$, we wish to maximize the alphabet
size (or rate) of $M_{1}$ such that the distribution $P_{M_{1}Z^{n}}$
induced by the code is equal to the target distribution $P_{M_{1}}P_{Z}^{n}$
and $M_{1}$ can be decoded correctly asymptotically when $n\to\infty$. 
\begin{defn}
The tuple $(R_{0},R_{1})$ is {\em $P_{Z}$-achievable} if there
exists a sequence of $\left(n,R_{0},R_{1}\right)$ secrecy codes with
induced distribution $P_{M_{0}M_{1}Z^{n}\widehat{M}_{0}\widehat{M}_{1}}$
such that 

{[}leftmargin={*}{]} 
\begin{enumerate}
\item Error constraint: 
\begin{equation}
\lim_{n\rightarrow\infty}\mathbb{P}\left\{ \left(M_{0},M_{1}\right)\neq(\widehat{M}_{0},\widehat{M}_{1})\right\} =0;\label{eq:-30-1}
\end{equation}
\item Secrecy constraint: 
\begin{equation}
{\displaystyle P_{M_{1}Z^{n}}=P_{M_{1}}P_{Z}^{n}}.\label{eq:-31}
\end{equation}
\end{enumerate}
\end{defn}
Here we assume $P_{Z}$ satisfies $\mathcal{P}\left(P_{Z|X},P_{Z}\right)\neq\emptyset$
($\mathcal{P}\left(P_{Z|X},P_{Z}\right)$ is defined in \eqref{eq:-35});
otherwise, \eqref{eq:-31} cannot be satisfied by any secrecy code. 
\begin{defn}
The {\em $P_{Z}$-admissible region} is defined as 
\begin{equation}
\mathcal{R}(P_{Z}):=\textrm{Closure}\left\{ (R_{0},R_{1}):(R_{0},R_{1})\textrm{ is \ensuremath{P_{Z}}-achievable}\right\} .
\end{equation}
The \emph{perfect stealth (or perfect covertness) capacity}{} is
defined as 
\begin{equation}
C_{0}(P_{Z}):=\max_{(R_{0},R_{1})\in\mathcal{R}(P_{Z})}R_{0}.
\end{equation}
The \emph{perfect stealth-secrecy capacity} is defined as 
\begin{equation}
C_{1}(P_{Z}):=\max_{(R_{0},R_{1})\in\mathcal{R}(P_{Z})}R_{1}.
\end{equation}
\end{defn}

There are two reasons we assume $M_{0},M_{1}$ follow uniform distributions.
Firstly, this assumption is consistent with the setting in traditional
communication problems. Secondly, even if the sources (or messages)
to be transmitted (denote them as $S_{0},S_{1}$) are not uniform,
for example, they are memoryless and follow $P_{S_{k}},k=0,1$, respectively,
then by Theorem \ref{thm:ExactIR} we know that for $k=0,1$,\emph{
}there exists $P_{S_{k}^{n}M_{k}}\in C(P_{S_{k}}^{n},P_{M_{k}}^{\mathrm{U}})$
such that $\lim_{n\to\infty}\max_{f_{n}}\mathbb{P}\left\{ M_{k}=f_{n}(S_{k}^{n})\right\} =1$
if the rate $R_{k}$ of $M_{k}$ satisfies $R_{k}>H(S_{k})$. Hence
using $P_{M_{k}|S_{k}^{n}},k=0,1$, we transform the sources into
two uniformly distributed messages. Moreover, for the error constraint,
if the legitimate user can recover $M_{0},M_{1}$, he can recover
$S_{0},S_{1}$ as well since $\lim_{n\to\infty}\max_{f_{n}}\mathbb{P}\left\{ M_{k}=f_{n}(S_{k}^{n})\right\} =1$.
For the secrecy constraint, ${\displaystyle P_{M_{1}Z^{n}}=P_{M_{1}}P_{Z}^{n}}$
implies ${\displaystyle P_{S_{1}Z^{n}}=P_{S_{1}}P_{Z}^{n}}$. Therefore,
the perfect stealth and secrecy communication of uniform messages
implies the perfect stealth and secrecy communication of non-uniform
messages if $R_{k}>H(S_{k}),k=0,1$. Obviously, the converse holds
if $R_{k}<H(S_{k}),k=0,1$. Therefore, the perfect stealth and secrecy
communication of non-uniform messages is feasible if and only if $(H(S_{0}),H(S_{1}))$
is $P_{Z}$-achievable. This ensures that we only need to consider
uniform messages.

\subsection{Main Result }

For full-rank channels, we completely characterize the admissible
region. 
\begin{thm}
\label{thm:FullRankChannels} If the wiretap channel $P_{Z|X}$ is
of full-rank (including additive channels and identity channels),
we have 
\begin{align}
\mathcal{R}(P_{Z}) & =\left\{ \begin{array}{l}
(R_{0},R_{1}):R_{0}\leq C_{\mathsf{GK}}(X;Y)\\
R_{1}=0
\end{array}\right\} ,\label{eq:-120-1}
\end{align}
where $P_{X}$ is the unique distribution that induces the target
distribution $P_{Z}$. That is, $C_{0}(P_{Z})=C_{\mathsf{GK}}(X;Y)$
and $C_{1}(P_{Z})=0$. 
\end{thm}
\begin{IEEEproof}
The achievability part follows from the result on channel capacity
with input distribution constraint (Theorem \ref{thm:capacity} in
the previous section). Now we prove the converse part.

Note that $\boldsymbol{P}_{Z}^{\otimes n}=\boldsymbol{P}_{Z^{n}|M_{1}=m_{1}}=\boldsymbol{P}_{Z|X}^{\otimes n}\boldsymbol{P}_{X^{n}|M_{1}=m_{1}}$
for any $m_{1}$, and $\boldsymbol{P}_{Z|X}^{\otimes n}$ is invertible.
Hence 
\begin{equation}
\boldsymbol{P}_{X^{n}|M_{1}=m_{1}}=\left(\boldsymbol{P}_{Z|X}^{\otimes n}\right)^{-1}\boldsymbol{P}_{Z}^{\otimes n}=\left(\boldsymbol{P}_{Z|X}^{-1}\boldsymbol{P}_{Z}\right)^{\otimes n}
\end{equation}
for any $m_{1}$. Note that $(\boldsymbol{P}_{Z|X}^{-1}\boldsymbol{P}_{Z})^{\otimes n}$
does not depend on $m_{1}$, hence $X^{n}$ is independent of $M_{1}$.
On the other hand, $M_{1}\to X^{n}\to Y^{n}$ forms a Markov chain,
hence $Y^{n}$ is independent of $M_{1}$. That is, $R_{1}=0$.

The converse part for $R_{0}\leq C_{\mathsf{GK}}(X;Y)$ follows from
the converse part of Theorem \ref{thm:capacity}. 
\end{IEEEproof}
For general channels, we derive an upper bound and a lower bound for
the perfect stealth-secrecy capacity. 
\begin{thm}
\label{thm:Perfect-stealth-secrecy-capacity} The perfect stealth
capacity and the perfect stealth-secrecy capacity are respectively
bounded as {} 
\begin{align}
 & \sup_{k\geq1}\frac{1}{k}\max_{P_{X^{k}}\in\mathcal{P}(P_{Z}^{k})}C_{\mathsf{GK}}(X^{k};Y^{k})\nonumber \\
 & \leq C_{0}(P_{Z})\\
 & \leq\max_{P_{X}\in\mathcal{P}(P_{Z})}I(X;Y),
\end{align}
and 
\begin{align}
 & \max_{P_{UTX}:U\to T\to Z,P_{X}\in\mathcal{P}(P_{Z})}I(U;Y)-I(U;T)\nonumber \\
 & \leq C_{1}(P_{Z})\label{eq:-55}\\
 & \leq\max_{P_{UX}:P_{X}\in\mathcal{P}(P_{Z})}\min_{\substack{P_{T|XZ}:P_{Z|T}\textrm{ is of full-rank},\\
X\to T\to Z
}
}I(U;Y)-I(U;T).\label{eq:-39}
\end{align}
\end{thm}
\begin{rem}
The lower bound for $C_{1}(P_{Z})$ can be further lower bounded by
$\max_{P_{UX}:U\perp Z,P_{X}\in\mathcal{P}(P_{Z})}I(U;Y).$ The upper
bound for $C_{1}(P_{Z})$ can be further upper bounded by $\max_{P_{UX}:P_{X}\in\mathcal{P}(P_{Z})}I(U;Y)-I(U;Z).$ 
\end{rem}
\begin{rem}
Wang \emph{et al.}{} \cite{wang2016fundamental} proved that if the
sender and the legitimate user share a sufficiently large rate of
secret key, then the covert capacity $C_{0}(P_{Z})=\max_{P_{X}\in\mathcal{P}(P_{Z})}I(X;Y)$.
\end{rem}
\begin{IEEEproof}
The achievability part for $C_{0}(P_{Z})$ follows from the result
on channel capacity with input distribution constraint (Theorem \ref{thm:capacity}
in the previous section). Conversely, $C_{0}(P_{Z})\leq\frac{1}{n}I(X^{n};Y^{n})\leq I(X_{Q};Y_{Q})\leq\max_{P_{X}\in\mathcal{P}(P_{Z})}I(X;Y)$,
where $Q\sim\mathrm{Unif}[1:n]$ denotes a time-sharing random variable,
independent of $X^{n},Y^{n}$. The last inequality follows since $P_{X_{Q}}\in\mathcal{P}(P_{Z})$.
Next we prove the lower and upper bounds for $C_{1}(P_{Z})$.

\emph{Achievability for $C_{1}(P_{Z})$: }Suppose $P_{UTX}$ is a
distribution such that $U\to T\to Z,P_{X}\in\mathcal{P}(P_{Z})$.
Then we use the following scheme to obtain the inner bound.

Codebook generation: Fix the conditional pmf $P_{U|T}$ and $P_{X|UT}$
and let $\widetilde{R}_{1}>R_{1}$. For each message $m_{1}\in[1:e^{nR_{1}}]$
generate a subcodebook $\mathcal{C}(m_{1})$ consisting of $e^{n(\widetilde{R}_{1}-R_{1})}$
randomly and independently generated sequences $u^{n}(l),l\in[(m_{1}-1)e^{n(\widetilde{R}_{1}-R_{1})}+1:m_{1}e^{n(\widetilde{R}_{1}-R_{1})}]$,
each according to $\prod_{i=1}^{n}P_{U}(u_{i})$.

Encoding: Generate a sequence $t^{n}$ according to $\prod_{i=1}^{n}P_{T}(t_{i})$.
Upon receiving message $m_{1}\in[1:e^{nR_{1}}]$ and sequence $t^{n}$,
the encoder chooses a sequence $u^{n}(l)\in\mathcal{C}(m_{1})$ such
that $(u^{n}(l),t^{n})\in\mathcal{T}_{\epsilon}^{(n)}$. If no such
sequence exists, it picks $l=1$. For brevity, denote $U^{n}=U^{n}(L)$.
Then upon $U^{n}=u^{n},T^{n}=t^{n}$, the encoder generates $x^{n}$
according to $\prod_{i=1}^{n}P_{X|UT}(x_{i}|u_{i},t_{i})$ and transmits
it.

Decoding: Let $\epsilon'>\epsilon$. Upon receiving $y^{n}$ , the
decoder declares that $\hat{m}_{1}\in[1:e^{nR_{1}}]$ is sent if it
is the unique message such that $(u^{n}(l),y^{n})\in\mathcal{T}_{\epsilon'}^{(n)}$
for some $u^{n}(l)\in\mathcal{C}(\hat{m}_{1})$; otherwise it declares
an error.

Analysis of Error Probability and Secrecy: If $T^{n}$ is considered
as a side information, then the achievability scheme above is also
a Gelfand-Pinsker code for the channel coding problem with non-causal
side information at the transmitter. By Gelfand-Pinsker's proof \cite[pp. 181]{Gamal},
we have that if $R_{1}<I(U;Y)-I(U;T)$ then 
\begin{equation}
\lim_{n\rightarrow\infty}\mathbb{P}\left\{ M_{1}\neq\widehat{M}_{1}\right\} =0.\label{eq:-30-1-1-2}
\end{equation}

Furthermore, 
\begin{align}
 & P_{Z^{n}|M_{1}}(z^{n}|m_{1})\nonumber \\
 & =\sum_{u^{n},t^{n}}P_{Z|UT}^{n}(z^{n}|u^{n},t^{n})P_{T}^{n}(t^{n})P_{U^{n}(L)|T^{n}M_{1}}(u^{n}|t^{n},m_{1})\\
 & =\sum_{u^{n},t^{n}}P_{Z|T}^{n}(z^{n}|t^{n})P_{T}^{n}(t^{n})P_{U^{n}(L)|T^{n}M_{1}}(u^{n}|t^{n},m_{1})\label{eq:-14-1}\\
 & =\sum_{t^{n}}P_{Z|T}^{n}(z^{n}|t^{n})P_{T}^{n}(t^{n})\\
 & =P_{Z}^{n}(z^{n}),
\end{align}
where \eqref{eq:-14-1} follows from $U\to T\to Z$.

\emph{Converse for $C_{1}(P_{Z})$:} Similar to the proof of Theorem
\ref{thm:FullRankChannels}, it can be shown that $T^{n}$ is independent
of $M$ with $T_{i}$ generated through a channel $P_{T_{i}|X_{i}Z_{i}}$
such that $P_{Z_{i}|T_{i}}$ is of full-rank and $X_{i}\to T_{i}\to Z_{i}$.
\begin{align}
 & nR_{1}\nonumber \\
 & \le I(Y^{n};M)\\
 & =I(Y^{n};M)-I(T^{n};M)\\
 & =\sum_{i=1}^{n}I(Y_{i};M|V_{i})-I(T_{i};M|V_{i})\label{eq:-34}\\
 & =\sum_{i=1}^{n}\sum_{v_{i}}P_{V_{i}}(v_{i})\left(I(Y_{i};M|V_{i}=v_{i})-I(T_{i};M|V_{i}=v_{i})\right)\label{eq:-38}\\
 & \leq n\max_{P_{UX}:P_{X}\in\mathcal{P}(P_{Z})}\min_{\substack{P_{T|XZ}:P_{Z|T}\textrm{ is of full-rank},\\
X\to T\to Z
}
}I(U;Y)-I(U;T),\label{eq:-36}
\end{align}
where $V_{i}:=Y^{i-1}T_{i+1}^{n}$, \eqref{eq:-34} follows from the
standard steps in the weak converse proof for the wiretap channel
\cite[pp. 555]{Gamal}, and \eqref{eq:-36} follows since $P_{T_{i}|X_{i}Z_{i}}$
is arbitrary such that $P_{Z_{i}|T_{i}}$ is of full-rank and $X_{i}\to T_{i}\to Z_{i}$
and \eqref{eq:-36} reduces to \eqref{eq:-38} if $U$ is set to $M$. 
\end{IEEEproof}
\begin{defn}
A function $f(X)$ is said to be a sufficient statistic relative to
$P_{Z|X}$ if $X$ is independent of $Z$ given $f(X)$ for any distribution
on $X$ (i.e., for any distribution on $X$, $X\to f(X)\to Z$ forms
a Markov chain). 
\end{defn}
The lower bound and upper bound in Theorem \ref{thm:Perfect-stealth-secrecy-capacity}
coincide for full-rank sufficient statistic channels. 
\begin{cor}[Full-rank Sufficient Statistic Channel]
\label{cor:sufficientstatistic} If there exists a sufficient statistic
$f(X)$ relative to $P_{Z|X}$ such that $P_{Z|f(X)}$ is full-rank,
then 
\begin{equation}
C_{1}(P_{Z})=\max_{P_{UX}:P_{f(X)}\in\mathcal{P}(P_{Z})}I(U;Y)-I(U;f(X)).
\end{equation}
\end{cor}
\begin{rem}
If $P_{Z|X}=P_{Z|X_{2}}$ with $X_{2}=f(X)$ for some function $f$
and for all input random variables $X$, and $P_{Z|X_{2}}$ is of
full-rank, then the perfect secrecy capacity $C_{1}(P_{Z})=\max_{P_{UX}:P_{X_{2}}\in\mathcal{P}(P_{Z})}I(U;Y)-I(U;X_{2})$. 
\end{rem}
\begin{rem}
Corollary \ref{cor:sufficientstatistic} is consistent with Theorem
\ref{thm:FullRankChannels}, since both of them imply $C_{1}(P_{Z})=0$
for full-rank channels. 
\end{rem}
As a special case of Corollary \ref{cor:sufficientstatistic}, we
have the following result. 
\begin{cor}[Gaussian Wiretap Channel]
If $X=(X_{1},X_{2})$, the channel satisfies $Y=X_{1}+X_{2}+E_{1},Z=X_{2}+E_{2}$,
with $E_{k}\sim\mathcal{N}(0,N_{k}),k=1,2$ and $\mathbb{E}X_{1}^{2}\le P_{1}$,
and $P_{Z}=\mathcal{N}(0,N_{Z})$ with $N_{Z}\ge N_{2}$, then the
perfect secrecy capacity 
\begin{equation}
C_{1}(P_{Z})=\frac{1}{2}\log\left(1+\frac{P_{1}}{N_{1}}\right).
\end{equation}
\end{cor}
Similar to Definition \ref{def:We-say-}, here we define $P_{Z}$-redundant
channel as follows. 
\begin{defn}
A channel $P_{YZ|X}$ is a \emph{$P_{Z}$-redundant channel} if there
exist two distributions $Q_{X}$ and $Q_{X}'$ that induce the same
$P_{Z}$ through $P_{Z|X}$ but induce two different distributions
of $Y$ through $P_{Y|X}$. 
\end{defn}
We give a sufficient and necessary condition for that the stealth-secrecy
capacity is positive. The proof of the following theorem is provided
in Appendix \ref{sec:Perfect-stealth-secrecy}. 
\begin{thm}
\label{thm:Perfect-stealth-secrecy} $C_{1}(P_{Z})>0$ if and only
if the channel $P_{YZ|X}$ is a $P_{Z}$-redundant channel. 
\end{thm}

\section{Conclusion and Future Work}

\label{sec:concl}In this paper, we studied asymptotics of several
coupling problems, including the problems of maximal coupling, minimum
distance coupling, maximal guessing coupling, and minimum entropy
coupling. We also applied these results to some information-theoretic
problems, including the problems of exact intrinsic randomness, exact
resolvability, and perfectly stealthy/covert and secret communication.

Our results generalize or extend several classical and recent results.
Firstly, our results on exact intrinsic randomness extend those by
Vembu and Verdú \cite{vembu1995generating} and Han \cite{Han03}
as we consider the scenario in which the output {\em exactly} follows
a uniform distribution. Secondly, our resolvability results extend
those by Han and Verdú \cite{Han}, by Hayashi \cite{Hayashi06,Hayashi11},
and by Yu and Tan \cite{Yu} as we consider the scenario in which
output {\em exactly} follows a target distribution. Finally, our
results for the wiretap channel extend those by Hou and Kramer \cite{hou2014effective},
by Yu and Tan \cite{Yu}, by Bash \emph{et al.} \cite{bash2012limits,bash2015quantum},
by Wang \emph{et al.} \cite{wang2016fundamental}, and by Bloch \cite{bloch2016covert},
as we measure the stealth (or effective secrecy ) or covertness using
an {\em exact} distribution constraint.

\subsection{Open Problems}

There are also some problems that remain to be solved. 

{[}leftmargin={*}{]} 
\begin{enumerate}
\item The optimal exponent of the minimum excess-distance probability coupling
problem for the case in which $d<{\cal D}(P_{X},P_{Y})$ has been
solved in this paper. However, the optimal exponent for the case in
which $d>{\cal D}(P_{X},P_{Y})$ is still unknown. Besides, the minimum
excess-distance probability and the corresponding optimal exponent
for the case $d={\cal D}(P_{X},P_{Y})$ are still unknown. 
\item In this paper, we characterized the limiting value of the maximal
guessing coupling problem for the case $H(X)\neq H(Y)$. However,
it is still open for the case $H(X)=H(Y)$. Furthermore, the optimal
exponent for this problem is still unknown. The same comment applies
to the optimal exponent for the minimum entropy coupling problem. 
\item Under the assumption of uniform distributions, we provided the necessary
and sufficient condition for the existence of a deterministic coupling
or an asymptotically deterministic coupling for two product marginal
distributions. However, the general case, stated in Conjectures \ref{conj:AsymptoticallyDeterministicCoupling}
and \ref{conj:DeterministicCoupling}, is still open. 
\item An achievability result on the minimum Rényi (conditional) entropy
coupling problem was provided in Corollary \ref{cor:MinRenyiEntropyCoupling}.
Other minimum Rényi entropy coupling (or maximum Rényi mutual information
coupling) problems are still open. 
\item We only characterized the exact channel resolvability rate for full-rank
channels. The complete characterization of the exact channel resolvability
rate for general channels is still open. 
\item We provided a sufficient and necessary condition in Theorem \ref{thm:Perfect-stealth-secrecy}
for the scenario in which the stealth-secrecy capacity is positive.
We also characterized the stealth-secrecy capacity for the full-rank
sufficient statistic channel in Corollary \ref{cor:sufficientstatistic}.
However, the complete characterization of the perfect stealth-secrecy
capacity for general channels is still open. 
\end{enumerate}

\appendices{}

\section{\label{sec:MinExcessDistProbCoupling}Proof of Proposition \ref{prop:MinExcessDistProbCoupling} }

Proof of Statement 1): Observe that $\max\left\{ D(Q_{X}\|P_{X}),D(Q_{Y}\|P_{Y})\right\} $
is continuous in $Q_{XY}\in\mathcal{P}(\mathcal{X\times Y})$. By
\cite[Lem. 5]{Yu}, we know 
\begin{align}
 & \lim_{n\to\infty}\min_{\substack{T_{XY}\in\mathcal{P}_{n}(\mathcal{X\times Y}):\\
\sum_{x,y}T_{XY}(x,y)d(x,y)\leq d
}
}\max\left\{ D(T_{X}\|P_{X}),D(T_{Y}\|P_{Y})\right\} \nonumber \\
= & \min_{\substack{Q_{XY}\in\mathcal{P}(\mathcal{X\times Y}):\\
\sum_{x,y}Q_{XY}(x,y)d(x,y)\leq d
}
}\max\left\{ D(Q_{X}\|P_{X}),D(Q_{Y}\|P_{Y})\right\} .\label{eq:-37}
\end{align}
Hence to prove Statement 1), we only need to show the exponent $\overline{\mathsf{E}}(d)$
is 
\begin{equation}
\lim_{n\to\infty}\min_{\substack{T_{XY}\in\mathcal{P}_{n}(\mathcal{X\times Y}):\\
\sum_{x,y}T_{XY}(x,y)d(x,y)\leq d
}
}\max\left\{ D(T_{X}\|P_{X}),D(T_{Y}\|P_{Y})\right\} .
\end{equation}
Next we prove this point.

First we prove the converse part. 
\begin{align}
 & \mathbb{P}\left\{ d(X^{n},Y^{n})\leq d\right\} \nonumber \\
 & =\mathbb{P}\left\{ \sum_{x,y}T_{X^{n}Y^{n}}(x,y)d(x,y)\leq d\right\} \\
 & =\sum_{x^{n},y^{n}}P_{X^{n}Y^{n}}(x^{n},y^{n})1\left\{ \sum_{x,y}T_{x^{n}y^{n}}(x,y)d(x,y)\leq d\right\} \\
 & =\sum_{T_{XY}}P_{X^{n}Y^{n}}(\mathcal{T}(T_{XY}))1\left\{ \sum_{x,y}T_{XY}(x,y)d(x,y)\leq d\right\} \\
 & \leq\sum_{T_{XY}}\min\{P_{X^{n}}(\mathcal{T}(T_{X})),P_{Y^{n}}(\mathcal{T}(T_{Y}))\}\nonumber \\
 & \qquad\times1\left\{ \sum_{x,y}T_{XY}(x,y)d(x,y)\leq d\right\} \\
 & \leq(n+1)^{|\mathcal{X}\|\mathcal{Y}|}\max_{T_{XY}}\min\{P_{X^{n}}(T_{X})),P_{Y^{n}}(\mathcal{T}(T_{Y}))\}\nonumber \\
 & \qquad\times1\left\{ \sum_{x,y}T_{XY}(x,y)d(x,y)\leq d\right\} \\
 & \doteq\max_{T_{XY}}\min\left\{ e^{-nD(T_{X}\|P_{X})},e^{-nD(T_{Y}\|P_{Y})}\right\} \nonumber \\
 & \qquad\times1\left\{ \sum_{x,y}T_{XY}(x,y)d(x,y)\leq d\right\} \\
 & =e^{-n\min_{T_{XY}:\sum_{x,y}T_{XY}(x,y)d(x,y)\leq d}\max\left\{ D(T_{X}\|P_{X}),D(T_{Y}\|P_{Y})\right\} }
\end{align}
The exponent $\overline{\mathsf{E}}(d)$ is lower bounded by 
\begin{equation}
\min_{T_{XY}:\sum_{x,y}T_{XY}(x,y)d(x,y)\leq d}\max\left\{ D(T_{X}\|P_{X}),D(T_{Y}\|P_{Y})\right\} .
\end{equation}

Next we prove the achievability part. First we note that finding a
coupling $P_{X^{n}Y^{n}}$ of $P_{X}^{n}$ and $P_{Y}^{n}$ that maximizes
$\mathbb{P}\left\{ d(X^{n},Y^{n})\leq d\right\} $ is equivalent to
finding a ``coupling'' $\left\{ P_{X^{n}Y^{n}}(\mathcal{T}(T_{XY}))\right\} _{T_{XY}}$
of $\left\{ P_{X^{n}}(\mathcal{T}(T_{X}))\right\} _{T_{X}}$ and $\left\{ P_{Y^{n}}(\mathcal{T}(T_{Y}))\right\} _{T_{Y}}$
that maximizes 
\begin{equation}
\sum_{T_{XY}}P_{X^{n}Y^{n}}(\mathcal{T}(T_{XY}))\cdot1\left\{ \mathbb{E}_{T_{XY}}d(X,Y)\leq d\right\} .
\end{equation}
This is because, on one hand, if we get a desired ``coupling'' $\left\{ P_{X^{n}Y^{n}}(\mathcal{T}(T_{XY}))\right\} _{T_{XY}}$,
and for each type $T_{XY}$, let the sequences in the type class $\mathcal{T}(T_{XY})$
uniformly share the total probability $P_{X^{n}Y^{n}}(\mathcal{T}(T_{XY}))$,
i.e., 
\begin{equation}
P_{X^{n}Y^{n}}(x^{n},y^{n})=\frac{P_{X^{n}Y^{n}}(\mathcal{T}(T_{XY}))}{|\mathcal{T}(T_{XY})|},\quad(x^{n},y^{n})\in\mathcal{T}(T_{XY}),
\end{equation}
then the marginal distributions are also uniform in each type class.
Moreover, the marginal distributions have the probabilities of the
type classes $\left\{ P_{X^{n}}(\mathcal{T}(T_{X}))\right\} _{T_{X}}$
and $\left\{ P_{Y^{n}}(\mathcal{T}(T_{Y}))\right\} _{T_{Y}}$. This
two points ensure that the marginal distributions are respectively
$P_{X}^{n}$ and $P_{Y}^{n}$.

Now we find a desired ``coupling'' $\left\{ P_{X^{n}Y^{n}}(\mathcal{T}(T_{XY}))\right\} _{T_{XY}}$
of $\left\{ P_{X^{n}}(\mathcal{T}(T_{X}))\right\} _{T_{X}}$ and $\left\{ P_{Y^{n}}(\mathcal{T}(T_{Y}))\right\} _{T_{Y}}$.
Denote $T_{XY}^{*}$ as a type that achieves 
\begin{equation}
\min_{T_{XY}:\sum_{x,y}T_{XY}(x,y)d(x,y)\leq d}\max\left\{ D(T_{X}\|P_{X}),D(T_{Y}\|P_{Y})\right\} .
\end{equation}
Obviously, $\sum_{x,y}T_{XY}^{*}(x,y)d(x,y)\leq d$. Without loss
of generality, we only consider the case of $D(T_{X}^{*}\|P_{X})\geq D(T_{Y}^{*}\|P_{Y})$.
We allocate $P_{X^{n}}(\mathcal{T}(T_{X}^{*}))$ to $P_{X^{n}Y^{n}}(\mathcal{T}(T_{XY}^{*}))$,
i.e., set $P_{X^{n}Y^{n}}(\mathcal{T}(T_{XY}^{*}))=P_{X^{n}}(\mathcal{T}(T_{X}^{*}))$
and $P_{X^{n}Y^{n}}(\mathcal{T}(T_{XY}))=0$ for all $T_{XY}$ with
$T_{X}=T_{X}^{*}$ but $T_{XY}\neq T_{XY}^{*}$. On the other hand,
there is no restriction for the probabilities of other joint types.
Hence we set $\big\{\sum_{T_{XY}\in C(T_{X},T_{Y})}P_{X^{n}Y^{n}}(\mathcal{T}(T_{XY}))\big\}_{T_{X},T_{Y}:T_{X}\neq T_{X}^{*}}$
to be any coupling of $\left\{ P_{X^{n}}(\mathcal{T}(T_{X}))\right\} _{T_{X}\neq T_{X}^{*}}$
and $\left\{ P_{Y^{n}}(\mathcal{T}(T_{Y}))\right\} _{T_{Y}}$. Then
for such a coupling $\left\{ P_{X^{n}Y^{n}}(\mathcal{T}(T_{XY}))\right\} _{T_{XY}}$,
we have 
\begin{align}
 & \mathbb{P}\left\{ d(X^{n},Y^{n})\leq d\right\} \nonumber \\
 & =\mathbb{P}\left\{ \sum_{x,y}T_{X^{n}Y^{n}}(x,y)d(x,y)\leq d\right\} \\
 & =\sum_{x^{n},y^{n}}P_{X^{n}Y^{n}}(x^{n},y^{n})1\left\{ \sum_{x,y}T_{x^{n}y^{n}}(x,y)d(x,y)\leq d\right\} \\
 & =\sum_{T_{XY}}P_{X^{n}Y^{n}}(\mathcal{T}(T_{XY}))1\left\{ \sum_{x,y}T_{XY}(x,y)d(x,y)\leq d\right\} \\
 & \geq P_{X^{n}Y^{n}}(\mathcal{T}(T_{XY}^{*}))1\left\{ \sum_{x,y}T_{XY}^{*}(x,y)d(x,y)\leq d\right\} \\
 & =P_{X^{n}}(\mathcal{T}(T_{X}^{*}))\doteq e^{-nD(T_{X}^{*}\|P_{X})}.
\end{align}
By symmetry, for the case of $D(T_{X}^{*}\|P_{X})\geq D(T_{Y}^{*}\|P_{Y})$,
we have 
\begin{align}
\mathbb{P}\left\{ d(X^{n},Y^{n})\leq d\right\}  & \dotge e^{-nD(T_{Y}^{*}\|P_{Y})}.
\end{align}
Therefore, 
\begin{align}
\overline{\mathsf{E}}(d) & \leq\max\left\{ D(T_{X}^{*}\|P_{X}),D(T_{Y}^{*}\|P_{Y})\right\} \\
 & =\min_{T_{XY}:\sum_{x,y}T_{XY}(x,y)d(x,y)\leq d}\nonumber \\
 & \qquad\max\left\{ D(T_{X}\|P_{X}),D(T_{Y}\|P_{Y})\right\} .
\end{align}
Invoking \eqref{eq:-37}, we complete the proof of Statement 1).

Proofs of Statements 2) and 3): Proof of the achievability by product
couplings: For the product coupling $P_{X^{n}Y^{n}}=P_{XY}^{n}$ where
$P_{XY}:=\arg\min_{P_{XY}\in C(P_{X},P_{Y})}\mathbb{E}d(X,Y)$, by
the large deviation theory, the exponents for the cases of Statement
1) and 2) are respectively $\max_{t\geq0}\left(-td-\log\mathbb{E}e^{-td(X,Y)}\right),$
and $\max_{t\geq0}\left(td-\log\mathbb{E}e^{td(X,Y)}\right),$ and
for the case of Statement 3), by the central limit theorem, 
\begin{equation}
\mathbb{P}\left\{ d(X^{n},Y^{n})\leq d\right\} =\frac{1}{2}+O\left(\frac{1}{\sqrt{n}}\right).
\end{equation}

\section{\label{sec:maxguesscoup}Proof of Theorem \ref{thm:MaxGuessCoup}}

Proof of Statement 1): From the soft-covering lemma or the distribution
approximation problem \cite[Theorem 2.1.1]{Han03} we know that if
$H(X)>H(Y)$, there exists a sequence of functions $f_{n}(x^{n})$
such that $|P_{f_{n}(X^{n})}-P_{Y^{n}}|\to0$ exponentially fast.
On the other hand, by the equivalence between the maximal guessing
coupling problem and the distribution approximation problem (Theorem
\ref{thm:equivalence}), 
\begin{equation}
\max_{P_{X^{n}Y^{n}}\in C(P_{X}^{n},P_{Y}^{n})}\max_{f_{n}}\mathbb{P}\left\{ Y^{n}=f_{n}(X^{n})\right\} \to1
\end{equation}
at least exponentially fast as $n\to\infty$. Furthermore, the lower
bound in \eqref{eqn:Eiid-2} is an exponent obtained by i.i.d. codes
\cite{Yu}. A different exponent can be obtained from \cite[Lemma 2.1.1]{Han03}.

Proof of Statement 2): Statement 2) can be obtained by combining Han's
result \cite[Theorem 2.1.1]{Han03} and our Theorem \ref{thm:equivalence},
and an exponent can be obtained from \cite[Lemma 2.1.2]{Han03}. But
in the following, we prove it using the method of types, which gives
us a different exponent. 
\begin{align}
 & \mathbb{P}\left\{ Y^{n}=f_{n}(X^{n})\right\} \nonumber \\
 & =\sum_{x^{n},y^{n}}P_{X^{n}Y^{n}}(x^{n},y^{n})1\left\{ y^{n}=f_{n}(x^{n})\right\} \\
 & =\sum_{x^{n},y^{n}}P_{X^{n}Y^{n}}(x^{n},y^{n})1\left\{ y^{n}=f_{n}(x^{n}),x^{n}\in\mathcal{T}_{\epsilon}^{n}(P_{X})\right\} \nonumber \\
 & \qquad+\sum_{x^{n},y^{n}}P_{X^{n}Y^{n}}(x^{n},y^{n})1\left\{ y^{n}=f_{n}(x^{n}),x^{n}\notin\mathcal{T}_{\epsilon}^{n}(P_{X})\right\} \\
 & \leq\sum_{x^{n},y^{n}}P_{X^{n}Y^{n}}(x^{n},y^{n})1\left\{ y^{n}\in\mathcal{A}\right\} \nonumber \\
 & \qquad+P_{X^{n}}\left((\mathcal{T}_{\epsilon}^{n}(P_{X}))^{c}\right)\\
 & =P_{Y^{n}}\left(\mathcal{A}\right)+P_{X^{n}}\left((\mathcal{T}_{\epsilon}^{n}(P_{X}))^{c}\right)\\
 & =P_{Y^{n}}\left(\mathcal{A}\cap\mathcal{T}_{\epsilon}^{n}(P_{Y})\right)+P_{Y^{n}}\left(\mathcal{A}\cap(\mathcal{T}_{\epsilon}^{n}(P_{Y}))^{c}\right)\nonumber \\
 & \qquad+P_{X^{n}}\left((\mathcal{T}_{\epsilon}^{n}(P_{X}))^{c}\right)\\
 & \leq|\mathcal{A}|e^{-n(1-\epsilon)H(Y)}+P_{Y^{n}}\left((\mathcal{T}_{\epsilon}^{n}(P_{Y}))^{c}\right)\nonumber \\
 & \qquad+P_{X^{n}}\left((\mathcal{T}_{\epsilon}^{n}(P_{X}))^{c}\right)\\
 & \dotle e^{-n\left((1-\epsilon)H(Y)-(1+\epsilon)H(X)\right)}+e^{-n\delta_{\epsilon}(P_{Y})}+e^{-n\delta_{\epsilon}(P_{X})}
\end{align}
where $\mathcal{A}:=\left\{ f_{n}(x^{n}):x^{n}\in\mathcal{T}_{\epsilon}^{n}(P_{X})\right\} $
with $\mathcal{T}_{\epsilon}^{n}(P_{X})$ denoting the $\epsilon$-typical
set, and for a set $\mathcal{B}$, $\mathcal{B}^{c}$ denotes the
complement of $\mathcal{B}$. Hence if $H(X)<H(Y)$, and $\epsilon>0$
is elected to be sufficiently small such that $(1-\epsilon)H(Y)-(1+\epsilon)H(X)>0$,
then in view of \eqref{eq:-45} it follows that $\mathcal{G}(P_{X}^{n},P_{Y}^{n})\to0$
exponentially fast as $n\to\infty$.

Proof of Statement 3): An optimal product coupling $P_{X^{n}Y^{n}}=P_{XY}^{n}$
with $P_{XY}$ achieving $\mathcal{G}(P_{X},P_{Y})$ achieves the
lower bound $\mathcal{G}^{n}(P_{X},P_{Y})$.

\section{\label{sec:Conjectures} Some Special Cases of Conjectures \ref{conj:AsymptoticallyDeterministicCoupling}
and \ref{conj:DeterministicCoupling}}

\subsection{\label{subsec:AsympDetCoupling} A Special Case of Conjecture \ref{conj:AsymptoticallyDeterministicCoupling} }
\begin{prop}[Asymptotically Deterministic Coupling with Uniform $P_{X}$ or $P_{Y}$]
Assume $\mathcal{X}=[1:M]$ and $P_{X}(x)=\frac{1}{M}$ for all $x\in[1:M]$
or $\mathcal{Y}=[1:M]$ and $P_{Y}(y)=\frac{1}{M}$ for all $y\in[1:M]$
for some $M\in\mathbb{N}$, and $H(X)=H(Y)=\log M$. Then $\mathcal{G}(P_{X}^{n},P_{Y}^{n})\rightarrow1$
if and only if $\mathcal{G}(P_{X},P_{Y})=1.$ That is, there exists
a (asymptotically deterministic) coupling $P_{X^{n}Y^{n}}\in C(P_{X}^{n},P_{Y}^{n})$
for which $Y^{n}$ is an asymptotic function of $X^{n}$, if and only
if there exists a (deterministic) coupling $P_{XY}\in C(P_{X},P_{Y})$
for which $Y$ is a function of $X$. 
\end{prop}
\begin{rem}
More explicitly, for the case that $P_{X}$ is uniform but $P_{Y}$
is not, we have $\mathcal{G}(P_{X}^{n},P_{Y}^{n})\leq\alpha_{n}$
where 
\begin{equation}
\alpha_{n}:=1-\frac{1}{2}\left(\Phi\Big(-\frac{1}{n}\log2\Big)-\frac{\eta(Y)}{\sqrt{nV^{3}(Y)}}\right)
\end{equation}
with $\Phi(\cdot)$ denotes the cumulative distribution function (cdf)
of the standard Gaussian distribution, and 
\begin{align}
V(Y) & :=\mathrm{Var}\left[\log P_{Y}(Y)\right]\\
\eta(Y) & :=\mathbb{E}_{P_{Y}}[|\log P_{Y}(Y)+H(P_{Y})|^{3}];
\end{align}
and for the case that $P_{Y}$ is uniform but $P_{X}$ is not, we
have $\mathcal{G}(P_{X}^{n},P_{Y}^{n})\leq\beta_{n}$ where 
\begin{align}
\beta_{n} & :=1-\sup_{\gamma\geq1}\frac{1}{2}\biggl(\left(1-\frac{1}{\gamma}\right)\Phi\Big(\frac{1}{n}\log\gamma\Big)\nonumber \\
 & \qquad-\left(1+\frac{1}{\gamma}\right)\frac{\eta(X)}{\sqrt{nV^{3}(X)}}\biggr).
\end{align}
Furthermore, $\lim_{n\to\infty}\alpha_{n}=\lim_{n\to\infty}\beta_{n}=\frac{3}{4}$. 
\end{rem}
\begin{IEEEproof}
If $\mathcal{G}(P_{X},P_{Y})=1,$ then $\mathcal{G}(P_{X}^{n},P_{Y}^{n})=1$
for any $n$, regardless of whether $P_{X}$ is uniform or $P_{Y}$
is uniform.

Next we focus on the other direction.

Case 1 ($P_{X}$ is uniform): If $\mathcal{G}(P_{X},P_{Y})<1,$ then
by the assumption $H(X)=H(Y)=\log M$, we know that $P_{X}$ is uniform
but $P_{Y}$ is not. For this case, we have 
\begin{align}
 & |P_{Y^{n}}-P_{f_{n}(X^{n})}|\nonumber \\
 & \geq\frac{1}{2}\sum_{y^{n}:P_{Y^{n}}(y^{n})<\frac{1}{2M^{n}}}P_{Y^{n}}(y^{n})\label{eq:-28}\\
 & =\frac{1}{2}P_{Y^{n}}\left\{ y^{n}:P_{Y^{n}}(y^{n})<\frac{1}{2M^{n}}\right\} \label{eq:-24}\\
 & =\frac{1}{2}P_{Y^{n}}\left\{ y^{n}:-\frac{1}{n}\sum_{i=1}^{n}\log P_{Y}(y_{i})>H({Y})+\frac{1}{n}\log2\right\} \\
 & \geq\frac{1}{2}\left(\Phi\Big(-\frac{1}{n}\log2\Big)-\frac{\eta(Y)}{\sqrt{nV^{3}(Y)}}\right),\label{eq:-13}
\end{align}
where \eqref{eq:-28} follows since $P_{f_{n}(X^{n})}(y^{n})\geq\frac{1}{M^{n}}$
or $P_{f_{n}(X^{n})}(y^{n})=0$ for every $y^{n}\in\mathcal{Y}^{n}$
and thus $|P_{Y^{n}}(y^{n})-P_{f_{n}(X^{n})}(y^{n})|\geq P_{Y^{n}}(y^{n})$
for every $y^{n}$ such that $P_{Y^{n}}(y^{n})<\frac{1}{2M^{n}}$,
and \eqref{eq:-13} follows from the Berry--Esseen theorem \cite[Sec. XVI.5]{feller2008introduction}.
Hence 
\begin{align}
 & \mathcal{G}(P_{X}^{n},P_{Y}^{n})\nonumber \\
 & =1-\min_{f_{n}}|P_{Y^{n}}-P_{f_{n}(X^{n})}|\\
 & \leq1-\frac{1}{2}\left(\Phi\Big(-\frac{1}{n}\log2\Big)-\frac{\eta(Y)}{\sqrt{nV^{3}(Y)}}\right)\\
 & \to\frac{3}{4}\textrm{ as }n\to\infty.
\end{align}

Case 2 ($P_{Y}$ is uniform): If $\mathcal{G}(P_{X},P_{Y})<1,$ then
by the assumption $H(X)=H(Y)=\log M$, we know that $P_{Y}$ is uniform
but $P_{X}$ is not. For this case, we have 
\begin{align}
 & |P_{Y^{n}}-P_{f_{n}(X^{n})}|\nonumber \\
 & \geq\sup_{\gamma\geq1}\frac{1}{2}\sum_{y^{n}:P_{f_{n}(X^{n})}(y^{n})\geq\frac{\gamma}{M^{n}}}\left(P_{f_{n}(X^{n})}(y^{n})-\frac{1}{M^{n}}\right)\label{eq:-29}\\
 & \geq\sup_{\gamma\geq1}\frac{1}{2}\sum_{x^{n}:P_{X^{n}}(x^{n})\geq\frac{\gamma}{M^{n}}}\left(P_{X^{n}}(x^{n})-\frac{1}{M^{n}}\right)\label{eq:-30}\\
 & \geq\sup_{\gamma\geq1}\frac{1}{2}\biggl(\Phi\Big(\frac{1}{n}\log\gamma\Big)-\frac{\eta(X)}{\sqrt{nV^{3}(X)}}\nonumber \\
 & \qquad-\frac{1}{\gamma}\left(\Phi(\frac{1}{n}\log\gamma)+\frac{\eta(X)}{\sqrt{nV^{3}(X)}}\right)\biggr)\label{eq:-32}\\
 & =\sup_{\gamma\geq1}\frac{1}{2}\left(\left(1-\frac{1}{\gamma}\right)\Phi\Big(\frac{1}{n}\log\gamma\Big)-\left(1+\frac{1}{\gamma}\right)\frac{\eta(X)}{\sqrt{nV^{3}(X)}}\right),
\end{align}
where \eqref{eq:-30} follows since to make \eqref{eq:-29} as small
as possible, the function $f_{n}$ must be injective on the set $\left\{ x^{n}:P_{X^{n}}(x^{n})\geq\frac{\gamma}{M^{n}}\right\} $,
\eqref{eq:-32} follows from the Berry--Esseen theorem \cite[Sec. XVI.5]{feller2008introduction}
and 
\begin{align}
 & \Phi\left(\frac{1}{n}\log\gamma\right)+\frac{\eta(X)}{\sqrt{nV^{3}(X)}}\nonumber \\
 & \geq\sum_{x^{n}:P_{X^{n}}(x^{n})>\frac{\gamma}{M^{n}}}P_{X^{n}}(x^{n})\\
 & \geq\frac{\gamma\left|\left\{ x^{n}:P_{X^{n}}(x^{n})\geq\frac{\gamma}{M^{n}}\right\} \right|}{M^{n}}.
\end{align}
Hence 
\begin{align}
\mathcal{G}(P_{X}^{n},P_{Y}^{n}) & =1-\min_{f_{n}}|P_{Y^{n}}-P_{f_{n}(X^{n})}|\\
 & \leq1-\sup_{\gamma\geq1}\frac{1}{2}\biggl(\left(1-\frac{1}{\gamma}\right)\Phi\Big(\frac{1}{n}\log\gamma\Big)\nonumber \\
 & \qquad-\left(1+\frac{1}{\gamma}\right)\frac{\eta(X)}{\sqrt{nV^{3}(X)}}\biggr)\\
 & \to\frac{3}{4}\textrm{ as }n\to\infty.
\end{align}
This completes the proof. 
\end{IEEEproof}

\subsection{\label{subsec:DetCoupling} Two Special Cases of Conjecture \ref{conj:DeterministicCoupling}}
\begin{prop}[Entropy Criterion of Deterministic Coupling]

{[}leftmargin={*}{]} We have the following claims: 
\begin{enumerate}
\item If $H_{\alpha}(X)<H_{\alpha}(Y)$ for some $\alpha\in[0,\infty]$,
then for any $n$, $\mathcal{G}(P_{X}^{n},P_{Y}^{n})<1.$ 
\item If $H_{\alpha}(X)=H_{\alpha}(Y)$ for some $\alpha\in[0,\infty]$,
then $\mathcal{G}(P_{X}^{n},P_{Y}^{n})=1$ if and only if $\mathcal{G}(P_{X},P_{Y})=1.$
That is, there exists a deterministic coupling $P_{X^{n}Y^{n}}\in C(P_{X}^{n},P_{Y}^{n})$
for which $Y^{n}$ is a function of $X^{n}$, if and only if there
exists a deterministic coupling $P_{XY}\in C(P_{X},P_{Y})$ for which
$Y$ is a function of $X$. This is also equivalent to the fact that
$P_{X}$ and $P_{Y}$ have the same set of probability values. 
\end{enumerate}
\end{prop}
\begin{IEEEproof}
We first prove Statement 1). Suppose $\mathcal{G}(P_{X}^{n},P_{Y}^{n})=1.$
Then by the definition \eqref{eq:-45}, if $\mathcal{G}(P_{X}^{n},P_{Y}^{n})=1$,
then there exists a coupling of $P_{X}^{n},P_{Y}^{n}$ such that $Y^{n}$
is a deterministic function of $X^{n}$. Therefore, we have 
\begin{align}
nH_{\alpha}(X) & =H_{\alpha}(X^{n})\\
 & =H_{\alpha}(X^{n}Y^{n})\label{eq:-23}\\
 & \geq H_{\alpha}(Y^{n})\label{eq:-14}\\
 & =nH_{\alpha}(Y).\label{eq:-18}
\end{align}
This contradicts the assumption $H_{\alpha}(X)<H_{\alpha}(Y)$.

We next prove Statement 2). Obviously if $\mathcal{G}(P_{X},P_{Y})=1$,
then $\mathcal{G}(P_{X}^{n},P_{Y}^{n})=1$. Next we prove that if
$\mathcal{G}(P_{X}^{n},P_{Y}^{n})=1$ then $\mathcal{G}(P_{X},P_{Y})=1.$

Since in \eqref{eq:-18}{} we show that $H_{\alpha}(X)\geq H_{\alpha}(Y)$,
and as assumed, $H_{\alpha}(X)=H_{\alpha}(Y)$, the inequality in
\eqref{eq:-14}{} is in fact an equality, i.e., $H_{\alpha}(X^{n}Y^{n})=H_{\alpha}(X^{n})=H_{\alpha}(Y^{n})$.
That is, $Y^{n}$ is a function of $X^{n}$, and $X^{n}$ is also
a function of $Y^{n}$. Hence the mapping between $X^{n}$ and $Y^{n}$
is bijective, which further implies that $P_{X}^{n}$ and $P_{Y}^{n}$
have the same set of probability values.

Since $P_{X}^{n}$ and $P_{Y}^{n}$ have the same number of positive
probability values, the support sizes of $P_{X}$ and $P_{Y}$ are
equal. Denote the size as $k$, i.e., $k:=\left|\textrm{supp}(P_{X})\right|=\left|\textrm{supp}(P_{Y})\right|$.
Suppose $p_{1}\geq p_{2}\geq...\geq p_{k}$ and $q_{1}\geq q_{2}\geq...\geq q_{k}$
are the positive probability values of $P_{X}$ and $P_{Y}$, respectively,
ordered in a non-increasing fashion. Then the positive probability
values of $P_{X}^{n}$ and $P_{Y}^{n}$ must be $p_{1}^{n}\geq p_{1}^{n-1}p_{2}\geq...\geq p_{k}^{n-1}p_{k-1}\geq p_{k}^{n}$
and $q_{1}^{n}\geq q_{1}^{n-1}q_{2}\geq...\geq q_{k}^{n-1}q_{k-1}\geq q_{k}^{n}$.
Hence $p_{1}=q_{1},p_{2}=q_{2},p_{k-1}=q_{k-1},p_{k}=q_{k}$. Next
we prove $p_{i}=q_{i},i\in[3:k-2]$.

Remove $p_{1}^{i}p_{2}^{n-i},i\in[0:n]$ and $q_{1}^{i}q_{2}^{n-i},i\in[0:n]$
from the lists $p_{1}^{n}\geq p_{1}^{n-1}p_{2}\geq...\geq p_{k}^{n-1}p_{k-1}\geq p_{k}^{n}$
and $q_{1}^{n}\geq q_{1}^{n-1}q_{2}\geq...\geq q_{k}^{n-1}q_{k-1}\geq q_{k}^{n}$,
respectively. Then the maximum values among the resulting lists are
respectively $p_{1}^{n-1}p_{3}$ and $q_{1}^{n-1}q_{3}$. They must
be equal. Hence $p_{3}=q_{3}$. In the same way, we can show $p_{i}=q_{i},i\in[3:k-2]$. 
\end{IEEEproof}
\begin{prop}[Deterministic Coupling with Uniform $P_{X}$]
\label{prop:DetCoupling} If $P_{X}(x)=\frac{1}{M}$ for all $x\in[1:M]$
for some $M\in\mathbb{N}$, then for any $P_{Y}$, $\mathcal{G}(P_{X}^{n},P_{Y}^{n})=1$
if and only if $\mathcal{G}(P_{X},P_{Y})=1.$ That is, there exists
a deterministic coupling $P_{X^{n}Y^{n}}\in C(P_{X}^{n},P_{Y}^{n})$
for which $Y^{n}$ is a function of $X^{n}$, if and only if there
exists a deterministic coupling $P_{XY}\in C(P_{X},P_{Y})$ for which
$Y$ is a function of $X$. 
\end{prop}
\begin{IEEEproof}
We split the proof into three cases.

Case 1: If $P_{Y}(y_{1})$ is irrational for some $y_{1}\in\mathcal{Y}$
and $P_{Y}(y_{2})$ is rational for other some $y_{2}\in\mathcal{Y}$,
then $\mathcal{G}(P_{X}^{n},P_{Y}^{n})=1$ only if $P_{Y}(y_{1})=a_{1}^{\frac{1}{n}}$
with $a_{1}$ rational. Consider the term $P_{Y}(y_{1})\left(P_{Y}(y_{2})\right)^{n-1}$.
It is irrational since $P_{Y}(y_{1})$ is irrational and $P_{Y}(y_{2})$
is rational. Hence for any $n$, $P_{Y}(y_{1})\left(P_{Y}(y_{2})\right)^{n-1}$
is not a multiple of the probability value $P_{X}^{n}(x^{n})=\frac{1}{M^{n}}$.

Case 2: If $P_{Y}(y)$ is irrational for all $y\in\mathcal{Y}$, then
$\mathcal{G}(P_{X}^{n},P_{Y}^{n})=1$ only if for any $y$, $P_{Y}(y)=a_{y}^{\frac{1}{n}}$
with $a_{y}$ rational. Consider the terms $P_{Y}(y_{i})\left(P_{Y}(y_{j})\right)^{n-1}$.
Next we prove that there must exist some $(i,j)$ such that $P_{Y}(y_{i})\left(P_{Y}(y_{j})\right)^{n-1}$
is irrational. Suppose $P_{Y}(y_{i})\left(P_{Y}(y_{j})\right)^{n-1}$
is rational for any $(i,j)$. Then $\frac{P_{Y}(y_{i})}{P_{Y}(y_{j})}$
is rational since $\left(P_{Y}(y_{j})\right)^{n}$ is rational. That
is, $P_{Y}(y_{i})=k_{i,j}P_{Y}(y_{j})$ for some rational $k_{i,j}$.
Therefore, $\sum_{y\in\mathcal{Y}}P_{Y}(y)=\sum_{i=1}^{|\mathcal{Y}|}k_{i,1}P_{Y}(y_{1})=P_{Y}(y_{1})\sum_{i=1}^{|\mathcal{Y}|}k_{i,1}$
is irrational, since $P_{Y}(y_{1})$ is irrational and $\sum_{i=1}^{|\mathcal{Y}|}k_{i,1}$
is rational. However this contradicts the fact that $\sum_{y\in\mathcal{Y}}P_{Y}(y)=1$
is rational. Therefore, $P_{Y}(y_{i})\left(P_{Y}(y_{j})\right)^{n-1}$
is irrational for some $(i,j)$, and hence it cannot be composited
by the probability values $P_{X}^{n}(x^{n})=\frac{1}{M^{n}}$ for
any $n$.

Case 3: If $P_{Y}(y)$ is rational for all $y\in\mathcal{Y}$, then
denote $P_{Y}(y)=\frac{a}{b}$ with $a,b$ coprime, and $\mathcal{G}(P_{X}^{n},P_{Y}^{n})=1$
implies $\left(\frac{a}{b}\right)^{n}=k\left(\frac{1}{M}\right)^{n},$
i.e., $\left(\frac{Ma}{b}\right)^{n}=k.$ Hence $b|M$, otherwise,
$\left(\frac{Ma}{b}\right)^{n}\notin\mathbb{N}$ since $\frac{Ma}{b}\in\mathbb{Q}$
and $\frac{Ma}{b}\notin\mathbb{N}$. Assume $M=k'b$. Then $P_{Y}(y)=\frac{a}{b}=\frac{k'a}{M}.$
Hence $\mathcal{G}(P_{X},P_{Y})=1.$ On the other hand, it is obvious
that $\mathcal{G}(P_{X},P_{Y})=1$ implies $\mathcal{G}(P_{X}^{n},P_{Y}^{n})=1.$
Therefore, the theorem holds for the case where $P_{Y}(y)$ is rational
for all $y\in\mathcal{Y}$.

Combining the above three cases completes the proof. 
\end{IEEEproof}

\section{\label{sec:GeneralSCResolvability}Proof of Theorem \ref{thm:General Source-Channel Resolvability-1}{} }

We first prove the upper bound in \eqref{eq:-47}. To this end, we
need the following one-shot achievability result due to Cuff. 
\begin{lem}
\label{lem:General Source-Channel Resolvability} \cite[Theorem VII.1]{Cuff}
Given a source distribution $P_{W}$, codebook distribution $P_{X|W}$,
and channel $P_{Y|WX}$, let ${\cal C}$ be a randomly generated collection
of channel inputs $x(w)\in{\cal X}$, $w\in{\cal W}$, each drawn
independently according to $P_{X|W}$, and let $P_{Y|\mathcal{C}}$
be the output distribution induced by applying the codebook. For any
$\tau>0$, we have 
\begin{equation}
\mathbb{E}_{\mathcal{C}}\left|P_{Y|\mathcal{C}}-P_{Y}\right|\leq P({\cal A}_{\tau})+\frac{1}{2}e^{\tau/2},\label{eq:-9}
\end{equation}
where the expectation is with respect to the random codebook, and
\begin{equation}
{\cal A}_{\tau}:=\left\{ (w,x,y)\;:\;\log\frac{P_{W}(w)P_{Y|WX}(y|w,x)}{P_{Y}(y)}>\tau\right\} .
\end{equation}
\end{lem}
We have 
\begin{align}
 & \min_{f}|P_{Y_{f}}-P_{Z}|\nonumber \\
 & \leq\min_{P_{X|W}}\mathbb{E}_{\mathcal{C}}\left|P_{Y|\mathcal{C}}-P_{Z}\right|\label{eq:-56}\\
 & \leq\min_{P_{X|W}}\left\{ \mathbb{E}_{\mathcal{C}}\left|P_{Y|\mathcal{C}}-P_{Y}\right|+\left|P_{Y}-P_{Z}\right|\right\} \label{eq:-17}\\
 & \leq\min_{P_{X|W}}\left\{ \left|P_{Y}-P_{Z}\right|+P({\cal A}_{\tau})\right\} +\frac{1}{2}e^{\tau/2},\label{eq:-49}
\end{align}
where \eqref{eq:-56} follows since $\min_{P_{X|W}}\mathbb{E}_{\mathcal{C}}\left|P_{Y|\mathcal{C}}-P_{Z}\right|\geq\min_{c}\left|P_{Y|\mathcal{C}=c}-P_{Z}\right|=\min_{f}|P_{Y_{f}}-P_{Z}|$,
\eqref{eq:-17} follows from the triangle inequality, and \eqref{eq:-49}
follows from Lemma \ref{lem:General Source-Channel Resolvability}.

We next prove the lower bound in \eqref{eq:-47}. Observe that 
\begin{align}
 & P_{WY_{f}}\left\{ (w,y):\frac{P_{W}(w)P_{Y|WX}(y|w,f(w))}{P_{Y_{f}}(y)}>1\right\} \nonumber \\
 & =P_{WY_{f}}\left\{ (w,y):\frac{P_{W}(w)P_{Y|WX}(y|w,f(w))}{\sum_{w}P_{W}(w)P_{Y|WX}(y|w,f(w))}>1\right\} \\
 & =0.\label{eq:-50}
\end{align}
We relax the deterministic function $f$ to a random mapping $P_{X|W}$.
Then we get 
\begin{equation}
\min_{f}|P_{Z}-P_{Y_{f}}|\ge\min_{P_{X|W}:P_{WXY}({\cal A}_{0})=0}\left|P_{Z}-P_{Y}\right|.
\end{equation}

We finally prove the lower bound in \eqref{eq:-48}. By the maximal
coupling equality (Lemma \ref{lem:MaxCoupling}), there exists a coupling
$P_{Y_{f}Z}\in C(P_{Y_{f}},P_{Z})$ such that 
\begin{equation}
\mathbb{P}\left\{ Y_{f}\neq Z\right\} =|P_{Y_{f}}-P_{Z}|.
\end{equation}
Consider the joint distribution $P_{W}(w)P_{Y|WX}(y|w,f(w))P_{Z|Y_{f}}(z|y)$.
We have 
\begin{equation}
\mathbb{P}\left\{ \left(W,Y_{f}\right)\neq\left(W,Z\right)\right\} =\mathbb{P}\left\{ Y_{f}\neq Z\right\} .
\end{equation}
On the other hand, again by the maximal coupling equality, we have
\begin{align}
 & \mathbb{P}\left\{ \left(W,Y_{f}\right)\neq\left(W,Z\right)\right\} \nonumber \\
 & \geq\min_{P_{(W',Y_{f}),(W,Z)}\in C(P_{W,Y_{f}},P_{W,Z})}\mathbb{P}\left\{ \left(W',Y_{f}\right)\neq\left(W,Z\right)\right\} \\
 & =|P_{W,Y_{f}}-P_{W,Z}|.
\end{align}
Therefore, 
\begin{equation}
|P_{W,Y_{f}}-P_{W,Z}|\leq|P_{Y_{f}}-P_{Z}|.\label{eq:-54}
\end{equation}
Observe that 
\begin{align}
 & |P_{W,Y_{f}}-P_{W,Z}|\nonumber \\*
 & \geq P_{WZ}\left\{ (w,y):\frac{P_{W}(w)P_{Y|WX}(y|w,f(w))}{P_{Y}(y)}>1\right\} \nonumber \\
 & \qquad-P_{WY}\left\{ (w,y):\frac{P_{W}(w)P_{Y|WX}(y|w,f(w))}{P_{Y}(y)}>1\right\} \\
 & =P_{WZ}\left\{ (w,y):\frac{P_{W}(w)P_{Y|WX}(y|w,f(w))}{P_{Y}(y)}>1\right\} \label{eq:-51}\\
 & =\mathbb{P}\left\{ \frac{P_{W}(W)P_{Y|WX}(Z|W,f(W))}{P_{Z}(Z)}\frac{P_{Z}(Z)}{P_{Y}(Z)}>1\right\} \\
 & \geq\mathbb{P}\biggl\{\log\frac{P_{W}(W)P_{Y|WX}(Z|W,f(W))}{P_{Z}(Z)}>\tau,\nonumber \\
 & \qquad\log\frac{P_{Z}(Z)}{P_{Y}(Z)}>-\tau\biggr\}\\
 & =\mathbb{P}\left\{ \log\frac{P_{W}(W)P_{Y|WX}(Z|W,f(W))}{P_{Z}(Z)}>\tau\right\} \nonumber \\*
 & \qquad-\mathbb{P}\biggl\{\log\frac{P_{W}(W)P_{Y|WX}(Z|W,f(W))}{P_{Z}(Z)}>\tau,\nonumber \\*
 & \qquad\log\frac{P_{Z}(Z)}{P_{Y}(Z)}\leq-\tau\biggr\}\\
 & \geq\mathbb{P}\left\{ \log\frac{P_{W}(W)P_{Y|WX}(Z|W,f(W))}{P_{Z}(Z)}>\tau\right\} \nonumber \\
 & \qquad-\mathbb{P}\left\{ \log\frac{P_{Z}(Z)}{P_{Y}(Z)}\leq-\tau\right\} \\
 & \geq\mathbb{P}\left\{ \log\frac{P_{W}(W)P_{Y|WX}(Z|W,f(W))}{P_{Z}(Z)}>\tau\right\} -e^{-\tau}\\
 & \geq\min_{P_{WZ}\in C(P_{W},P_{Z})}P_{WZ}\biggl\{(w,z):\nonumber \\
 & \qquad\log\frac{P_{W}(w)P_{Y|WX}(z|w,f(w))}{P_{Z}(z)}>\tau\biggr\}-e^{-\tau}\label{eq:-52}\\
 & \geq\min_{P_{X|W}}\min_{P_{WZ}\in C(P_{W},P_{Z})}P_{WXZ}({\cal B}_{\tau})-e^{-\tau},\label{eq:-53}
\end{align}
where \eqref{eq:-51} follows from \eqref{eq:-50}, \eqref{eq:-52}
follows since we relax the distribution $P_{WZ}$ to any coupling
in $C(P_{W},P_{Z})$, and \eqref{eq:-53} follows since we relax the
deterministic function $f$ to a random mapping $P_{X|W}$.

Combining \eqref{eq:-54} and \eqref{eq:-53} gives us that for any
$f$, 
\begin{equation}
|P_{Y_{f}}-P_{Z}|\ge\min_{P_{WZ}\in C(P_{W},P_{Z})}\min_{P_{X|W}}P_{WXZ}({\cal B}_{\tau})-e^{-\tau}.
\end{equation}
This implies the lower bound in \eqref{eq:-48}.

\section{\label{sec:MinRenyiEntropyCoupling}Proof of Corollary \ref{cor:MinRenyiEntropyCoupling} }

\begin{figure}
\centering\includegraphics[width=0.5\textwidth]{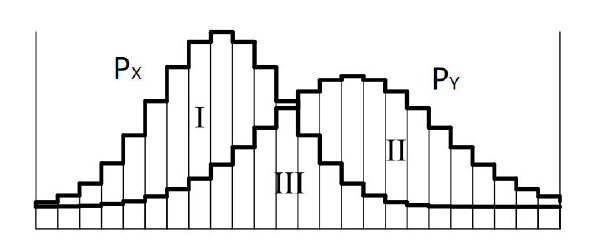}\caption{\label{fig:A-standard-maximal} A standard maximal coupling $P_{XY}$
of $(P_{X},P_{Y})$ for which $P_{Y|X}$ keeps the region III unchanged
and transfers probability mass from the region I to the region II
\cite{levin2017markov}.}
\end{figure}

We only need consider $\alpha<1$ case, since $H_{\alpha}(Y^{n}|X^{n})$
is decreasing in $\alpha$. By Theorem \ref{thm:equivalence}, we
can construct a maximal guessing coupling of $P_{X}^{n}$ and $P_{Y}^{n}$,
which cascades a probability distribution approximation code $f_{n}(x^{n})$
with a maximal coupling code $P_{Y^{n}|f_{n}(X^{n})}$. Here we adopt
a standard maximal coupling code (see Fig. \ref{fig:A-standard-maximal}).
The ``diagonal'' probabilities satisfy 
\begin{align}
 & P_{Y^{n}|f_{n}(X^{n})}(y^{n}|y^{n})\nonumber \\
 & =\begin{cases}
1, & P_{f_{n}(X^{n})}(y^{n})\leq P_{Y}^{n}(y^{n});\\
\frac{P_{Y}^{n}(y^{n})}{P_{f_{n}(X^{n})}(y^{n})}, & P_{f_{n}(X^{n})}(y^{n})>P_{Y}^{n}(y^{n}),
\end{cases}
\end{align}
for any $y^{n}\in\mathcal{Y}^{n}$, while the ``non-diagonal'' probabilities
can take on any value.

Then by Theorem \ref{thm:MaxGuessCoup}, we know 
\begin{equation}
\mathcal{G}(P_{X}^{n},P_{Y}^{n})\to1\label{eq:-15}
\end{equation}
at least exponentially fast as $n\to\infty$. The optimal exponent
is denoted as $\overline{\mathsf{E}}\left(P_{X},P_{Y}\right)$.

Denote $Z^{n}:=f_{n}(X^{n})$, 
\begin{align}
p & :=P_{Y^{n}}\left\{ z^{n}:P_{Y^{n}}(z^{n})<P_{Z^{n}}(z^{n})\right\} ,\quad\mbox{and}\\
\delta & :=|P_{Z^{n}}-P_{Y^{n}}|.
\end{align}
Then 
\begin{equation}
P_{Z^{n}}\left\{ z^{n}:P_{Y^{n}}(z^{n})<P_{Z^{n}}(z^{n})\right\} =\delta+p.
\end{equation}
Therefore, we have 
\begin{align}
 & e^{(1-\alpha)H_{\alpha}(Y^{n}|X^{n})}\nonumber \\
 & \leq e^{(1-\alpha)H_{\alpha}(Y^{n}|f_{n}(X^{n}))}\label{eqn:213}\\
 & =\sum_{z^{n},y^{n}}P_{Z^{n}}(z^{n})P_{Y^{n}|Z^{n}}^{\alpha}(y^{n}|z^{n})\\
 & =\sum_{z^{n}}P_{Z^{n}}(z^{n})\biggl\{1\left\{ P_{Y^{n}}(z^{n})\geq P_{Z^{n}}(z^{n})\right\} \nonumber \\
 & \qquad+\bigg(\Big(\frac{P_{Y^{n}}(z^{n})}{P_{Z^{n}}(z^{n})}\Big)^{\alpha}+\sum_{y^{n}\neq z^{n}}P_{Y^{n}|Z^{n}}^{\alpha}(y^{n}|z^{n})\bigg)\nonumber \\
 & \qquad\times1\left\{ P_{Y^{n}}(z^{n})<P_{Z^{n}}(z^{n})\right\} \biggr\}\\
 & \leq\sum_{z^{n}}P_{Z^{n}}(z^{n})\biggl\{1\left\{ P_{Y^{n}}(z^{n})\geq P_{Z^{n}}(z^{n})\right\} \nonumber \\
 & \qquad+\bigg(\Big(\frac{P_{Y^{n}}(z^{n})}{P_{Z^{n}}(z^{n})}\Big)^{\alpha}+\left(|\mathcal{Y}|^{n}-1\right)\bigg(\frac{1-\frac{P_{Y^{n}}(z^{n})}{P_{Z^{n}}(z^{n})}}{|\mathcal{Y}|^{n}-1}\bigg)^{\alpha}\bigg)\nonumber \\
 & \qquad\times1\left\{ P_{Y^{n}}(z^{n})<P_{Z^{n}}(z^{n})\right\} \biggr\}\label{eq:-19}\\
 & =\sum_{z^{n}}P_{Z^{n}}(z^{n})1\left\{ P_{Y^{n}}(z^{n})\geq P_{Z^{n}}(z^{n})\right\} \nonumber \\
 & \qquad+\sum_{z^{n}}P_{Z^{n}}(z^{n})\left(\frac{P_{Y^{n}}(z^{n})}{P_{Z^{n}}(z^{n})}\right)^{\alpha}1\left\{ P_{Y^{n}}(z^{n})<P_{Z^{n}}(z^{n})\right\} \nonumber \\
 & \qquad+\sum_{z^{n}}P_{Z^{n}}(z^{n})\left(|\mathcal{Y}|^{n}-1\right)^{1-\alpha}\left(1-\frac{P_{Y^{n}}(z^{n})}{P_{Z^{n}}(z^{n})}\right)^{\alpha}\nonumber \\
 & \qquad\times1\left\{ P_{Y^{n}}(z^{n})<P_{Z^{n}}(z^{n})\right\} \\
 & \leq P_{Z^{n}}\left\{ z^{n}:P_{Y^{n}}(z^{n})\geq P_{Z^{n}}(z^{n})\right\} \nonumber \\
 & \qquad+P_{Z^{n}}\left\{ z^{n}:P_{Y^{n}}(z^{n})<P_{Z^{n}}(z^{n})\right\} \nonumber \\
 & \qquad\times\left(\frac{\sum_{z^{n}}P_{Z^{n}}(z^{n})1\left\{ P_{Y^{n}}(z^{n})<P_{Z^{n}}(z^{n})\right\} }{P_{Z^{n}}\left\{ z^{n}:P_{Y^{n}}(z^{n})<P_{Z^{n}}(z^{n})\right\} }\frac{P_{Y^{n}}(z^{n})}{P_{Z^{n}}(z^{n})}\right)^{\alpha}\nonumber \\
 & \qquad+|\mathcal{Y}|^{\left(1-\alpha\right)n}P_{Z^{n}}\left\{ z^{n}:P_{Y^{n}}(z^{n})<P_{Z^{n}}(z^{n})\right\} \nonumber \\
 & \qquad\times\left(\frac{\sum_{z^{n}}P_{Z^{n}}(z^{n})1\left\{ P_{Y^{n}}(z^{n})<P_{Z^{n}}(z^{n})\right\} }{P_{Z^{n}}\left\{ z^{n}:P_{Y^{n}}(z^{n})<P_{Z^{n}}(z^{n})\right\} }\right)^{\alpha}\nonumber \\
 & \qquad\times\left(1-\frac{P_{Y^{n}}(z^{n})}{P_{Z^{n}}(z^{n})}\right)^{\alpha}\label{eq:-20}\\
 & =P_{Z^{n}}\left\{ z^{n}:P_{Y^{n}}(z^{n})\geq P_{Z^{n}}(z^{n})\right\} \nonumber \\
 & \qquad+P_{Z^{n}}^{1-\alpha}\left\{ z^{n}:P_{Y^{n}}(z^{n})<P_{Z^{n}}(z^{n})\right\} \nonumber \\
 & \qquad\times P_{Y^{n}}^{\alpha}\left\{ z^{n}:P_{Y^{n}}(z^{n})<P_{Z^{n}}(z^{n})\right\} \nonumber \\
 & \qquad+|\mathcal{Y}|^{\left(1-\alpha\right)n}P_{Z^{n}}^{1-\alpha}\left\{ z^{n}:P_{Y^{n}}(z^{n})<P_{Z^{n}}(z^{n})\right\} \nonumber \\
 & \qquad\times\left(\!\sum_{z^{n}}\!\!\left(P_{Z^{n}}(z^{n})\!-\!P_{Y^{n}}(z^{n})\right)\!1\left\{ P_{Y^{n}}(z^{n})\!<\!P_{Z^{n}}(z^{n})\right\} \!\right)^{\alpha}\\
 & =1-(\delta+p)+\left(\delta+p\right)^{1-\alpha}p^{\alpha}\nonumber \\
 & \qquad+|\mathcal{Y}|^{\left(1-\alpha\right)n}\left(\delta+p\right)^{1-\alpha}\delta^{\alpha}\label{eq:-21}\\
 & =1-(\delta+p)+\left(\delta+p\right)\left(\frac{p}{\delta+p}\right)^{\alpha}\nonumber \\
 & \qquad+|\mathcal{Y}|^{\left(1-\alpha\right)n}\left(\delta+p\right)^{1-\alpha}\delta^{\alpha}\\
 & \leq1+|\mathcal{Y}|^{\left(1-\alpha\right)n}\delta^{\alpha}\\
 & \leq1+|\mathcal{Y}|^{\left(1-\alpha\right)n}e^{-\alpha n\overline{\mathsf{E}}\left(P_{X},P_{Y}\right)}\\
 & \to1\quad\textrm{as }n\to\infty,
\end{align}
where \eqref{eq:-19} follows since $\sum_{y^{n}\neq z^{n}}P_{Y^{n}|Z^{n}}^{\alpha}(y^{n}|z^{n})$
for $\alpha<1$ is maximized by the uniform distribution 
\begin{equation}
P_{Y^{n}|Z^{n}}(y^{n}|z^{n})=\frac{1-\frac{P_{Y^{n}}(z^{n})}{P_{Z^{n}}(z^{n})}}{|\mathcal{Y}|^{n}-1}
\end{equation}
for $y^{n}\neq z^{n}$ (this point is similar to the fact that the
uniform distribution maximizes the Rényi entropy), and \eqref{eq:-20}
follows since $x^{\alpha}$ with $0<\alpha<1$ is concave in $x$.


\section{\label{sec:Resolvability}Proof of Proposition \ref{prop:Resolvability} }

According to the definition of $\overline{G}_{\mathsf{E}}(P_{Y|X},P_{Y})$
and Remark \ref{rem:It-is-easy}, we have $\overline{G}_{\mathsf{E}}(P_{Y|X},P_{Y})=\inf_{P_{\boldsymbol{X}}\in\mathcal{P}(P_{\boldsymbol{Y}|\boldsymbol{X}},P_{\boldsymbol{Y}})}\overline{H}(\boldsymbol{X})$.
Next we prove $\overline{G}_{\mathsf{E}}(P_{Y|X},P_{Y})=\lim_{n\to\infty}\min_{P_{X^{n}}\in\mathcal{P}(P_{Y|X}^{n},P_{Y}^{n})}\frac{1}{n}H(X^{n})$.

First it is easy to lower bound $\overline{G}_{\mathsf{E}}(P_{Y|X},P_{Y})$
as 
\begin{align}
\overline{G}_{\mathsf{E}}(P_{Y|X},P_{Y}) & =\inf_{P_{\boldsymbol{X}}\in\mathcal{P}(P_{\boldsymbol{Y}|\boldsymbol{X}},P_{\boldsymbol{Y}})}\overline{H}(\boldsymbol{X})\\
 & \geq\limsup_{n\to\infty}\inf_{P_{X^{n}}\in\mathcal{P}(P_{Y|X}^{n},P_{Y}^{n})}\frac{1}{n}H(X^{n})\label{eq:-57}
\end{align}
where \eqref{eq:-57} follows since $\overline{H}(\boldsymbol{X})\ge\limsup_{n\to\infty}\frac{1}{n}H(P_{X^{n}})$
for any $\boldsymbol{X}$ with a finite alphabet $\mathcal{X}$ (see
\cite[Theorem 1.7.2]{Han03}).

Assume $n=mk+l$ where $l<k$ with a fixed number $k$. For the first
$mk$ symbols, we use the code $f_{mk}$ in \cite{Han} to exactly
synthesize $P_{X^{k}}^{m}$ with $P_{X^{k}}\in\mathcal{P}(P_{Y|X}^{k},P_{Y}^{k})$.
By Corollary \ref{cor:ExactIR-1-1}, we have that if the code rate
$R^{(1)}>\frac{1}{k}H(X^{k})$ 
\begin{equation}
\lim_{m\to\infty}\mathbb{P}\left\{ X^{mk}=f_{mk}(M_{mk}^{(1)})\right\} =1,
\end{equation}
where $M_{mk}^{(1)}\sim\mathsf{Unif}[1:e^{mkR^{(1)}}]$. On the other
hand, for each of the last $l$ symbols, we use a single-letter code
$f_{1}$ to approximately synthesize $P_{X}$ with $P_{X}\in\mathcal{P}(P_{Y|X},P_{Y})$.
Here we assume $f_{1}$ satisfies $\left|P_{X}-P_{f_{1}(M^{(2)})}\right|\leq|\mathcal{X}|e^{-R_{m}^{(2)}}$
where $M^{(2)}\sim\mathsf{Unif}[1:e^{R_{m}^{(2)}}]$. Obviously, there
exists at least one code $f_{1}$ satisfying this condition. By the
equivalence \eqref{eq:-11}, we know that there exists a coupling
$P_{M^{(2)}X}\in C(P_{M^{(2)}},P_{X})$ satisfying $\min_{f_{1}}\mathbb{P}\left\{ X\neq f_{1}(M^{(2)})\right\} \leq|\mathcal{X}|e^{-R_{m}^{(2)}}$.

For this concatenated code, we have that the overall code rate is
$\frac{mkR^{(1)}+lR_{m}^{(2)}}{mk+l}$, and the overall minimum guessing
error probability is upper bounded as 
\begin{align}
 & \min_{f_{n}}\mathbb{P}\left\{ X^{n}\neq f_{n}(M_{n})\right\} \nonumber \\
 & \leq\mathbb{P}\left\{ X^{mk}\neq f_{mk}(M_{mk}^{(1)})\right\} +l\min_{f_{1}}\mathbb{P}\left\{ X\neq f_{1}(M^{(2)})\right\} \\
 & \leq\mathbb{P}\left\{ X^{mk}\neq f_{mk}(M_{mk}^{(1)})\right\} +k|\mathcal{X}|e^{-R_{m}^{(2)}},
\end{align}
where $M_{n}=(M_{mk}^{(1)},(M^{(2)})^{l})$. We choose $R^{(1)},R_{m}^{(2)}$
such that $R^{(1)}>\frac{1}{k}H(X^{k})$ for some $P_{X^{k}}\in\mathcal{P}(P_{Y|X}^{k},P_{Y}^{k})$,
$\lim_{n\to\infty}R_{m}^{(2)}=\infty$ and $R_{m}^{(2)}=o(m)$ (e.g.,
$R_{m}^{(2)}=\sqrt{m}$). Then for fixed $k$, the overall rate $\lim_{n\to\infty}\frac{mkR^{(1)}+lR_{m}^{(2)}}{mk+l}=R^{(1)}$,
and the overall minimum guessing error probability 
\begin{align}
 & \lim_{n\to\infty}\min_{f_{n}}\mathbb{P}\left\{ X^{n}\neq f_{n}(M_{n})\right\} \nonumber \\
 & \leq\lim_{m\to\infty}\mathbb{P}\left\{ X^{mk}\neq f_{mk}(M_{mk}^{(1)})\right\} +\lim_{m\to\infty}k|\mathcal{X}|e^{-R_{m}^{(2)}}\\
 & =0.
\end{align}
This implies we get a channel resolvability code with rate $\min_{P_{X^{k}}\in\mathcal{P}(P_{Y|X}^{k},P_{Y}^{k})}\frac{1}{k}H(X^{k})$.
Since $k$ is arbitrary, we have 
\begin{align}
\overline{G}_{\mathsf{E}}(P_{Y|X},P_{Y})\leq & \liminf_{k\to\infty}\min_{P_{X^{k}}\in\mathcal{P}(P_{Y|X}^{k},P_{Y}^{k})}\frac{1}{k}H(X^{k}).\label{eq:-40-1-2}
\end{align}
Combining \eqref{eq:-57} and \eqref{eq:-40-1-2}, we have $\overline{G}_{\mathsf{E}}(P_{Y|X},P_{Y})=\lim_{n\to\infty}\min_{P_{X^{n}}\in\mathcal{P}(P_{Y|X}^{n},P_{Y}^{n})}\frac{1}{n}H(X^{n})$.

\section{\label{sec:Perfect-stealth-secrecy}Proof of Theorem \ref{thm:Perfect-stealth-secrecy} }

Proof of ``if'': Suppose that $Q_{X}$ and $Q_{X}'$ induce the
same $P_{Z}$ through $P_{Z|X}$ but induce two different distributions
of $Y$ through $P_{Y|X}$. Define $P_{X}^{(B)}:=BQ_{X}+(1-B)Q_{X}'$,
where $B\in[0,1]$ with distribution $P_{B}$ such that $P_{B}(0)P_{B}(1)>0$.
Consider a new wiretap channel $P_{YZ|B}=P_{Y|B}P_{Z}$. 
\begin{equation}
C_{1}(P_{Z})\geq\max_{P_{UX}:U\perp Z,P_{X}\in\mathcal{P}(P_{Z})}I(U;Y)\geq I(B;Y)>0.
\end{equation}
The last inequality follows the following argument via contradiction.
Suppose $I(B;Y)=0$, then $B\perp Y$. Hence $P_{Y|B=0}=P_{Y|B=1}=P_{Y}$.
This contradicts with the assumption that $Q_{X}$ and $Q_{X}'$ induce
two different distributions of $Y$ through $P_{Y|X}$.

Proof of ``only if'': We prove this by contradiction. That is, we
need to show if for any two distributions $Q_{X}$ and $Q_{X}'$ that
induce the same $P_{Z}$ through $P_{Z|X}$, they must induce a same
distribution of $Y$ through $P_{Y|X}$, then the perfect stealth-secrecy
capacity is zero.

Suppose $\boldsymbol{P}_{Z|X}\boldsymbol{Q}_{X}=\boldsymbol{P}_{Z}$
has infinitely many solutions; otherwise, by Lemma \ref{lem:Properties:},
$P_{Z|X}$ is a full-rank channel or $\boldsymbol{P}_{Z|X}\boldsymbol{Q}_{X}=\boldsymbol{P}_{Z}$
has a single unique solution $P_{X}$ which is a degenerate distribution.
For the former case, by Theorem \ref{thm:FullRankChannels} we know
that the perfect stealth-secrecy capacity is zero. For the latter
case, since $nR_{1}\le I(Y^{n};X^{n})=0$, the perfect stealth-secrecy
capacity is also zero. So we only need to consider the case that $\boldsymbol{P}_{Z|X}\boldsymbol{Q}_{X}=\boldsymbol{P}_{Z}$
has infinitely many solutions.

In addition, note that we also only need to consider the case that
there exists a solution to $\boldsymbol{P}_{Z|X}\boldsymbol{Q}_{X}=\boldsymbol{P}_{Z}$
which is an interior point of the probability simplex $\left\{ \boldsymbol{P}_{X}:\sum_{x}P_{X}(x)=1,P_{X}(x)\ge0\right\} $.
This is because if all the solutions to $\boldsymbol{P}_{Z|X}\boldsymbol{Q}_{X}=\boldsymbol{P}_{Z}$
are at the boundary of the probability simplex $\left\{ \boldsymbol{P}_{X}:\sum_{x}P_{X}(x)=1,P_{X}(x)\ge0\right\} $,
then there exists a set $\mathcal{X}_{0}$ such that the solutions
satisfy $Q_{X}(x_{0})=0$ for any $x_{0}\in\mathcal{X}_{0}$. Hence
remove the corresponding columns of $\boldsymbol{P}_{Z|X}$ and the
corresponding rows of $\boldsymbol{P}_{X}$, and denote the resulting
matrix and vector as $\overline{\boldsymbol{P}}_{Z|X}$ and $\overline{\boldsymbol{Q}}_{X}$
respectively, then we get equation $\overline{\boldsymbol{P}}_{Z|X}\overline{\boldsymbol{Q}}_{X}=\boldsymbol{P}_{Z}$.
For this new equation, there exists a solution which is an interior
point of the probability simplex $\left\{ \boldsymbol{P}_{X}\in\mathcal{P}(\mathcal{X}\backslash\mathcal{X}_{0}):\sum_{x}P_{X}(x)=1,P_{X}(x)\ge0\right\} $.

Suppose $P_{Y}$ is the distribution induced by $Q_{X}'$ through
$P_{Y|X}$ where $Q_{X}'$ is a distribution inducing $P_{Z}$ through
$P_{Z|X}$. By subtracting $\boldsymbol{Q}_{X}'$ from the solutions
to $\boldsymbol{P}_{Z|X}\boldsymbol{Q}_{X}=\boldsymbol{P}_{Z}$ and
$\boldsymbol{P}_{Y|X}\boldsymbol{Q}_{X}=\boldsymbol{P}_{Y}$, we get
the equation $\boldsymbol{P}_{Z|X}\widehat{\boldsymbol{Q}}=0$ and
$\boldsymbol{P}_{Y|X}\widehat{\boldsymbol{Q}}=0$ (here $\widehat{\boldsymbol{Q}}$
denotes $\boldsymbol{Q}_{X}-\boldsymbol{Q}_{X}'$). Denote $\mathcal{S}_{Z}$
as the set of solutions to $\boldsymbol{P}_{Z|X}\widehat{\boldsymbol{Q}}=0$
and $\mathcal{S}_{Y}$ as the set of solutions to $\boldsymbol{P}_{Y|X}\widehat{\boldsymbol{Q}}=0$.
Then by assumption, $\mathcal{S}_{Z}\subseteq\mathcal{S}_{Y}$.

Note that the set of solutions to $\boldsymbol{P}_{Z|X}\boldsymbol{Q}_{X}=\boldsymbol{P}_{Z}$
(with $\boldsymbol{Q}_{X}$ constrained to be a probability distribution)
is the intersection of the set of solutions to $\boldsymbol{P}_{Z|X}\boldsymbol{Q}=\boldsymbol{P}_{Z}$
without the probability constraint on $\boldsymbol{Q}$ and the probability
simplex $\left\{ \boldsymbol{P}_{X}:\sum_{x}P_{X}(x)=1,P_{X}(x)\ge0\right\} $.
If there exists a solution $\boldsymbol{Q}_{X}^{*}$ to $\boldsymbol{P}_{Z|X}\boldsymbol{Q}_{X}=\boldsymbol{P}_{Z}$
which is an interior point of the probability simplex $\left\{ \boldsymbol{P}_{X}:\sum_{x}P_{X}(x)=1,P_{X}(x)\ge0\right\} $,
then the subspace of $\mathbb{R}^{|\mathcal{X}|}$ spanned by the
set $\mathcal{S}_{Z}$ is the same to the orthogonal complement of
the subspace of $\mathbb{R}^{|\mathcal{X}|}$ spanned by the rows
of $\boldsymbol{P}_{Z|X}$, and also the same to the set of the solutions
to $\boldsymbol{P}_{Z|X}\boldsymbol{Q}=0$ (without the probability
constraint). Since $\mathcal{S}_{Z}\subseteq\mathcal{S}_{Y}$ (or
equivalently, $\left(\mathcal{S}_{Z}+\boldsymbol{Q}_{X}'\right)\subseteq\left(\mathcal{S}_{Y}+\boldsymbol{Q}_{X}'\right)$),
$\boldsymbol{Q}_{X}^{*}$ is also a solution to $\boldsymbol{P}_{Y|X}\boldsymbol{Q}_{X}=\boldsymbol{P}_{Y}$.
Since $\boldsymbol{Q}_{X}^{*}$ is an interior point of the probability
simplex, similarly, we have that the subspace of $\mathbb{R}^{|\mathcal{X}|}$
spanned by the set $\mathcal{S}_{Y}$ is the same to the set of the
solutions to $\boldsymbol{P}_{Y|X}\boldsymbol{Q}=0$ (without the
probability constraint). Denote $\mathcal{S}_{Z}'$ as the set of
solutions to $\boldsymbol{P}_{Z|X}\boldsymbol{Q}=0$ (without the
probability constraint) and $\mathcal{S}_{Y}'$ as the set of solutions
to $\boldsymbol{P}_{Y|X}\boldsymbol{Q}=0$ (without the probability
constraint). Then $\mathcal{S}_{Z}'\subseteq\mathcal{S}_{Y}'$. 

A vector $\boldsymbol{Q}$ is a solution to $\boldsymbol{P}_{Z|X}\boldsymbol{Q}=0$
(without probability constraint) if and only if it lies in the orthogonal
complement of the subspace of $\mathbb{R}^{|\mathcal{X}|}$ spanned
by the rows of $\boldsymbol{P}_{Z|X}$. Hence $\mathcal{S}_{Z}'\subseteq\mathcal{S}_{Y}'$
means that the orthogonal complement of the row space of $\boldsymbol{P}_{Z|X}$
is a subset of that of the row space of $\boldsymbol{P}_{Y|X}$. It
means that the row space of $\boldsymbol{P}_{Y|X}$ is a subset of
the row space of $\boldsymbol{P}_{Z|X}$. Hence every row of $\boldsymbol{P}_{Y|X}$
is a linear combination of the rows of $\boldsymbol{P}_{Z|X}$. Thus,
$\boldsymbol{P}_{Y|X}=\boldsymbol{A}\boldsymbol{P}_{Z|X}$ for some
matrix $\boldsymbol{A}$. On the other hand, observe that $\boldsymbol{A}\boldsymbol{P}_{Z|X}\boldsymbol{Q}_{X}=\boldsymbol{A}\boldsymbol{P}_{Z}$,
$\boldsymbol{A}\boldsymbol{P}_{Z|X}=\boldsymbol{P}_{Y|X}$, and $\boldsymbol{P}_{Y|X}\boldsymbol{Q}_{X}=\boldsymbol{P}_{Y}$.
Hence $\boldsymbol{A}\boldsymbol{P}_{Z}=\boldsymbol{P}_{Y}$.

Now we prove the following property for any $n$: for all distributions
$Q_{X^{n}}$ that induce $P_{Z}^{n}$ through $P_{Z|X}^{n}$, they
must induce the same distribution of $Y^{n}$ through $P_{Y|X}^{n}$.
Consider the equation 
\begin{equation}
\boldsymbol{P}_{Z|X}^{\otimes n}\boldsymbol{Q}_{X^{n}}=\boldsymbol{P}_{Z}^{\otimes n}.
\end{equation}
Multiply $\boldsymbol{A}^{\otimes n}$ at both sides, then we get
\begin{equation}
\boldsymbol{A}^{\otimes n}\boldsymbol{P}_{Z|X}^{\otimes n}\boldsymbol{Q}_{X^{n}}=\boldsymbol{A}^{\otimes n}\boldsymbol{P}_{Z}^{\otimes n}
\end{equation}
which is equivalent to 
\begin{equation}
\left(\boldsymbol{A}\boldsymbol{P}_{Z|X}\right)^{\otimes n}\boldsymbol{Q}_{X^{n}}=\left(\boldsymbol{A}\boldsymbol{P}_{Z}\right)^{\otimes n}.
\end{equation}
Substituting $\boldsymbol{A}\boldsymbol{P}_{Z|X}=\boldsymbol{P}_{Y|X}$
and $\boldsymbol{A}\boldsymbol{P}_{Z}=\boldsymbol{P}_{Y}$, we get
\begin{equation}
\boldsymbol{P}_{Y|X}^{\otimes n}\boldsymbol{Q}_{X^{n}}=\boldsymbol{P}_{Y}^{\otimes n}.
\end{equation}
Observe $\boldsymbol{A},P_{Z}$ are fixed, hence $P_{Y}$ is fixed
as well. This means for all distributions $Q_{X^{n}}$ that induce
$P_{Z}^{n}$ through $P_{Z|X}^{n}$, they must induce the same distribution
$P_{Y}^{n}$ through $P_{Y|X}^{n}$.

Using on the property above, we return to proving that the perfect
stealth-secrecy capacity is zero. Note that by the secrecy constraint,
\begin{equation}
P_{Z}^{n}(\cdot)=P_{Z^{n}|M}(\cdot|m)=\sum_{x^{n}}P_{Z|X}^{n}(\cdot|x^{n})P_{X^{n}|M}(x^{n}|m)\label{eqn:prop}
\end{equation}
for any $m$. Hence for any $m$, $P_{X^{n}|M}(\cdot|m)$ is a distribution
that induces $P_{Z}^{n}$ through $P_{Z|X}^{n}$. By the property
stated in \eqref{eqn:prop}, we have that for different $m$, $P_{X^{n}|M}(\cdot|m)$
induces the same distribution of $Y^{n}$ through $P_{Y|X}^{n}$,
i.e., $P_{Y^{n}|M}(\cdot|m)=\sum_{x^{n}}P_{Y|X}^{n}(\cdot|x^{n})P_{X^{n}|M}(x^{n}|m)$
does not depend on $m$. Consequently, $Y^{n}$ is independent of
$M$, i.e., 
\begin{align}
nR_{1} & \le I(Y^{n};M)=0.
\end{align}

\subsection*{Acknowledgments}

The authors would like to thank Prof. Igal Sason for pointing out
reference \cite{prelov2015coupling}. The authors also thank the reviewers
and the editor for their suggestions to improve the quality of the
paper.

 \bibliographystyle{unsrt}
\bibliography{ref}

\begin{IEEEbiographynophoto}{Lei Yu}
received the B.E. and Ph.D. degrees, both in electronic engineering,
from University of Science and Technology of China (USTC) in 2010
and 2015, respectively. From 2015 to 2017, he was a postdoctoral researcher
at the Department of Electronic Engineering and Information Science
(EEIS), USTC. Currently, he is a research fellow at the Department
of Electrical and Computer Engineering, National University of Singapore.
His research interests include information theory, probability theory,
and security. 
\end{IEEEbiographynophoto}

\begin{IEEEbiographynophoto}{Vincent Y.\ F.\ Tan}
(S'07-M'11-SM'15) was born in Singapore in 1981. He is currently
an Associate Professor in the Department of Electrical and Computer
Engineering and the Department of Mathematics at the National University
of Singapore (NUS). He received the B.A.\ and M.Eng.\ degrees in
Electrical and Information Sciences from Cambridge University in 2005
and the Ph.D.\ degree in Electrical Engineering and Computer Science
(EECS) from the Massachusetts Institute of Technology (MIT) in 2011.
His research interests include information theory, machine learning,
and statistical signal processing.

Dr.\ Tan received the MIT EECS Jin-Au Kong outstanding doctoral thesis
prize in 2011, the NUS Young Investigator Award in 2014, the NUS Engineering
Young Researcher Award in 2018, and the Singapore National Research
Foundation (NRF) Fellowship (Class of 2018). He is also an IEEE Information
Theory Society Distinguished Lecturer. He has authored a research
monograph on {\em ``Asymptotic Estimates in Information Theory
with Non-Vanishing Error Probabilities''} in the Foundations and
Trends in Communications and Information Theory Series (NOW Publishers).
He is currently an Associate Editor of the IEEE Transactions on Signal
Processing. 
\end{IEEEbiographynophoto}

\end{document}